# Autonomous Threat Hunting: A Future Paradigm for AI-Driven Threat Intelligence

Siva Raja Sindiramutty

*Taylors University, Subang Jaya Malaysia*

siva.sindiramutty@taylors.edu.my

**ABSTRACT**

*The evolution of cybersecurity has spurred the emergence of autonomous threat hunting as a pivotal paradigm in the realm of AI-driven threat intelligence. This review navigates through the intricate landscape of autonomous threat hunting, exploring its significance and pivotal role in fortifying cyber defense mechanisms. Delving into the amalgamation of artificial intelligence (AI) and traditional threat intelligence methodologies, this paper delineates the necessity and evolution of autonomous approaches in combating contemporary cyber threats. Through a comprehensive exploration of foundational AI-driven threat intelligence, the review accentuates the transformative influence of AI and machine learning on conventional threat intelligence practices. It elucidates the conceptual framework underpinning autonomous threat hunting, spotlighting its components, and the seamless integration of AI algorithms within threat hunting processes. Furthermore, the review scrutinizes state-of-the-art AI techniques deployed in autonomous threat hunting, ranging from machine learning models (supervised, unsupervised, and reinforcement learning) to natural language processing (NLP), sentiment analysis, and deep learning architectures. Insightful discussions on challenges encompassing scalability, interpretability, and ethical considerations in AI-driven models enrich the discourse. Moreover, through illuminating case studies and evaluations, this paper showcases real-world implementations, underscoring success stories and lessons learned by organizations adopting AI-driven threat intelligence. In conclusion, this review consolidates key insights, emphasizing the substantial implications of autonomous threat hunting for the future of cybersecurity. It underscores the significance of continual research and collaborative efforts in harnessing the potential of AI-driven approaches to fortify cyber defenses against evolving threats.*

## 1. INTRODUCTION

**Background And Motivation for Autonomous Threat Hunting**

The landscape of cybersecurity has undergone substantial evolution with the proliferation of sophisticated threats targeting systems and networks. Traditional cybersecurity measures often struggle to keep pace with the rapidly advancing threat landscape, prompting the emergence of autonomous threat hunting as a proactive defense mechanism. This approach involves leveraging artificial intelligence (AI) and machine learning (ML) algorithms to autonomously detect, analyze, and mitigate potential threats in real time.

The escalating complexity and frequency of cyber threats necessitate a more proactive stance in cybersecurity defense mechanisms [1, 2, 497]. Manual threat detection methods have proven insufficient, leading to delays in identifying and responding to emerging threats. The need for rapid threat identification and mitigation underscores the importance of autonomous threat hunting in fortifying cyber defenses [3, 4, 498]. Furthermore, the dynamic nature of cyber threats requires continuous monitoring and analysis, a task that surpasses human capabilities alone [5, 6]. Autonomous threat hunting systems excel in processing vast amounts of data, identifying patterns, and discerning anomalies that may signal potential threats, thereby enhancing overall threat intelligence [7, 499]. Another crucial aspect motivating the adoption of autonomous threat hunting is the imperative need to minimize response times in cyber incidents [8,9]. Swift identification and containment of threats are critical in preventing widespread damage and minimizing the impact of cyber-attacks on organizations [10,11, 500]. Autonomous systems, equipped with advanced algorithms, can significantly reduce response times, thereby limiting the potential fallout of cyber incidents [12, 13, 501] Moreover, the evolving nature of cyber threats demands a shift from a reactive to a proactive cybersecurity posture [14, 15, 502] Traditional security approaches primarily focus on responding to known threats, leaving systems vulnerable to novel and emerging risks. Autonomous threat hunting systems proactively seek out potential threats, enabling organizations to stay ahead of adversaries and anticipate their tactics [16, 503]. The integration of autonomous threat hunting aligns with the concept of continuous monitoring and assessment, a fundamental principle in modern cybersecurity frameworks [17, 504]. By employing AI-driven systems, organizations can achieve a continuous and comprehensive evaluation of their security posture, enabling prompt identification and mitigation of vulnerabilities and potential threats [18, 505].

In conclusion, the escalating complexity and sophistication of cyber threats, coupled with the limitations of traditional cybersecurity approaches, underscore the crucial need for autonomous threat hunting. Leveraging AI and ML technologies, these systems provide proactive, real-time threat detection, thereby bolstering cybersecurity defenses and enabling organizations to stay ahead in the ever-evolving landscape of cyber threats.

**Evolution Of Threat Intelligence and The Role Of AI**

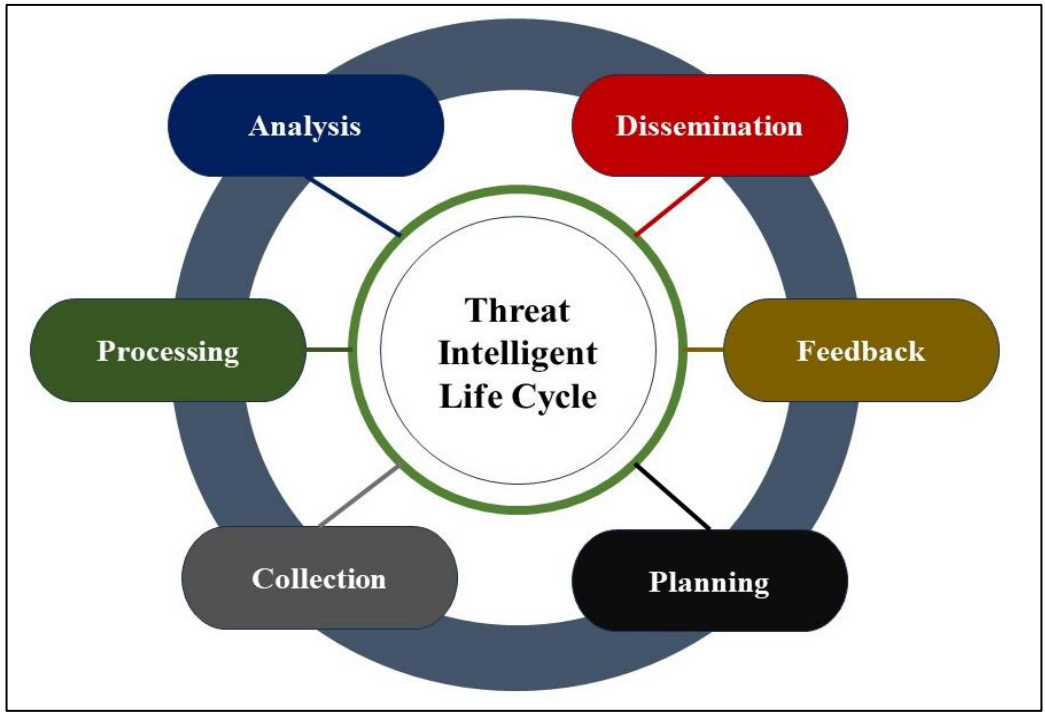

Figure 1.0: Threat Intelligent Life Cycle

Threat intelligence has significantly evolved over the years, transitioning from manual data analysis to leveraging advanced technologies, prominently Artificial Intelligence (AI). The evolution of threat intelligence denotes a progression from a reactive to a proactive approach in cybersecurity [19, 506] Initially, threat intelligence relied heavily on human analysts sifting through data, but the exponential growth of data made this approach insufficient and time-consuming [20, 21, 507] With the advent of AI, this landscape has witnessed a remarkable transformation. Figure 1.0 shows threat intelligent life cycle.

AI plays a pivotal role in threat intelligence by enabling the processing of vast amounts of data at unparalleled speeds. Machine learning algorithms can identify patterns and anomalies in data that might go unnoticed by human analysts [22, 23, 508]. This assists in the early detection and mitigation of potential threats, offering a proactive stance against cyber-attacks. Furthermore, AI-powered threat intelligence systems continuously learn and adapt, improving their efficacy over time [24, 25, 509]. One of the significant advantages of AI in threat intelligence is its ability to automate various tasks, thereby freeing up human analysts to focus on more complex and strategic activities [26, 27, 510]. AI-driven tools can perform repetitive tasks such as data collection, analysis, and correlation more efficiently, allowing analysts to concentrate on decision-making and crafting better security strategies [28, 29, 511]. This collaborative effort between AI and human analysts maximizes the effectiveness of threat intelligence operations. Moreover, AI augments threat intelligence by providing predictive capabilities. Through historical data analysis, AI models can forecast potential threats and vulnerabilities, enabling organizations to proactively strengthen their defenses [30, 31,32, 512]. This proactive approach helps in preemptively mitigating risks before they escalate into significant security breaches. However, the integration of AI in threat intelligence also poses challenges such as the potential for adversarial attacks targeting AI models [33, 34, 513] Adversaries can manipulate AI algorithms, leading to false identifications or evading detection. Thus, ensuring the security and robustness of AI systems in threat intelligence remains an ongoing concern [35, 36, 37, 514].

In conclusion, the evolution of threat intelligence has been greatly influenced by the integration of AI technologies. AI-driven capabilities enable proactive threat detection, automation of tasks, and predictive analysis, significantly enhancing the effectiveness of cybersecurity measures. Nonetheless, ensuring the resilience of AI systems against potential adversarial attacks remains a critical focus in leveraging AI for threat intelligence.

**Statement Of the Research Problem and The Need For Autonomous Approaches**

The cybersecurity landscape has become increasingly complex and dynamic, with threats evolving in sophistication and scale. Traditional threat intelligence methodologies often struggle to keep pace with these rapid advancements, leading to a critical research problem: the inability to detect, analyze, and mitigate emerging threats in real-time swiftly and effectively. The need for autonomous approaches in threat hunting arises from this persistent challenge. Human-operated systems are limited in their capacity to handle the vast amount of data generated by diverse sources and to discern nuanced patterns indicative of potential threats. Moreover, the time-sensitive nature of cyber threats demands a proactive and automated response. Autonomous threat hunting aims to bridge this gap by leveraging the capabilities of AI-driven systems. These systems can autonomously collect, process, and analyze large volumes of data, thereby enabling the identification of subtle indicators of compromise and previously unseen attack vectors. The urgency to develop autonomous approaches is underscored by the growing necessity for adaptive, scalable, and swift threat detection and mitigation mechanisms. Therefore, the research problem revolves around the inefficiency of conventional methods in coping with the speed and complexity of modern cyber threats, emphasizing the pressing need for autonomous approaches empowered by AI to fortify cybersecurity measures.

**Objectives And Research Contribution of The Review Paper**

The primary objectives of this review paper are threefold: Firstly, to comprehensively elucidate the evolving landscape of threat intelligence, spotlighting the pivotal role AI plays in shaping its trajectory. The paper seeks to outline the historical progression of threat intelligence methodologies and articulate how AI and machine learning are revolutionizing conventional paradigms. Secondly, this paper aims to delineate the conceptual framework of autonomous threat hunting, offering a definitive definition, and elucidating its crucial components. It endeavors to delve into the integration of AI algorithms within threat hunting processes, detailing the intricate framework and its operational dynamics. Lastly, this review paper endeavors to highlight the state-of-the-art AI techniques pertinent to autonomous threat hunting, presenting an in-depth analysis of machine learning models, natural language processing, sentiment analysis, and deep learning architectures. It aims to provide insights into the practical application of these techniques in fortifying cybersecurity landscapes. The research contribution of this paper lies in its synthesis of existing knowledge, providing a comprehensive and structured overview of AI-driven threat intelligence. By amalgamating foundational theories with contemporary advancements, it aspires to offer a holistic understanding of autonomous threat hunting. Additionally, the paper strives to identify challenges, present real-world case studies, propose evaluation metrics, and forecast future trends, thereby offering a robust foundation for further research and practical implementation in cybersecurity domains.

**Research Paper Organization**

This review paper is structured to comprehensively delve into the transformative realm of autonomous threat hunting facilitated by AI-driven approaches. It is divided into distinct sections to systematically explore the evolving landscape of cybersecurity through the lens of autonomous threat detection and mitigation.

**1. Introduction:** The paper commences with an introduction outlining the fundamental motivations behind the evolution of autonomous threat hunting. It traces the trajectory of threat intelligence methodologies and underscores the pivotal role of AI in revolutionizing these practices. Additionally, it poses the research problem, emphasizing the necessity of autonomous approaches, sets the objectives, and outlines the contribution of this review paper.

**2. Foundations of AI-Driven Threat Intelligence:** This section provides a foundational understanding by elucidating traditional threat intelligence methodologies, introducing AI and machine learning in cybersecurity, and delineating the transformative role of AI in reshaping conventional threat intelligence practices.

**3. Autonomous Threat Hunting: Conceptual Framework:** The subsequent section intricately explores the concept of autonomous threat hunting. It defines its scope, dissects the essential components of an autonomous threat hunting system, explicates the integration of AI algorithms within threat hunting processes, and delves into the framework/process in detail.

**4. State-of-the-Art AI Techniques in Autonomous Threat Hunting:** This segment scrutinizes cutting-edge AI techniques utilized in autonomous threat hunting, encompassing machine learning models, natural language processing (NLP), sentiment analysis, and deep learning architectures, explicating their applications in threat detection and intelligence extraction.

**5. Challenges in Autonomous Threat Hunting:** Addressing the multifaceted landscape of autonomous threat hunting, this section illuminates the challenges encompassing scalability, interpretability, ethical considerations, and potential biases in AI algorithms.

**6. Case Studies and Applications:** Highlighting real-world implementations, success stories, and lessons learned, this section illustrates the practical applications and efficacy of AI-driven threat intelligence through case studies of organizations.

**7. Evaluation Metrics and Performance Benchmarks:** Focusing on assessing effectiveness, this section delineates metrics and conducts a comparative analysis between AI-driven systems and traditional methods.

**8. Future Directions and Emerging Trends:** This segment explores forthcoming advancements in autonomous threat hunting, highlights emerging technologies, and identifies potential challenges, laying the groundwork for future research avenues.

**9. Conclusion:** The paper culminates by summarizing key insights, implications of autonomous threat hunting on cybersecurity, and advocating for further research and implementation.

## 2. FOUNDATIONS OF AI-DRIVEN THREAT INTELLIGENCE

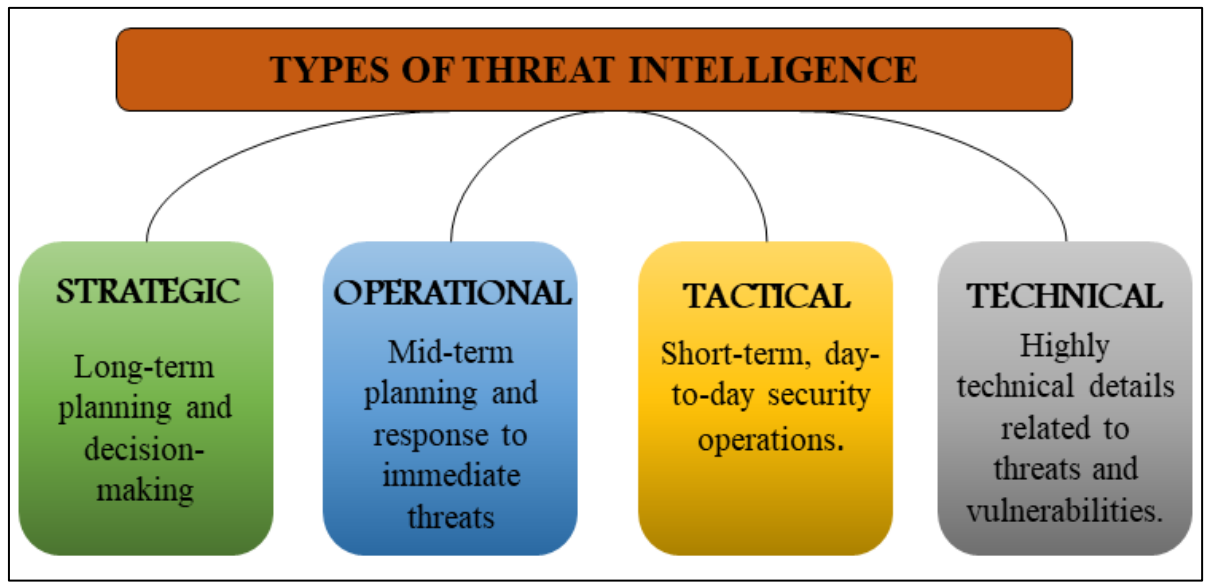

Figure 2: Types of Threat Intelligence

## Overview Of Traditional Threat Intelligence Methodologies

Traditional methodologies for gathering threat intelligence encompass a range of techniques aimed at identifying, analyzing, and mitigating risks. These methodologies draw from diverse sources to gather information and comprehend the landscape of cyber threats. Initially, threat intelligence relied on Open-Source Intelligence (OSINT), encompassing publicly available information from various sources like social media, forums, and websites [38, 39, 515]. OSINT enables analysts to grasp adversaries' tactics, techniques, and procedures (TTPs) and understand potential vulnerabilities in systems. Additionally, Human Intelligence (HUMINT) plays a pivotal role by involving individuals with specialized knowledge or access to gather insider information on potential threats [40, 41. 516] This methodology involves building relationships with informants or insiders to obtain valuable insights into adversaries' plans or activities. Furthermore, Technical Intelligence (TECHINT) involves analyzing technical data from malware samples, network logs, and system configurations [42, 517]. TECHINT focuses on understanding the technical aspects of cyber threats, such as malware behavior and network traffic patterns, aiding in the identification and mitigation of potential risks Figure 2 shows type of threats intelligent.

Another crucial methodology is Signals Intelligence (SIGINT), which involves intercepting and analyzing electronic signals, including communication between threat actors [43, 44, 518]. SIGINT helps in uncovering potential threats by monitoring and interpreting electronic transmissions, offering insights into adversaries' communications and intentions. Moreover, Geopolitical Intelligence encompasses analyzing geopolitical events and factors that may influence cyber threats [45, 519]. Understanding political climates, international relations, and regional conflicts aids in predicting potential cyber threats originating from state-sponsored actors or politically motivated groups. Additionally, Tactical Intelligence involves real-time analysis of ongoing threats and incidents to provide immediate responses [46, 520]. This methodology focuses on quickly identifying, analyzing, and responding to cyber threats to minimize their impact on systems and networks. Furthermore, Strategic Intelligence involves long-term planning and analysis to understand the broader trends and patterns of cyber threats [47, 521]. Strategic intelligence assists in devising comprehensive security strategies to counter evolving and persistent threats effectively.

Traditional threat intelligence methodologies encompass a combination of these approaches, enabling organizations to gather comprehensive information and insights into potential cyber threats. However, these methodologies face challenges in the rapidly evolving threat landscape, requiring continuous adaptation and integration of new technologies and intelligence gathering techniques. In conclusion, traditional threat intelligence methodologies encompass diverse approaches, from open-source information to technical analysis and geopolitical factors. However, the dynamic nature of cyber threats necessitates ongoing innovation and adaptation within these methodologies to effectively combat emerging risks and vulnerabilities.

**Introduction To AI And Machine Learning in Cybersecurity**

In the digital age, cybersecurity stands as a critical shield against evolving threats, where Artificial Intelligence (AI) and Machine Learning (ML) play pivotal roles in fortifying defense mechanisms. AI, a technology mimicking human intelligence, coupled with ML, a subset of AI enabling systems to learn and improve from data, revolutionizes cybersecurity by enhancing threat detection, response, and mitigation strategies [48, 49, 522] This amalgamation empowers security systems to decipher intricate patterns within vast datasets at unprecedented speeds, fortifying preemptive and responsive measures against cyber threats [50, 523]. One of the primary applications of AI and ML in cybersecurity lies in threat detection. Through the analysis of historical data, these technologies identify anomalous behavior patterns that may signal potential cyber-attacks [51, 52, 524]. Utilizing algorithms capable of recognizing deviations from normal system operations, AI-equipped systems bolster defense mechanisms by swiftly pinpointing and neutralizing threats before they manifest [53, 525].

Moreover, AI-driven systems contribute significantly to vulnerability management by identifying and patching system weaknesses [54, 55, 526]. ML algorithms can scrutinize system vulnerabilities and anticipate potential entry points for cybercriminals, allowing cybersecurity professionals to proactively reinforce defenses [56, 527]. Additionally, AI and ML play a pivotal role in refining incident response strategies [57, 528]. By swiftly analyzing and categorizing incidents based on their severity, origin, and patterns, AI-enhanced systems expedite response times, minimizing the impact of cyber breaches [58, 59, 529]. Furthermore, these technologies enable predictive analysis, allowing cybersecurity teams to foresee potential threats and strategize preemptive measures. Nonetheless, the integration of AI and ML in cybersecurity isn't devoid of challenges. Issues surrounding data privacy and ethics emerge due to the reliance on extensive data sets for training algorithms [60, 531]. Ensuring the ethical use of AI in cybersecurity remains an ongoing concern, demanding stringent regulations and guidelines [61, 62, 493, 530].

In conclusion, the integration of AI and ML in cybersecurity stands as a paradigm shift, empowering defense mechanisms with proactive threat detection, vulnerability management, and enhanced incident response strategies. Despite challenges surrounding data privacy and ethics, the continual advancements in AI and ML technologies promise a fortified cybersecurity landscape against the ever-evolving threat landscape.

**Role Of AI In Transforming Traditional Threat Intelligence**

Threat intelligence has undergone a profound transformation owing to the integration of Artificial Intelligence (AI) technologies. The advent of AI has revolutionized the way organizations perceive, analyze, and respond to cyber threats, enhancing the efficacy of traditional threat intelligence methods. This section aims to elucidate the pivotal role played by AI in transforming conventional threat intelligence paradigms.

AI technologies such as machine learning and natural language processing (NLP) have catalyzed the automation and analysis of vast amounts of data, providing a more comprehensive and proactive approach to identifying potential threats [63, 494, 532]. Machine learning algorithms enable the rapid processing of massive datasets, facilitating the detection of anomalies and patterns indicative of potential cyber threats [64, 65, 533]. Furthermore, NLP-based AI systems assist in contextualizing and understanding unstructured data from diverse sources, thereby enhancing the accuracy and speed of threat detection [66, 67, 534]. The incorporation of AI algorithms into traditional threat intelligence methodologies has significantly bolstered the capability to detect and predict evolving threats [68, 69, 535]. AI-driven predictive analytics leverage historical data to forecast potential threats, enabling organizations to preemptively fortify their cyber defenses [70, 536]. Moreover, AI-powered threat detection systems continuously learn and adapt to new threat landscapes, mitigating the challenges posed by rapidly evolving cyber threats [71, 537]. AI augments human decision-making processes by providing actionable insights gleaned from complex data analysis [72, 538]. AI-driven threat intelligence empowers organizations to make informed decisions swiftly, thereby enhancing response times to mitigate potential risks [73, 539]. Additionally, AI-enabled automation assists in orchestrating response mechanisms, minimizing manual intervention and streamlining incident response workflows [74, 540]. Despite the transformative potential of AI in threat intelligence, challenges such as adversarial attacks on AI systems and ethical concerns surrounding AI decision-making persist [75, 76, 541]. The future of AI-driven threat intelligence involves addressing these challenges while further harnessing AI's potential through improved algorithms and ethical frameworks [77, 78, 542].

In conclusion, the integration of AI technologies has revolutionized traditional threat intelligence, offering unprecedented capabilities in threat detection, prediction, decision-making, and response. However, addressing challenges and embracing ethical considerations remains crucial in maximizing the potential of AI-driven threat intelligence for a more secure digital landscape.

## 3. AUTONOMOUS THREAT HUNTING: CONCEPTUAL FRAMEWORK

### Definition And Scope of Autonomous Threat Hunting

Autonomous Threat Hunting (ATH) refers to the proactive identification and mitigation of potential cyber threats using automated processes and intelligent algorithms [79, 80, 543]. It involves the utilization of advanced technologies like machine learning, artificial intelligence (AI), and big data analytics to detect, analyze, and respond to potential security incidents across diverse networks and systems [81, 82, 544]. ATH operates on the premise of continuous monitoring and analysis, aiming to detect anomalies, potential vulnerabilities, and suspicious activities that might indicate a security breach or impending threat [83, 84, 545] This approach emphasizes the proactive nature of threat detection, enabling organizations to stay ahead of potential cyber adversaries.

The scope of ATH encompasses various aspects of cybersecurity, primarily focusing on threat intelligence gathering, data analysis, and response mechanisms. Threat intelligence serves as a foundational element, providing ATH systems with real-time information about emerging cyber threats, malware signatures, and attack patterns [85, 86, 546]. Leveraging this intelligence, ATH systems employ sophisticated algorithms to analyze large volumes of data collected from network logs, endpoints, and other sources [87, 88, 547]. By correlating disparate data points, these systems can identify subtle indicators of compromise or potential security breaches that may elude traditional security measures [89, 90, 548]. Figure 3 shows traditional threats intelligent.

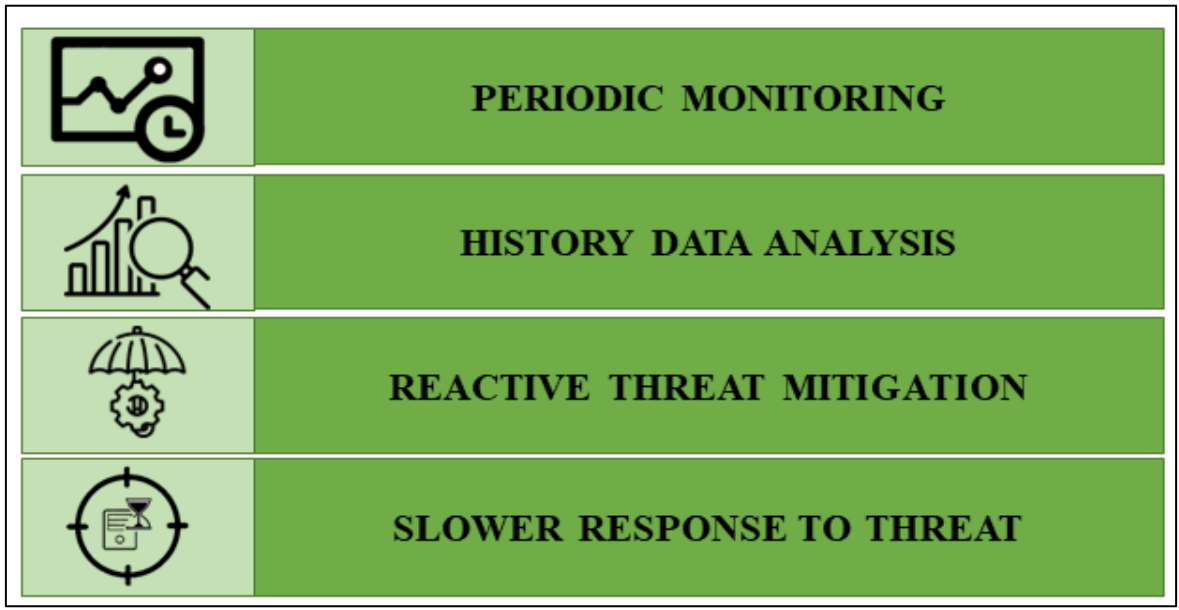

Figure 3: Traditional Threat Intelligence

The autonomy in ATH refers to the ability of these systems to operate with minimal human intervention. While human expertise remains essential for oversight and decision-making, ATH platforms utilize automation to handle routine tasks such as data collection, analysis, and initial response actions [91, 92, 549]. This autonomous capability significantly accelerates threat detection and response times, enabling organizations to mitigate potential risks more effectively [93, 550]. Additionally, the autonomous nature of ATH reduces the strain on cybersecurity teams, allowing them to focus on more strategic tasks, such as refining threat models and developing proactive defense strategies [94, 95, 551]. The significance of ATH lies in its ability to enhance the overall cybersecurity posture of organizations by providing a proactive and adaptive approach to threat detection and response [96, 552]. It complements traditional security measures by filling gaps in detection capabilities and addressing the evolving nature of cyber threats [97, 98]. As cyber threats continue to evolve in complexity and frequency, the adoption of autonomous threat hunting becomes increasingly crucial for organizations seeking to bolster their resilience against potential cyber-attacks [99, 100, 553].

In conclusion, Autonomous Threat Hunting encompasses the proactive use of advanced technologies and automated processes to detect, analyze, and respond to potential cyber threats. Its scope includes leveraging threat intelligence, data analysis, and autonomous capabilities to enhance cybersecurity defenses and enable organizations to stay ahead of evolving threats.

**Key Components of An Autonomous Threat Hunting System**

**Data Collection and Aggregation:** An autonomous threat hunting system relies on a diverse range of data sources, including logs, network traffic, endpoint activities, and external threat intelligence feeds [101]. These sources provide a comprehensive view of the environment, aiding in the identification of anomalies and potential threats.

**Machine Learning Algorithms:** Machine learning algorithms play a pivotal role in autonomously sifting through vast datasets to discern patterns and anomalies [102]. These algorithms enable the system to learn from historical data and adapt to new and emerging threats without human intervention.

**Behavioral Analytics:** Behavioral analytics techniques enable the system to establish a baseline of normal behavior within the network and endpoints [103]. Any deviations from this baseline can be flagged as potential threats, allowing for proactive investigation.

**Threat Intelligence Integration:** Integration of threat intelligence feeds from various sources enhances the system's ability to detect and respond to known and emerging threats [104]. This integration provides contextual information about malicious activities observed elsewhere in the cybersecurity landscape.

**Automated Response Mechanisms:** Autonomous threat hunting systems often include automated response mechanisms [105, 106] These mechanisms can take predefined actions to mitigate threats, such as isolating compromised endpoints or blocking suspicious network traffic.

**Continuous Monitoring and Real-time Analysis:** Continuous monitoring of network and endpoint activities combined with real-time analysis capabilities enables rapid threat detection and response [107]. This proactive approach reduces the dwell time of threats within the system.

**Human-Machine Collaboration:** While these systems operate autonomously, human expertise remains invaluable. Collaboration between automated systems and human analysts allows for the validation of identified threats and the fine-tuning of algorithms [108].

**Scalability and Flexibility:** Scalability and flexibility are essential components, enabling the system to adapt to changing network architectures and increasing data volumes [109, 110]. This ensures the system's efficacy as networks grow and evolve.

**Compliance and Regulatory Considerations**: Integration of compliance and regulatory considerations within the system ensures adherence to industry standards and legal requirements [111]. This is vital for maintaining the integrity and legality of threat hunting activities.

**Incident Reporting and Documentation:** Comprehensive incident reporting and documentation capabilities facilitate post-incident analysis and aid in improving the system's capabilities over time [112].

**Integration Of AI Algorithms in Threat Hunting Processes**

With the increasing complexity and volume of cyber threats, the integration of artificial intelligence (AI) algorithms has become crucial in enhancing threat hunting processes. AI empowers security analysts by automating repetitive tasks, analyzing vast amounts of data, and identifying patterns that might indicate potential threats [113, 554]. This section explores the impact of integrating AI algorithms in threat hunting and its significance in modern cybersecurity practices.

AI algorithms, including ML and deep learning (DL), play a pivotal role in threat hunting processes by enabling the analysis of large datasets to uncover anomalies and potential security risks [114, 115, 116, 555]. ML algorithms, such as supervised and unsupervised learning, aid in identifying known and unknown threats by learning from historical data patterns [117, 118, 556]. DL algorithms, particularly neural networks, excel in recognizing complex patterns and behaviors within network traffic, allowing for the detection of sophisticated threats like zero-day attacks [119, 120, 557]. Furthermore, the integration of AI-driven threat intelligence platforms enhances the efficiency of threat hunting by continuously updating and correlating threat data from various sources [121, 122, 558]. These platforms leverage natural language processing (NLP) algorithms to analyze unstructured data from threat feeds, security blogs, and reports, providing actionable insights to security analysts [123, 124, 496, 559].

The integration of AI algorithms in threat hunting processes offers numerous benefits, including increased accuracy in threat detection, faster response times, and the ability to handle a vast amount of data in real-time [125, 126, 560]. AI-powered systems can assist human analysts by reducing false positives, allowing them to focus on more critical tasks that require human intuition and decision-making skills [127, 128, 561]. However, challenges persist, such as the potential for AI models to be susceptible to adversarial attacks, where malicious actors manipulate algorithms to evade detection [129, 130, 562] Additionally, the reliance on AI-driven systems may lead to over-reliance or complacency among analysts, diminishing their proactive threat hunting capabilities [131, 563].

In conclusion, the integration of AI algorithms in threat hunting processes significantly enhances cybersecurity capabilities by enabling proactive identification and mitigation of potential threats. ML, DL, NLP, and AI-driven threat intelligence platforms collectively contribute to improving the efficiency and effectiveness of threat hunting operations. While the benefits are substantial, addressing challenges related to adversarial attacks and maintaining human involvement for critical decision-making remains essential in maximizing the potential of AI in threat hunting.

**Discussion Of the Process in Detail**

Threat hunting is a proactive cybersecurity approach aimed at identifying and mitigating potential threats that may have bypassed traditional security measures. It involves a systematic and continuous process of searching, detecting, and neutralizing sophisticated threats within an organization's network. This framework comprises several key steps, providing a structured methodology for security teams to enhance their defensive capabilities. The first step in the threat hunting process involves setting clear objectives and defining the scope of the hunt. This aligns with the concept that "a clear mission is key to successful threat hunting" [132, 564] ensuring focus on specific areas and threats within the network. Subsequently, data collection from various sources such as logs, network traffic, and endpoint devices is essential for thorough analysis [133, 134, 565] The collected data undergoes comprehensive analysis, which involves utilizing both automated tools and human expertise. According to [135] "the combination of automated tools and human intelligence is crucial in effective threat hunting". This analysis aims to detect anomalies, unusual patterns, or indicators of compromise (IoCs) that might signify potential threats lurking within the system [136, 566].

Upon the identification of potential threats, the next step is investigation and validation. This stage requires meticulous examination and verification of flagged activities or behaviors to determine their nature and severity [137, 138, 567] The investigation may involve correlating data across different sources and conducting in-depth forensic analysis to understand the attack's tactics, techniques, and procedures (TTPs) Once the threat is confirmed, containment and eradication strategies are initiated to mitigate the impact and prevent further spread. Swift action is crucial to limit the threat's reach and minimize potential damage [139, 568]. This aligns with the notion that rapid response and containment significantly reduce the impact of cyber threat. Post-incident analysis and documentation form the final phase, focusing on lessons learned and improvements for future threat hunting endeavors [140, 569]. This step ensures continuous enhancement of the threat hunting process through the integration of new insights and strategies to bolster cyber defenses [141, 142, 570].

In conclusion, the threat hunting framework is a proactive and systematic approach that enables organizations to detect and mitigate potential threats before they cause significant damage. By following a structured process encompassing objectives setting, data collection, analysis, investigation, containment, and post-incident analysis, security teams can effectively strengthen their cybersecurity posture.

## 4. STATE-OF-THE-ART AI TECHNIQUES IN AUTONOMOUS THREAT HUNTING

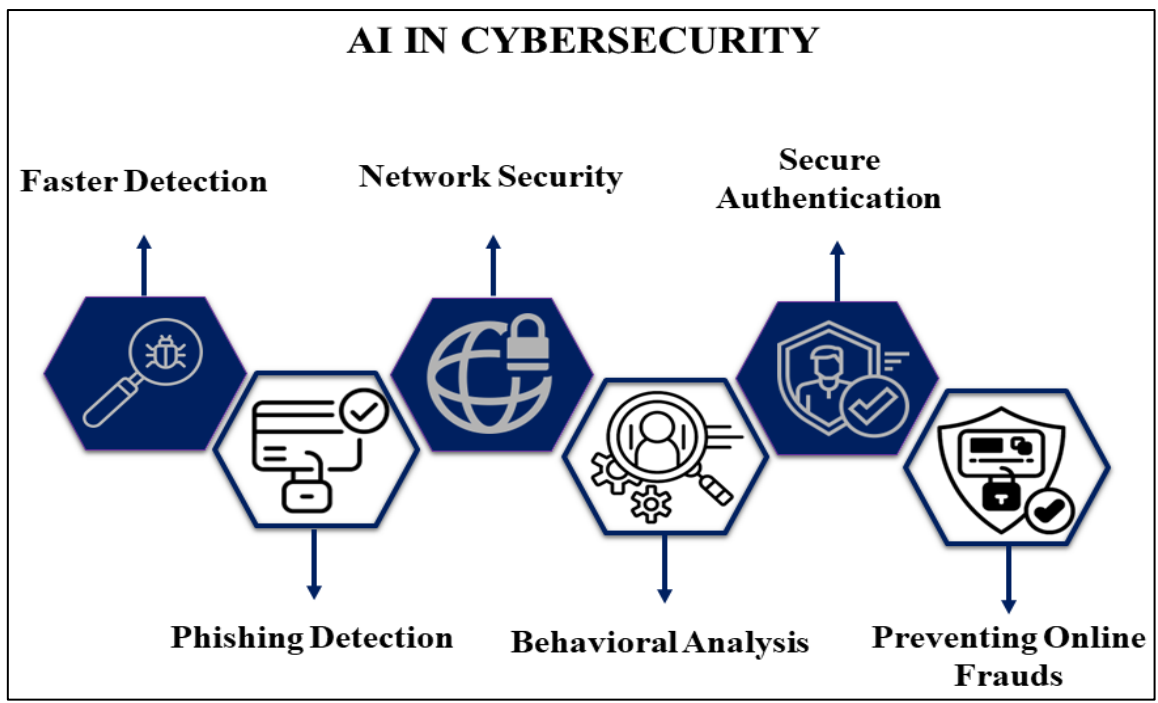

Figure 4: AI in Cybersecurity

### 4.1 Machine Learning Models

**Supervised Learning for Threat Detection**
In the realm of threat hunting, the use of supervised learning techniques has gained prominence for enhancing detection capabilities [143, 571]. Supervised learning involves training a model on labeled data, allowing it to recognize patterns and anomalies associated with known threats, enabling more effective threat identification and response [144, 572]. This section explores the application of supervised learning in threat detection within the context of threat hunting. Supervised learning leverages historical data, utilizing labeled examples to train algorithms to recognize specific patterns indicative of threats [145, 146, 573]. This method allows for the creation of models capable of distinguishing between normal and abnormal behavior, aiding in the identification of potential threats [147, 489, 574]. By employing techniques like classification and regression, supervised learning algorithms such as Support Vector Machines (SVM) and Random Forests can classify and predict threats based on predefined features extracted from data [148, 149]. The implementation of supervised learning models contributes significantly to threat detection by improving accuracy and efficiency [150, 151]. These models can identify complex threat patterns that might be challenging for traditional rule-based systems to detect [152, 153]. Moreover, they facilitate quicker response times by swiftly pinpointing potential threats within vast datasets. Additionally, supervised learning enables the continuous learning and adaptation of models to evolving threats, enhancing the overall cybersecurity posture [154,155]. Figure 4 shows AI in cybersecurity.

In conclusion, supervised learning techniques play a crucial role in threat detection within the realm of threat hunting. They offer improved capabilities in identifying and mitigating cybersecurity threats by utilizing labeled data to train models that recognize patterns indicative of malicious activities. However, challenges such as data availability, model interpretability, and the dynamic nature of threats necessitate ongoing advancements and adaptations in supervised learning approaches for robust threat detection.

**Unsupervised Learning for Anomaly Detection**

Anomaly detection in threat hunting plays a pivotal role in cybersecurity, aiming to identify irregular patterns or behaviors that deviate from expected norms within a system [156, 490]. With the evolving landscape of cyber threats, unsupervised learning techniques have emerged as a potent tool for anomaly detection. Unsupervised learning algorithms, devoid of labeled data, have the capability to discern anomalies by identifying patterns that differ significantly from the norm [157, 158]. Unsupervised learning methods like clustering, autoencoders, and principal component analysis (PCA) are instrumental in anomaly detection due to their ability to detect patterns in data without prior labeling [159, 160]. Clustering algorithms, such as k-means and DBSCAN, group similar data points together, aiding in the identification of outliers or anomalies [161, 162]. Autoencoders, a type of neural network, reconstruct input data and highlight anomalies by exhibiting higher reconstruction errors for irregular instances [163]. PCA helps in dimensionality reduction, aiding in the identification of unusual data points lying outside the expected variance [164,]. Unsupervised learning methods find application in diverse cybersecurity domains. In network security, anomaly detection techniques help in identifying suspicious network traffic or intrusions [165, 166]. Behavioral analysis using unsupervised learning assists in identifying abnormal user behaviors that might indicate insider threats or unauthorized access [167,168] Additionally, anomaly detection in system logs aids in identifying potential malware activities or system breaches [169, 170].

In conclusion, unsupervised learning techniques play a vital role in anomaly detection for threat hunting in cybersecurity. They offer efficient means to detect anomalies without relying on labeled data, enabling the identification of potential threats and vulnerabilities. Despite existing challenges, ongoing research and advancements in unsupervised learning promise improved accuracy and robustness in detecting anomalies, contributing significantly to cybersecurity efforts.

**Reinforcement Learning for Decision-Making in Threat Hunting**

Reinforcement learning (RL) has emerged as a promising framework for enhancing decision-making in threat hunting within cybersecurity [171]. This approach enables computer systems to learn optimal strategies by interacting with their environment, thus making it suitable for tackling the complexities and dynamic nature of cyber threats [172, 173]. Threat hunting involves actively seeking potential security breaches or abnormalities within a network to thwart cyber-attacks [174, 175]. RL algorithms, inspired by behavioral psychology, function on the principle of learning from trial and error by rewarding correct actions and penalizing mistakes [176].

In the realm of cybersecurity, RL presents a novel approach for autonomous decision-making in threat detection and mitigation. It aids in automating the decision-making process by allowing systems to learn from historical data, adapt to new threats, and optimize responses in real-time [177]. RL models, such as deep Q-networks (DQNs) and policy gradient methods, have shown efficacy in learning complex patterns from massive datasets, enabling swift threat identification [178, 179]. One significant advantage of RL in threat hunting is its adaptability to evolving threats. Traditional rule-based systems struggle to keep pace with rapidly changing attack methodologies, while RL algorithms can continuously learn and update their strategies based on new threat intelligence [180, 181] Moreover, RL's ability to explore different actions and exploit the most effective ones assists in optimizing resource allocation and minimizing false positives [182, 183]. By iteratively learning and refining its decision-making process, RL facilitates more accurate and efficient threat detection and response mechanisms [184, 185].

In conclusion, reinforcement learning offers a promising avenue for improving decision-making in threat hunting within cybersecurity. Its adaptability to evolving threats, capability to learn from vast datasets, and potential for automating complex decision-making processes mark it as a valuable tool for enhancing cyber defense mechanisms. However, challenges related to computational demands and ethical considerations must be addressed to harness its full potential in securing digital infrastructures.

## 4.2 Natural Language Processing (NLP) and Sentiment Analysis

**Leveraging NLP For Extracting Threat Intelligence from Unstructured Data.**

To effectively mitigate security threats, extracting meaningful intelligence from this unstructured data has become crucial. Natural Language Processing (NLP) has emerged as a pivotal tool in the realm of cybersecurity for extracting threat intelligence from unstructured data sources such as social media, dark web forums, and incident reports [186, 187]. NLP's role in cybersecurity involves various techniques to parse, understand, and categorize unstructured data. Named Entity Recognition (NER) and sentiment analysis are among the primary NLP techniques utilized to identify and extract entities like names, organizations, locations, and sentiments from text data [187, 189, 495]. Through NLP, cybersecurity analysts can process massive volumes of text-based data efficiently, enabling the identification of potential threats and vulnerabilities within an organization's digital infrastructure [190, 191]. Moreover, NLP-powered algorithms facilitate the extraction of indicators of compromise (IOCs) from unstructured data sources, aiding in the identification of potential cyber threats [192, 193].These IOCs, such as IP addresses, malware signatures, and suspicious URLs, play a critical role in proactively securing networks against cyberattacks [194, 195]. By applying NLP techniques, cybersecurity professionals can swiftly identify and respond to emerging threats, thus enhancing an organization's overall security posture [196, 197].

In conclusion, NLP stands as a powerful ally in the fight against cybersecurity threats by enabling the extraction of actionable intelligence from vast amounts of unstructured data. Its application in identifying entities, sentiments, and indicators of compromise enhances the efficiency of cybersecurity operations. Despite challenges, ongoing advancements in NLP methodologies hold immense promise for bolstering threat intelligence capabilities, thereby fortifying digital defense mechanisms against evolving cyber threats.

**Sentiment Analysis for Contextualizing Threat Information**

Sentiment analysis, an essential component of natural language processing (NLP), plays a pivotal role in contextualizing threat information by deciphering the emotions, opinions, and attitudes expressed within textual data [198,199]. In recent years, its application has significantly expanded across various domains due to advancements in machine learning and AI technologies. This section aims to explore the significance of sentiment analysis in understanding and contextualizing threat information.

Sentiment analysis involves the use of algorithms and linguistic techniques to determine the sentiment or emotional tone within text data [200, 201]. It categorizes sentiments into positive, negative, or neutral, providing valuable insights into the underlying emotions conveyed within the text [202, 203]. Leveraging machine learning models, sentiment analysis helps in identifying and quantifying opinions, allowing for a nuanced understanding of the text's meaning [204, 205]. In the realm of security and threat assessment, sentiment analysis plays a crucial role. By analyzing social media posts, news articles, or forum discussions, sentiment analysis aids in gauging public perception and reactions toward security threats [206]. It enables authorities to comprehend the severity of a situation by identifying panic, fear, or misinformation circulating in the public domain [207, 208]. Such insights assist in tailoring appropriate responses and crisis management strategies. Sentiment analysis contributes to enhancing decision-making processes within threat assessment. By analyzing sentiments expressed in intelligence reports or communication channels, security analysts can gauge the potential impact of a threat and its dissemination among the populace [209, 210]. Integrating sentiment analysis with other data analytics tools provides a comprehensive understanding, aiding in the prioritization of threats and resource allocation [211, 212]. Sentiment analysis stands as a vital tool in understanding and contextualizing threat information by unraveling the emotional underpinnings within textual data. Its application assists in proactive threat assessment, crisis management, and informed decision-making. Continued advancements and ethical considerations will further solidify its role in shaping effective security strategies. Figure 5 shows threat hunting framework.

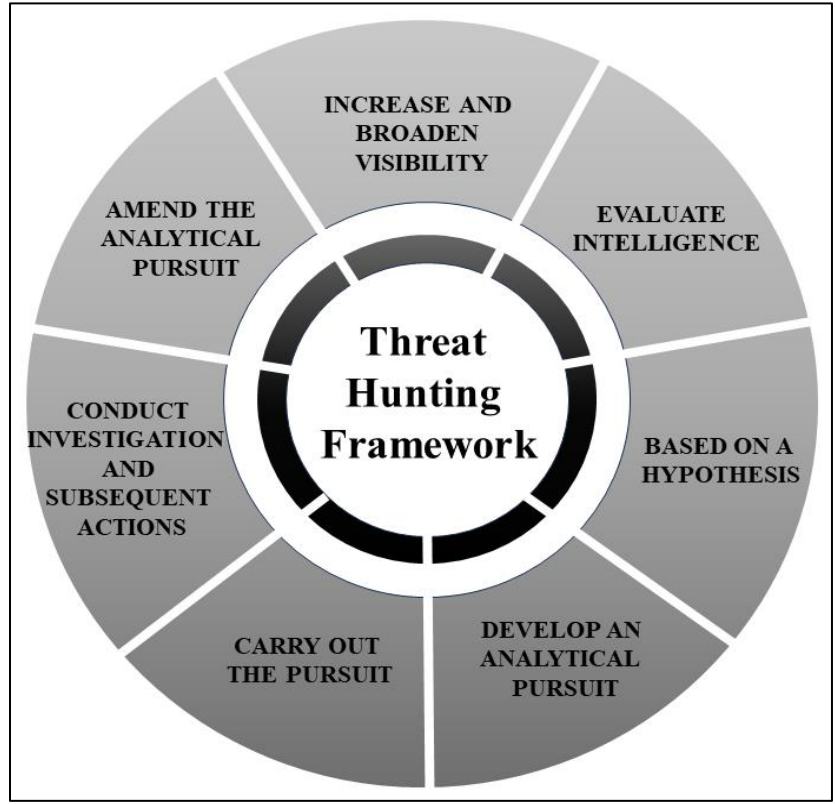

Figure 5: Threat Hunting Framework

### 4.3 Deep Learning Architectures

**Convolutional Neural Networks (CNNs) for image-based threat detection**

Convolutional Neural Networks (CNNs) have emerged as a powerful tool for image-based threat detection due to their ability to analyze visual data effectively [213, 214]. These neural networks are designed to automatically recognize patterns and features within images, making them highly suitable for detecting threats in various contexts, including security, healthcare, and environmental monitoring [215, 216]. CNNs consist of multiple layers, including convolutional layers, pooling layers, and fully connected layers [217]. Convolutional layers apply filters to input images to extract distinct features, while pooling layers down sample the extracted features, reducing computational complexity [218]. Fully connected layers analyze the extracted features and make predictions regarding the presence of threats within the images [219, 220]. The effectiveness of CNNs in threat detection heavily relies on the training process [221]. These networks learn to identify threats through exposure to a vast dataset of labeled images, where they adjust their parameters to minimize errors and enhance accuracy [222]. Continuous learning through feedback loops helps CNNs improve their threat detection capabilities over time [223]. CNNs find extensive application in security and defense sectors for threat detection in various scenarios [224]. For instance, they are used in airport security systems to identify concealed weapons or suspicious items in luggage scans [225]. Similarly, in military contexts, CNNs aid in detecting threats within satellite images or surveillance footage, enhancing situational awareness and aiding decision-making processes [226, 227]

Convolutional Neural Networks have revolutionized image-based threat detection due to their ability to automatically learn and identify patterns within visual data. With advancements in training methodologies and ongoing research addressing challenges, CNNs continue to evolve as a vital tool in enhancing security measures across various domains.

**Recurrent Neural Networks (RNNs) For Sequential Data in Threat Intelligence**

Recurrent Neural Networks (RNNs) have emerged as a potent tool in threat intelligence due to their ability to process sequential data efficiently [228]. RNNs are a class of neural networks designed to handle sequential information by preserving and utilizing previous inputs [229, 230]. This section aims to elucidate the significance of RNNs in threat intelligence for analyzing sequential data, highlighting their applications, strengths, and limitations.

RNNs find multifaceted applications in threat intelligence, such as in anomaly detection, malware classification, and network traffic analysis [231]. They excel in recognizing patterns within sequential data, making them effective for identifying irregularities in network traffic or detecting deviations from normal behavior, indicating potential threats [232, 233]. Additionally, RNNs demonstrate proficiency in processing time-series data, enabling the analysis of historical patterns to predict potential cyber threats [234, 235]. The strength of RNNs lies in their ability to capture temporal dependencies and contextual information from sequential data [236]. Unlike feedforward neural networks, RNNs possess memory cells that retain information about previous inputs, enabling them to consider the entire sequence when making predictions [237, 238]. This recurrent nature empowers RNNs to handle variable-length sequences, making them adaptable to diverse threat intelligence datasets [239, 240]. However, RNNs are not devoid of limitations. The vanishing and exploding gradient problems can hinder their performance when dealing with long sequences, impacting their ability to retain crucial information over time [241]. Moreover, RNNs might struggle with capturing long-term dependencies, leading to issues in accurately predicting threats based on historical data [242].

In conclusion, Recurrent Neural Networks (RNNs) play a pivotal role in threat intelligence by efficiently processing sequential data. Their ability to capture temporal dependencies and analyze patterns makes them instrumental in anomaly detection, malware classification, and predicting potential cyber threats. Despite their strengths, RNNs face challenges like the vanishing gradient problem and difficulties in capturing long-term dependencies. Addressing these limitations through advancements in network architectures or incorporating hybrid models can further enhance the efficacy of RNNs in threat intelligence.

## 5. CHALLENGES IN AUTONOMOUS THREAT HUNTING

**Scalability And Adaptability Of AI-Driven Models**

The scalability and adaptability of AI-driven models have become crucial factors in determining their efficacy in addressing complex real-world challenges. Scalability refers to an AI model's capability to handle increasing data volumes or computational demands without compromising performance [243]. On the other hand, adaptability denotes the AI system's capacity to learn from new data and adjust its functionality accordingly [244].

The scalability of AI-driven models is pivotal for their widespread implementation. Models like deep learning neural networks have demonstrated commendable scalability, efficiently handling vast datasets [245]. Architectural advancements, such as distributed computing frameworks like TensorFlow and PyTorch [246], contribute significantly to enhancing scalability by distributing computations across multiple processors or devices. Additionally, the emergence of cloud computing platforms enables easy access to scalable computational resources, facilitating the deployment of AI models in various applications [247, 248]. The adaptability of AI models is crucial in ensuring their relevance in dynamic environments. Techniques like transfer learning allow models to leverage knowledge gained from one domain and apply it to another, enhancing adaptability by reducing the need for extensive labeled data in new domains [249, 250]. Furthermore, continual learning methods enable AI models to learn incrementally from streaming data, ensuring continuous improvement and adaptation to evolving scenarios [251, 252].

In conclusion, the scalability and adaptability of AI-driven models are pivotal for their successful integration into diverse applications. Advances in distributed computing, cloud platforms, and innovative learning techniques have contributed significantly to enhancing these characteristics. However, challenges persist, necessitating ongoing research efforts to overcome hurdles and pave the way for more robust, scalable, and adaptable AI systems.

**Explainability And Interpretability of Autonomous Threat Hunting Decisions**

Autonomous threat hunting systems leverage sophisticated algorithms and machine learning techniques to analyze vast volumes of data, enabling proactive identification of potential security threats [253]. However, as these systems become more autonomous and complex, ensuring the explainability and interpretability of their decisions becomes crucial. Explainability refers to the system's capability to clarify how and why specific decisions are made, while interpretability concerns the human comprehension of these decisions [254, 255].

Explainability and interpretability are pivotal aspects ensuring transparency and trust in autonomous systems [256, 257, 258]. In the context of cybersecurity, these facets become paramount as autonomous threat hunting systems make critical decisions affecting organizational security posture [259, 260]. Transparent decision-making processes foster trust among cybersecurity professionals and stakeholders, allowing them to comprehend and validate the actions taken by these autonomous systems [261, 262]. Achieving explainability and interpretability in autonomous threat hunting systems presents challenges due to the complexity of machine learning models (Carvalho et al., 2020). Complex models, such as deep neural networks, often operate as "black boxes," hindering the understanding of decision-making rationale [263, 264]. Ensuring the interpretability of these systems is critical to enable cybersecurity experts to intervene and validate decisions, thereby improving the system's overall efficacy [265]. Moreover, the lack of explainability and interpretability in autonomous systems may lead to unintended consequences and biases [266, 267]. Biases embedded in training data or algorithmic processes can amplify inferences and decisions, potentially leading to erroneous threat identifications or false positives [268]. These issues pose substantial risks to cybersecurity as they could lead to oversight or misinterpretation of critical security threats. However, integrating explainable AI techniques, such as feature importance analysis and model-agnostic methods, can enhance the interpretability of autonomous threat hunting decisions [296]. By employing these techniques, cybersecurity professionals can gain insights into the decision-making processes of these systems and identify potential vulnerabilities or biases [270, 271].

In conclusion, the explainability and interpretability of autonomous threat hunting decisions are fundamental in ensuring trust, transparency, and effectiveness in cybersecurity operations. Addressing these aspects is crucial to mitigate risks associated with opaque decision-making, biases, and unintended consequences. Implementing transparent and interpretable AI techniques will empower cybersecurity experts to validate decisions, improve system reliability, and fortify organizational security against evolving threats.

**Ethical Considerations and Potential Biases in AI Algorithms**

The ethical considerations in AI algorithms encompass diverse facets, including issues of privacy, transparency, accountability, and fairness [272, 273]. Privacy concerns arise due to the extensive collection and utilization of personal data for training AI models [274, 275]. Transparency becomes an ethical concern as complex algorithms often lack interpretability, making it challenging to understand the decision-making process [276, 277]. This lack of transparency can lead to a lack of accountability, making it difficult to assign responsibility when errors or biases occur [278]. Moreover, ensuring fairness and preventing biases in AI systems is crucial to avoid perpetuating discrimination and societal inequalities [279].

AI algorithms can inherit biases from the data used for their training, resulting in biased outcomes. For instance, historical biases present in datasets, such as race or gender biases, can be perpetuated and amplified by AI systems [280]. This bias amplification poses significant ethical challenges, especially in domains like criminal justice, where biased predictions can lead to unfair treatment or sentencing [281, 282]. Additionally, algorithmic biases can exacerbate societal disparities by affecting access to resources and opportunities, further deepening existing inequalities [283]. Addressing these ethical concerns and biases requires comprehensive mitigation strategies. Implementing diverse and representative datasets for training AI models is crucial to minimize biases [284]. Encouraging interdisciplinary collaboration involving ethicists, social scientists, and technologists can foster discussions on ethical guidelines and frameworks for AI development [285, 286]. Furthermore, promoting transparency by designing AI systems that provide explanations for their decisions can enhance accountability and trust [287, 288].

In conclusion, the ethical considerations and potential biases embedded within AI algorithms present multifaceted challenges that necessitate urgent attention. Adhering to ethical principles, ensuring transparency, and mitigating biases in AI systems are imperative for their responsible and equitable deployment across various domains. Collaborative efforts involving policymakers, technologists, and ethicists are pivotal in developing robust frameworks to address these ethical concerns and foster the trustworthy and fair advancement of AI technology. Figure 6 show threats hunting strategy.

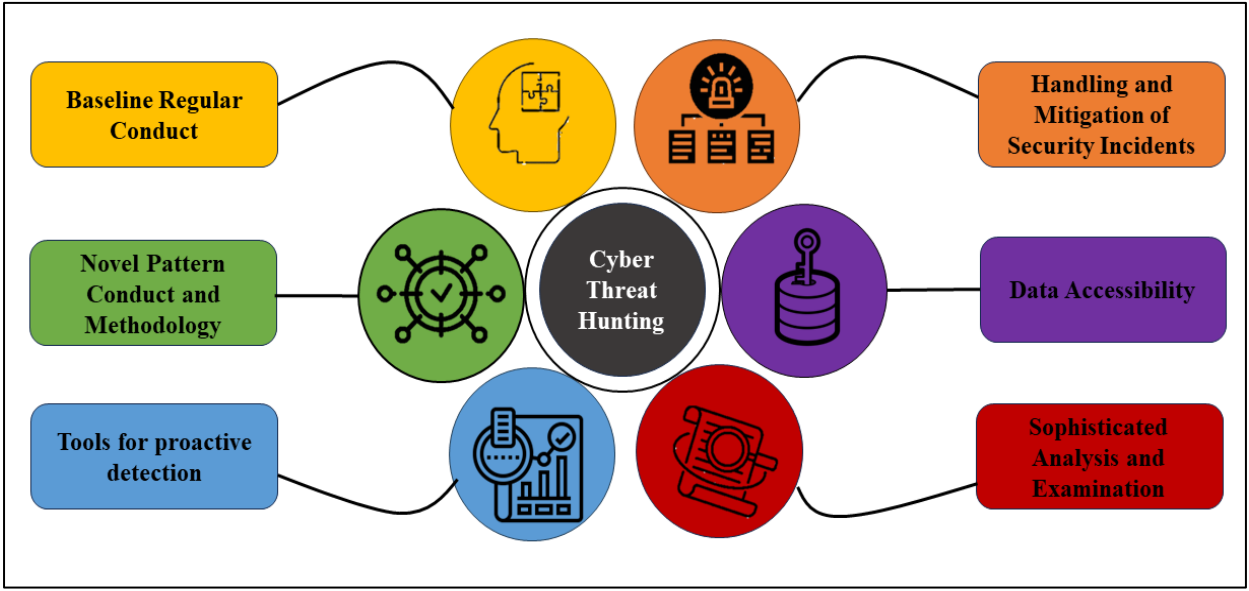

Figure 6: Threat Hunting Strategy

# 6. CASE STUDIES AND APPLICATIONS

**Highlighting Real-World Implementations of Autonomous Threat Hunting**

Autonomous threat hunting, an advanced cybersecurity approach, employs automated tools and technologies to proactively detect and mitigate potential cyber threats. This case study explores real-world applications of autonomous threat hunting, showcasing its significance in contemporary cybersecurity strategies.

The integration of machine learning and artificial intelligence (AI) algorithms has revolutionized threat hunting [289, 290]. These technologies enable the identification of anomalous patterns and potential threats in vast datasets [291, 292]. Machine learning models, such as neural networks and decision trees, analyze network traffic, behaviors, and system logs, aiding in the early detection of cyber threats [293]. Behavioral analytics plays a pivotal role in autonomous threat hunting by establishing baseline behaviors and identifying deviations that may indicate a potential threat. By leveraging user behavior analytics (UBA) and entity behavior analytics (EBA), organizations can swiftly detect anomalies and suspicious activities, enhancing threat identification capabilities [294, 295]. Real-time threat intelligence feeds from reputable sources equip autonomous threat hunting systems with the latest information on emerging threats and attack patterns [296]. Integration of threat intelligence into automated systems enhances their ability to recognize and respond to evolving threats promptly [297, 298]. Automated response mechanisms streamline incident response processes [299, 300]. Through orchestration platforms, security teams can automate containment actions, minimizing the impact of detected threats and reducing response time [301]. Autonomous threat hunting relies on continuous monitoring and adaptive learning mechanisms. Solutions such as SIEM (Security Information and Event Management) systems continuously gather and analyze data to adapt and improve threat detection accuracy [302].

Real-world implementations of autonomous threat hunting leverage cutting-edge technologies like machine learning, behavioral analytics, threat intelligence feeds, automation, and continuous monitoring to bolster cybersecurity defenses. These approaches enable proactive threat detection, rapid response, and mitigation, crucial in today's ever-evolving threat landscape.

**Success Stories and Lessons Learned from Organizations Adopting AI-Driven Threat Intelligence.**

Several success stories showcase the efficacy of AI-driven threat intelligence in bolstering cybersecurity defenses. For instance, IBM's Watson for Cyber Security employs AI to analyze vast volumes of security data, enabling faster threat detection and response [303]. Similarly, Darktrace's AI-powered platform detected and stopped a significant ransomware attack on a European energy company, preventing potential data breaches [304]. Another notable example is CrowdStrike, whose AI algorithms swiftly identified and neutralized a sophisticated cyberattack on a healthcare organization, safeguarding sensitive patient information [305, 306]. Organizations adopting AI-driven threat intelligence have gleaned valuable lessons. Firstly, the need for continuous learning and adaptation is paramount. AI models must evolve to counter evolving threats effectively [307, ]. Secondly, context is crucial; AI tools should not merely identify threats but also provide context and actionable insights for cybersecurity teams to respond effectively [308]. Additionally, maintaining human oversight and intervention remains essential; while AI augments capabilities, human expertise remains crucial for decision-making [309, 310]. Furthermore, collaboration and sharing threat intelligence across sectors enhance collective defense against cyber threats [311, 312]. Lastly, ethical considerations and transparency in AI usage are crucial to maintain trust and uphold ethical standards [313, 314].

In conclusion, the integration of AI-driven threat intelligence has led to significant advancements in cybersecurity, evidenced by numerous success stories across various industries. However, successful implementation relies not only on cutting-edge technology but also on strategic planning, continuous learning, human expertise, collaboration, and ethical considerations. These lessons serve as guiding principles for organizations aiming to harness AI's potential effectively in fortifying their cybersecurity defenses.

## 7. EVALUATION METRICS AND PERFORMANCE BENCHMARKS

**Metrics For Assessing the Effectiveness of Autonomous Threat Hunting**

As organizations increasingly adopt autonomous threat hunting technologies, assessing their effectiveness becomes paramount. Metrics serve as essential tools for gauging the success and impact of these autonomous systems. This section explores the key metrics used to evaluate the effectiveness of autonomous threat hunting methods and their significance in enhancing cybersecurity posture.

Detection Rate metric measures the accuracy of the system in identifying actual threats [325]. It calculates the percentage of genuine threats that the system successfully recognizes out of the total threats present in the network [316]. A high detection rate is crucial to ensure that the system can effectively pinpoint and address potential risks without missing significant threats [317]. False Positive Rate metric evaluates the number of false alarms or incorrectly flagged activities in relation to the total number of alerts generated by the system [318]. Lowering the false positive rate is essential as it minimizes the distractions caused by unnecessary alerts, enabling security teams to focus on legitimate threats rather than false alarms [319]. Mean Time to Detect (MTTD) measures the average time taken by the system to identify a threat from its initiation within the network [320]. A lower MTTD indicates a swifter response to potential risks, allowing security teams to take proactive measures promptly upon threat detection. Mean Time to Respond (MTTR) is a metric quantifies the average time required to mitigate or resolve a detected threat [321]. Reducing MTTR is critical as it minimizes the impact of threats and decreases the window of vulnerability, enabling organizations to effectively contain and neutralize security incidents swiftly [322]. Coverage assesses the breadth and depth of threats detected by the system across the entire threat landscape [323]. A comprehensive threat detection mechanism with higher coverage ensures a wider scope of threat identification, including known, unknown, and emerging threats [324]. Attack Surface Reduction metric evaluates the system's effectiveness in reducing the organization's attack surface, which refers to the various points where the network could be vulnerable to cyber threats [325, 326]. A successful autonomous threat hunting system minimizes vulnerabilities, making it more challenging for attackers to exploit weaknesses within the network [327]. Automated Response Efficiency measures how effectively the system responds to identified threats without the need for human intervention [328, 329]. An efficiently automated system can swiftly neutralize or contain threats, reducing reliance on manual intervention and ensuring timely response to security incidents [330, 331]. Adaptability to New Threats is a metric which assesses the system's capability to evolve and adapt to emerging threats, including new attack patterns, techniques, or malware [332, 333]. An adaptable system remains up to date with the evolving threat landscape, enabling it to identify and respond to novel threats effectively [334, 335]. Resource Utilization is a metric that evaluates the efficiency of resource consumption by the autonomous threat hunting system [336]. Optimizing computational power, storage, and other resources ensures that the system operates effectively without excessive resource consumption, thereby enhancing its overall performance [337, 338]. Figure 7 shows key metrics for assessing hunting effectiveness.

These metrics collectively serve as a comprehensive framework for organizations to evaluate the efficiency, efficacy, and overall performance of their autonomous threat hunting systems. By continuously monitoring and optimizing these metrics, organizations can strengthen their cybersecurity defenses against evolving and sophisticated threats.

**Comparative Analysis Of AI-Driven Systems Against Traditional Methods**

Artificial Intelligence (AI) has revolutionized various sectors, offering advancements in technology, decision-making, and problem-solving. This report aims to compare AI-driven systems against traditional methods in different domains to assess their effectiveness, limitations, and potential impacts on society. AI systems encompass machine learning algorithms, neural networks, and deep learning, while traditional methods rely on manual processes and rule-based algorithms.

- **Comparison in Healthcare**

AI-driven systems have significantly impacted healthcare, particularly in medical imaging analysis [339, 340]. Machine learning models and deep learning algorithms have shown exceptional promise in the interpretation of medical images like X-rays, MRIs, and CT scans [341, 342, 343]. These AI systems exhibit impressive accuracy and efficiency in detecting anomalies or diseases, aiding clinicians in making quicker and more precise diagnoses [344, 345]. On the other hand, traditional methods in healthcare heavily rely on the expertise of human practitioners to interpret medical images [346, 347]. However, these methods can be prone to errors due to human subjectivity and fatigue. Additionally, traditional approaches may lead to delays in diagnosis and treatment, which can impact patient outcomes negatively [348, 349].

- **In Finance**

AI-driven systems in finance, especially machine learning-based algorithms, have been instrumental in fraud detection and risk management [350, 351]. These systems can analyze enormous volumes of financial data in real-time, identifying patterns indicative of fraudulent activities with high accuracy [352, 353]. Moreover, AI models can adapt and evolve, improving their ability to detect new and sophisticated fraud schemes [354, 355]. Conversely, traditional methods in finance rely on predefined rules and human intervention to detect anomalies or fraudulent transactions [356, 357]. While effective to some extent, these methods might struggle to keep up with the constantly evolving nature of financial fraud, leading to higher rates of false positives and negatives [358].

- **Education Sector**

AI-driven systems have introduced personalized learning approaches in the education sector [359, 360]. These systems utilize adaptive algorithms to tailor educational content based on individual student needs, learning styles, and pace [361, 362, 363]. By analyzing student performance data, AI systems can offer customized learning experiences, potentially improving student engagement and academic outcomes [364, 365, 366]. Conversely, traditional teaching methods typically follow a uniform curriculum and teaching approach, which might not cater to the diverse learning needs of students [367]. This one-size-fits-all approach can overlook individual strengths, weaknesses, and learning styles, potentially hindering overall student performance.

- **Customer Service**

AI-driven systems, such as chatbots and virtual assistants, have transformed customer service experiences [368, 369]. These systems provide instant and automated responses to customer queries, offering round-the-clock assistance [370, 371]. They can handle multiple customer interactions simultaneously, ensuring prompt and consistent service [372]. In contrast, traditional customer service relies on human agents to address customer queries and issues. While human interaction can provide a personal touch, it's constrained by agents' availability and handling capacity, often resulting in longer wait times and inconsistent service experiences [373, 374].

AI-driven systems demonstrate superior performance in various domains compared to traditional methods, offering faster, more accurate solutions. However, ethical considerations and the potential for biases highlight the need for responsible AI development and regulation to ensure fair and transparent outcomes.

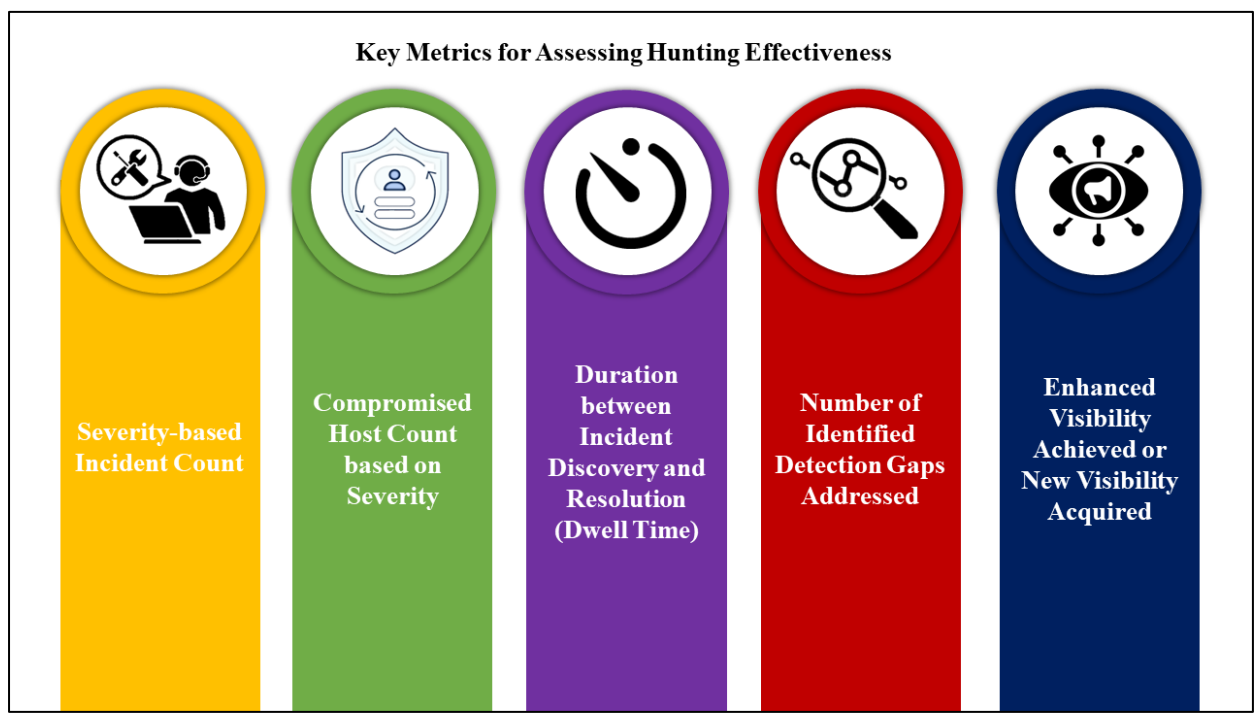

Figure 7: Key Metrics for Assessing Hunting Effectiveness

## 8. FUTURE DIRECTIONS AND EMERGING TRENDS

**Exploration of Potential Advancements in Autonomous Threat Hunting**

Autonomous threat hunting, an evolving field in cybersecurity, involves leveraging advanced technologies like artificial intelligence (AI) and machine learning (ML) to proactively identify and neutralize potential security threats. With the rapid evolution of cyber threats, the need for more sophisticated and autonomous threat hunting mechanisms has become crucial to protect digital assets and sensitive information. This section explores the potential advancements in autonomous threat hunting, focusing on the integration of AI, ML, and other cutting-edge technologies.

AI plays a pivotal role in autonomous threat hunting by enabling systems to learn from vast amounts of data and identify anomalous patterns that might indicate potential threats. According to [375] AI-driven algorithms enhance threat detection by continuously analyzing network behaviors and swiftly detecting deviations from normal patterns. Moreover, AI-based threat hunting tools like deep learning algorithms significantly improve detection accuracy [376]. ML algorithms are integral in automating threat hunting processes [377, 378, 379]. By employing unsupervised ML techniques, such as clustering and anomaly detection, security systems can identify unusual activities within networks [380, 381, 382]. This sentiment is echoed [383, 384] who emphasize the importance of ML-driven approaches in enabling autonomous systems to adapt and recognize emerging threats without human intervention. Predictive analytics enable proactive threat mitigation by predicting potential attack vectors based on historical data [385, 386]. Behavioral analysis, on the other hand, involves monitoring user behaviors to detect any deviations from standard patterns, which could signify security breaches [387, 388]. These advancements aid in early threat identification and mitigation. The utilization of big data analytics helps in processing and analyzing large volumes of security data in real-time [389, 390, 391]. Simultaneously, integrating threat intelligence feeds from various sources enhances the system's ability to recognize and counteract evolving threats [392].

In conclusion, the evolution of autonomous threat hunting through the integration of AI, ML, predictive analytics, behavioral analysis, big data, and threat intelligence feeds marks a significant advancement in cybersecurity. These technologies collectively enable systems to autonomously detect, analyze, and mitigate potential threats in real-time, thereby bolstering defenses against the constantly evolving landscape of cyber threats.

**Emerging Technologies Shaping the Future Of AI-Driven Threat Intelligence**

ML algorithms form the backbone of AI-driven threat intelligence. These algorithms, including supervised, unsupervised, and reinforcement learning, enable systems to process vast amounts of data, identify patterns, and autonomously improve over time without explicit programming [393, 394, 395]. They are crucial for anomaly detection, clustering similar threat behaviors, and predicting potential cyber threats based on historical data analysis [396, 397, 398]. For instance, algorithms like Random Forest, Support Vector Machines (SVM), and neural networks aid in identifying subtle patterns indicative of malicious activities within network traffic or system logs [399, 400].

Predictive analytics leverages historical data and statistical algorithms to forecast future cyber threats [401, 402, 403]. By analyzing patterns and trends, predictive models can anticipate potential security risks, enabling organizations to proactively strengthen their defenses before a threat manifests [404, 405]. This proactive approach helps in preemptively mitigating vulnerabilities and implementing measures to thwart anticipated attacks. Threat hunting automation involves the utilization of AI-driven tools to actively search for potential threats within an organization's network [406, 407]. These automated systems continuously monitor network traffic, endpoints, and system logs, identifying suspicious behaviors or indicators of compromise. By automating this process, security analysts can focus on investigating and responding to verified threats, enhancing the efficiency and effectiveness of threat detection and response. Blockchain's decentralized and immutable nature offers enhanced security in storing and sharing threat intelligence data [408, 409, 410] By utilizing blockchain, security professionals can ensure data integrity, traceability, and secure sharing of threat information across multiple entities [411, 412, 413]. This technology helps in establishing a trusted environment for exchanging sensitive threat data while preventing unauthorized alterations or tampering of information [414, 415, 416, 417, 418]. The potential of quantum computing in cybersecurity lies in its ability to exponentially accelerate computation, significantly impacting encryption and decryption techniques [419, 420,421, 422]. While quantum computing holds promise in developing more robust cryptographic solutions, it also poses challenges by potentially rendering current encryption methods vulnerable [423,424, 425, 426]. Research in quantum-resistant algorithms becomes crucial to prepare for the security implications posed by quantum computing advancements [427].

The convergence of AI with Cyber-Physical Systems (CPS) involves securing interconnected systems that integrate physical components with digital networks [428, 429, 430]. This integration spans critical infrastructure, industrial control systems, and IoT devices. AI augments threat intelligence in CPS by continuously monitoring and analyzing data from these systems, detecting anomalies or potential attacks that could disrupt operations or cause physical harm, thereby enhancing overall security measures [431, 432]. Explainable AI (XAI) focuses on making AI models transparent and understandable to humans [433, 434, 435]. In the context of cybersecurity, XAI allows security professionals to interpret AI-generated insights, understand the reasoning behind threat assessments, and validate the credibility of AI-based security decisions [436, 437]. This transparency is crucial for building trust and ensuring effective collaboration between AI systems and human experts [438, 439]. Inspired by collective behavior in nature, swarm intelligence algorithms facilitate collaborative threat detection and response [440, 441]. These algorithms enable interconnected security systems to share threat intelligence, collectively analyze data, and make decisions in real-time [442, 443]. Swarm intelligence enhances the overall resilience of security ecosystems by fostering cooperative defense mechanisms against evolving threats [444].

These technologies collectively drive the evolution of AI-driven threat intelligence, empowering organizations to bolster their cybersecurity posture amidst increasingly sophisticated and dynamic threats. Continuing research and innovation in these areas are pivotal in staying ahead of adversaries and safeguarding digital assets and infrastructure.

**Anticipated Challenges and Areas for Further Research**

One significant challenge lies in the abundance of data sources, leading to information overload. The vastness and diversity of these sources complicate the aggregation and analysis of data, potentially resulting in information gaps and inconsistencies [445, 446]. Moreover, the dynamic nature of cyber threats poses a challenge in maintaining the relevance and timeliness of threat intelligence [447, 448]. Another pressing concern involves the lack of standardization in threat intelligence sharing and reporting. Varying formats and classification systems hinder effective collaboration among entities, limiting the seamless exchange of critical threat information [449, 450]. Additionally, the issue of data quality and accuracy persists, as the authenticity and reliability of sources remain questionable [451, 452]. Ensuring the trustworthiness and validity of collected data is imperative to prevent erroneous threat assessments and subsequent misinformed actions [453, 454]. Further challenges arise from the need for improved automation and artificial intelligence (AI) integration within threat intelligence systems [455, 456]. While automation can streamline processes, the complexity of threat landscapes demands advanced AI capabilities to detect and respond to evolving threats in real-time [457, 458]. Developing AI-driven solutions that adapt to emerging threat patterns is crucial for proactive threat mitigation [459]. Additionally, there is a growing necessity for enhanced threat intelligence sharing across international borders [460, 461]. The global nature of cyber threats necessitates cooperation among nations, yet legal and geopolitical barriers hinder seamless cross-border information exchange [462, 463]. Addressing these barriers requires in-depth research to devise frameworks that facilitate secure and efficient international collaboration in threat intelligence. Moreover, the emergence of new technologies like the Internet of Things (IoT) and 5G introduces novel threat vectors that demand thorough exploration [464, 465]. Understanding the unique risks associated with these technologies and devising effective countermeasures are imperative to preempt potential vulnerabilities [466, 467].

In conclusion, addressing these anticipated challenges in threat intelligence necessitates focused research efforts. Overcoming issues related to data overload, standardization, data quality, automation, international cooperation, and emerging technologies will fortify the effectiveness of threat intelligence in combating evolving cyber threats. Figure 8 shows top 10 emerging cyber security for 2023.

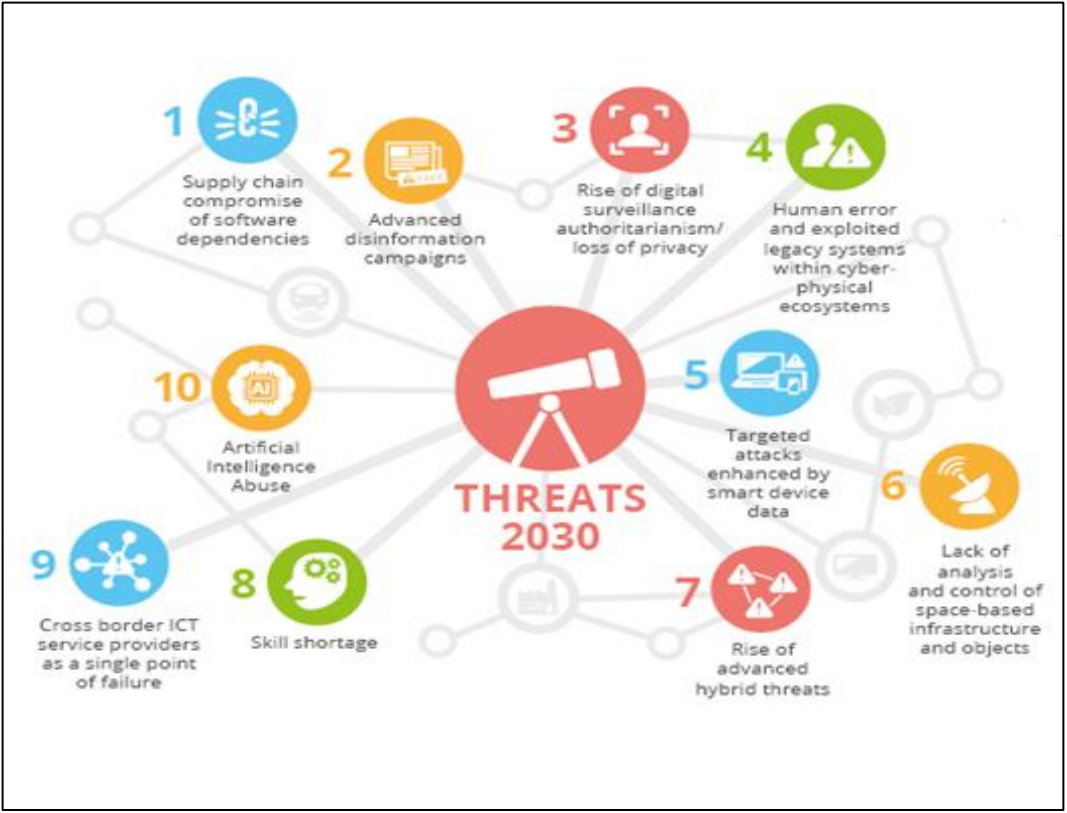

Figure 8: Top 10 emerging cyber security threats for 2023 [575]

## 9. CONCLUSION

**Summarization of key insights and contributions**

In this comprehensive review paper, the focus centers on the burgeoning realm of autonomous threat hunting propelled by AI-driven methodologies. Beginning with a profound exploration into the historical landscape of threat intelligence, the paper meticulously navigates through the evolutionary trajectory that led to the imperative integration of AI in this domain. The core contribution lies in elucidating the conceptual framework of autonomous threat hunting, delineating its components, and expounding on the pivotal role of AI algorithms within this paradigm. By examining state-of-the-art AI techniques, from machine learning models to deep learning architectures and NLP applications, the paper showcases the breadth and depth of AI's impact on threat detection, anomaly identification, and contextual threat information extraction.

However, amidst this technological marvel, the paper astutely highlights the pertinent challenges that accompany autonomous threat hunting. The scalability, interpretability, and ethical dimensions of AI-driven models emerge as crucial focal points warranting attention. Real-world case studies substantiate these discussions, providing invaluable insights and lessons learned from practical implementations. Furthermore, the paper offers a structured approach to evaluating the performance of autonomous systems, establishing a set of metrics for comparison against conventional methods. It transcends mere analysis, delving into future directions and emerging trends, envisioning the trajectory of AI-driven threat intelligence. It beckons researchers and practitioners alike to embrace this evolution, while concurrently flagging potential challenges and uncharted territories that merit further exploration. Ultimately, the review paper culminates in a concise yet powerful conclusion, encapsulating the transformative implications of autonomous threat hunting for cybersecurity. It beckons a call to action, inviting stakeholders to engage in further research and implementation, underscoring the paper's paramount contributions to this dynamic field.

**Implications Of Autonomous Threat Hunting for The Future of Cybersecurity**

Autonomous threat hunting, empowered by AI algorithms, enhances threat detection capabilities by continuously analyzing vast amounts of data [468]. It employs predictive analytics to identify anomalous patterns indicative of potential threats [469, 470] By utilizing historical attack data and real-time information, these systems can swiftly identify and mitigate emerging threats, reducing the dwell time of attackers within networks [471, 472]. Automation in threat hunting expedites incident response [473, 474]. Autonomous systems can swiftly investigate and contain threats, minimizing the impact of cyber-attacks [475, 476]. By autonomously orchestrating responses, such systems alleviate the burden on human analysts, allowing them to focus on more complex tasks requiring human intuition and decision-making [477]. Autonomous threat hunting does not replace human expertise but rather augments it [478, 479]. Human analysts can collaborate with AI systems to validate findings, interpret complex threats, and make strategic decisions [480, 481]. This synergy between human intelligence and machine learning algorithms optimizes efficiency and accuracy in threat identification and response [482, 483, 484]. The rise of autonomous threat hunting also raises ethical concerns, including privacy implications and the potential for algorithmic [485, 486]. Ensuring transparency in AI decision-making and implementing ethical guidelines are imperative to mitigate these issues [487, 488]. Striking a balance between autonomy and human oversight is crucial to uphold ethical standards in cybersecurity practices.

Autonomous threat hunting represents a paradigm shift in cybersecurity, offering enhanced threat detection, faster incident response, and workforce augmentation. However, ethical considerations surrounding privacy and biases necessitate careful monitoring and regulation. The future of cybersecurity lies in the collaboration between human expertise and autonomous systems to fortify defenses against evolving cyber threats.

**Call To Action for Researchers and Practitioners**

In embracing the realm of autonomous threat hunting powered by AI-driven intelligence, researchers and practitioners stand at the precipice of a transformative era in cybersecurity. This call to action heralds a collective responsibility to fortify our digital landscapes against evolving threats. We urge researchers to delve deeper, pushing the boundaries of AI algorithms, seeking more robust, explainable, and ethical frameworks. For practitioners, the call to action resonates in implementation. Embrace and integrate autonomous threat hunting systems into organizational security architectures. Drive initiatives to amalgamate human expertise with AI's prowess, ensuring symbiotic coexistence that enhances threat detection, response, and mitigation. Collaboration becomes our linchpin. Cross-disciplinary collaborations between academia, industry, and policymakers must flourish. This collaboration fosters the exchange of ideas, best practices, and ethical standards, nurturing a collective intelligence to tackle cybersecurity challenges. As stewards of technological advancement, let's remain vigilant against potential biases and ethical dilemmas. Let's champion transparency and ethical conduct in deploying AI-driven solutions. Together, let's forge a future where autonomous threat hunting not only safeguards our digital ecosystems but also upholds the principles of security, transparency, and ethical responsibility. This call to action implores us to unite in securing our digital future." This call to action seeks an active and collective involvement from researchers and practitioners, emphasizing collaboration, ethics, and transparency in advancing autonomous threat hunting for the greater good of cybersecurity.

# REFERENCES

[1] M. K. Aiden, S. M. Sabharwal, S. Chhabra, and M. A. Al-Asadi, "AI and Blockchain for Cyber Security in Cyber-Physical System," in *Springer eBooks*, 2023, pp. 203–230. doi: 10.1007/978-3-031-31952-5_10. Available: https://doi.org/10.1007/978-3-031-31952-5_10

[2] M. F. Safitra, M. Lubis, and H. Fakhrurroja, "Counterattacking Cyber Threats: A framework for the Future of Cybersecurity," *Sustainability*, vol. 15, no. 18, p. 13369, Sep. 2023, doi: 10.3390/su151813369. Available: https://doi.org/10.3390/su151813369

[3] S. Kumar, U. Gupta, A. Singh, and A. K. Singh, "Artificial intelligence," *Journal of Computers Mechanical and Management*, vol. 2, no. 3, pp. 31–42, Aug. 2023, doi: 10.57159/gadl.jcmm.2.3.23064. Available: https://doi.org/10.57159/gadl.jcmm.2.3.23064

[4] M. S. Rich, "Cyberpsychology: A longitudinal analysis of cyber adversarial tactics and techniques," *Analytics*, vol. 2, no. 3, pp. 618–655, Aug. 2023, doi: 10.3390/analytics2030035. Available: https://doi.org/10.3390/analytics2030035

[5] A.Iyer and K. S. Umadevi, "Role of AI and its impact on the development of cyber security applications," in *Advanced technologies and societal change*, 2023, pp. 23–46. doi: 10.1007/978-981-99-2115-7_2. Available: https://doi.org/10.1007/978-981-99-2115-7_2

[6] P. Dhoni and R. Kumar, "Synergizing Generative AI and Cybersecurity: Roles of Generative AI Entities, Companies, Agencies, and Government in Enhancing Cybersecurity," *TechRxiv*, Aug. 2023, doi: 10.36227/techrxiv.23968809.v1. Available: https://doi.org/10.36227/techrxiv.23968809.v1

[7] A.B. Z. Mahmoodi, S. Sheikhi, E. Peltonen, and P. Kostakos, "Autonomous federated learning for distributed intrusion detection systems in public networks," *IEEE Access*, vol. 11, pp. 121325–121339, Jan. 2023, doi: 10.1109/access.2023.3327922. Available: https://doi.org/10.1109/access.2023.3327922

[8] Θ. Θεοδωρόπουλος *et al.*, "Security in Cloud-Native Services: A survey," *Journal of Cybersecurity and Privacy*, vol. 3, no. 4, pp. 758–793, Oct. 2023, doi: 10.3390/jcp3040034. Available: https://doi.org/10.3390/jcp3040034

[9] M. M. Nair, A. Deshmukh, and A. K. Tyagi, "Artificial Intelligence for Cyber Security," *Wiley*, pp. 83–114, Nov. 2023, doi: 10.1002/9781394213948.ch5. Available: https://doi.org/10.1002/9781394213948.ch5

[10] E. M. Kala, "The impact of Cyber security on business: How to protect your business," *Open Journal of Safety Science and Technology*, vol. 13, no. 02, pp. 51–65, Jan. 2023, doi: 10.4236/ojsst.2023.132003. Available: https://doi.org/10.4236/ojsst.2023.132003

[11] R. Shawe and I. McAndrew, "Increasing threats to United States of America infrastructure based on Cyber-Attacks," *Journal of Software Engineering and Applications*, vol. 16, no. 10, pp. 530–547, Jan. 2023, doi: 10.4236/jsea.2023.1610027. Available: https://doi.org/10.4236/jsea.2023.1610027

[12] A.N. Lone, S. Mustajab, and M. Alam, "A comprehensive study on cybersecurity challenges and opportunities in the IoT world," *Security and Privacy*, vol. 6, no. 6, Apr. 2023, doi: 10.1002/spy2.318. Available: https://doi.org/10.1002/spy2.318

[13] J. M. Qurashi, K. Jambi, F. Alsolami, F. Eassa, M. Khemakhem, and A. Basuhail, "Resilient countermeasures against Cyber-Attacks on Self-Driving car architecture," *IEEE Transactions on Intelligent Transportation Systems*, pp. 1–30, Jan. 2023, doi: 10.1109/tits.2023.3288192. Available: https://doi.org/10.1109/tits.2023.3288192

[14] N. Sun *et al.*, "Cyber Threat Intelligence Mining for Proactive Cybersecurity Defense: A survey and New Perspectives," *IEEE Communications Surveys and Tutorials*, vol. 25, no. 3, pp. 1748–1774, Jan. 2023, doi: 10.1109/comst.2023.3273282. Available: https://doi.org/10.1109/comst.2023.3273282

[15] N. Joseph, "The growing threat of supply chain cyberattacks: Resilience strategies," *Social Science Research Network*, Jan. 2023, doi: 10.2139/ssrn.4643716. Available: https://doi.org/10.2139/ssrn.4643716

[16] P. Kumar *et al.*, "Explainable artificial intelligence envisioned security mechanism for cyber threat hunting," *Security and Privacy*, Mar. 2023, doi: 10.1002/spy2.312. Available: https://doi.org/10.1002/spy2.312

[17] M. A. Lozano, I. P. Llopis, and M. Domingo, "Threat hunting system for protecting critical infrastructures using a machine learning approach," *Mathematics*, vol. 11, no. 16, p. 3448, Aug. 2023, doi: 10.3390/math11163448. Available: https://doi.org/10.3390/math11163448

[18] H. Allioui, "Unleashing the potential of AI: Investigating Cutting-Edge technologies that are transforming businesses," Aug. 16, 2023. Available: https://www.ijceds.com/ijceds/article/view/59

[19] J. Kotsias, A. Ahmad, and R. Scheepers, "Adopting and integrating cyber-threat intelligence in a commercial organisation," *European Journal of Information Systems*, vol. 32, no. 1, pp. 35–51, Jul. 2022, doi: 10.1080/0960085x.2022.2088414. Available: https://doi.org/10.1080/0960085x.2022.2088414

[20] A.S. George, S.Sagayarajan, T. Baskar, and A. S. H. George, "Extending Detection and Response: How MXDR Evolves Cybersecurity," *Zenodo (CERN European Organization for Nuclear Research)*, Aug. 2023, doi: 10.5281/zenodo.8284342. Available: https://zenodo.org/record/8284342

[21] R. Ghioni, M. Taddeo, and L. Floridi, "Open source intelligence and AI: a systematic review of the GELSI literature," *AI & SOCIETY*, Jan. 2023, doi: 10.1007/s00146-023-01628-x. Available: https://doi.org/10.1007/s00146-023-01628-x

[22] N. Pocher, M. Zichichi, F. Merizzi, M. Z. Shafiq, and S. Ferretti, "Detecting anomalous cryptocurrency transactions: An AML/CFT application of machine learning-based forensics," *Electronic Markets*, vol. 33, no. 1, Jul. 2023, doi: 10.1007/s12525-023-00654-3. Available: https://doi.org/10.1007/s12525-023-00654-3

[23] U. Rauf, Z. Wei, and F. Mohsen, "Employee Watcher: A Machine Learning-based Hybrid Insider Threat Detection Framework," *IEEE*, Oct. 2023, doi: 10.1109/csnet59123.2023.10339777. Available: https://doi.org/10.1109/csnet59123.2023.10339777

[24] S. A. Jawaid, "Artificial Intelligence with Respect to Cyber Security," *Preprint*, Apr. 2023, doi: 10.20944/preprints202304.0923.v1. Available: https://doi.org/10.20944/preprints202304.0923.v1

[25] V. V. Vegesna, "Enhancing cyber resilience by integrating AI-Driven threat detection and mitigation strategies," *Vegesna | Transactions on Latest Trends in Artificial Intelligence*, Dec. 04, 2023. Available: https://ijsdcs.com/index.php/TLAI/article/view/396A.

[26] J. Varma *et al.*, "A roadmap for SMEs to adopt an AI based cyber threat intelligence," in *Studies in computational intelligence*, 2023, pp. 1903–1926. doi: 10.1007/978-3-031-12382-5_105. Available: https://doi.org/10.1007/978-3-031-12382-5_105

[27] R. Kunduru, "Cloud BPM Application (APPIAN) Robotic Process Automation capabilities," *Asian Journal of Research in Computer Science*, vol. 16, no. 3, pp. 267–280, Aug. 2023, doi: 10.9734/ajrcos/2023/v16i3361. Available: https://doi.org/10.9734/ajrcos/2023/v16i3361

[28] N. Rane, "Role of ChatGPT and similar generative artificial intelligence (AI) in construction industry," *Social Science Research Network*, Jan. 2023, doi: 10.2139/ssrn.4598258. Available: https://doi.org/10.2139/ssrn.4598258

[29] N. Rane, "Integrating Leading-Edge Artificial Intelligence (AI), Internet of things (IoT), and big Data technologies for smart and Sustainable Architecture, Engineering and Construction (AEC) industry: challenges and future directions," *Social Science Research Network*, Jan. 2023, doi: 10.2139/ssrn.4616049. Available: https://doi.org/10.2139/ssrn.4616049

[30] Y. Mirsky *et al.*, "The threat of offensive AI to organizations," *Computers & Security*, vol. 124, p. 103006, Jan. 2023, doi: 10.1016/j.cose.2022.103006. Available: https://doi.org/10.1016/j.cose.2022.103006

[31] A.J. G. De Azambuja, C. Plesker, K. Schützer, R. Anderl, B. Schleich, and V. R. Almeida, "Artificial Intelligence-Based Cyber Security in the context of Industry 4.0—A survey," *Electronics*, vol. 12, no. 8, p. 1920, Apr. 2023, doi: 10.3390/electronics12081920. Available: https://doi.org/10.3390/electronics12081920

[32] N. K. E. Al, "AI in Cybersecurity: Threat Detection and Response with Machine Learning," *Tuijin Jishu/Journal of Propulsion Technology*, vol. 44, no. 3, pp. 38–46, Sep. 2023, doi: 10.52783/tjjpt.v44.i3.237. Available: https://doi.org/10.52783/tjjpt.v44.i3.237

[33] A.Habbal, M. Ali, and M. A. Abuzaraida, "Artificial Intelligence Trust, Risk and Security Management (AI TRiSM): Frameworks, applications, challenges and future research directions," *Expert Systems With Applications*, vol. 240, p. 122442, Apr. 2024, doi: 10.1016/j.eswa.2023.122442. Available: https://doi.org/10.1016/j.eswa.2023.122442

[34] M. Schmitt and I. Fléchais, "Digital Deception: Generative artificial intelligence in social engineering and phishing," *arXiv (Cornell University)*, Oct. 2023, doi: 10.48550/arxiv.2310.13715. Available: https://arxiv.org/abs/2310.13715

[35] Y. Mirsky *et al.*, "The threat of offensive AI to organizations," *Computers & Security*, vol. 124, p. 103006, Jan. 2023, doi: 10.1016/j.cose.2022.103006. Available: https://doi.org/10.1016/j.cose.2022.103006

[36] A.Bajaj and D. K. Vishwakarma, "Evading text based emotion detection mechanism via adversarial attacks," *Neurocomputing*, vol. 558, p. 126787, Nov. 2023, doi: 10.1016/j.neucom.2023.126787. Available: https://doi.org/10.1016/j.neucom.2023.126787

[37] T. M. Ali, A. Eleyan, and T. Bejaoui, "Detecting Conventional and Adversarial Attacks Using Deep Learning Techniques: A Systematic Review," *IEEE*, Oct. 2023, doi: 10.1109/isncc58260.2023.10323872. Available: https://doi.org/10.1109/isncc58260.2023.10323872

[38] G. M. Félix, "Opportunities, risks and applications of open source Intelligence in cybersecurity and cyberdefence," Oct. 16, 2023. Available: https://digitum.um.es/digitum/handle/10201/134863

[39] V. Jesus, B. Bains, and V. Chang, "Sharing is Caring: Hurdles and Prospects of Open, Crowd-Sourced Cyber Threat Intelligence," *IEEE Transactions on Engineering Management*, pp.

1–20, Jan. 2023, doi: 10.1109/tem.2023.3279274. Available: https://doi.org/10.1109/tem.2023.3279274

[40] P. Martin, *Insider risk and personnel security: An introduction*. Taylor & Francis, 2023.

[41] A.N. Shulsky and G. J. Schmitt, *Silent Warfare: Understanding the World of Intelligence*. Potomac Books, Inc., 2011.

[42] Prezelj and T. T. Ristevska, "Intelligence scandals: a comparative analytical model and lessons learned from the test case of North Macedonia," *Intelligence & National Security*, vol. 38, no. 1, pp. 143–170, Apr. 2022, doi: 10.1080/02684527.2022.2065616. Available: https://doi.org/10.1080/02684527.2022.2065616

[43] Alsmadi, "Cyber Intelligence Analysis," in *Springer eBooks*, 2023, pp. 85–129. doi: 10.1007/978-3-031-21651-0_6. Available: https://doi.org/10.1007/978-3-031-21651-0_6

[44] Dehghantanha, M. Conti, and T. Dargahi, *Cyber Threat intelligence*. Springer, 2018.

[45] F. Saâdaoui and S. B. Jabeur, "Analyzing the influence of geopolitical risks on European power prices using a multiresolution causal neural network," *Energy Economics*, vol. 124, p. 106793, Aug. 2023, doi: 10.1016/j.eneco.2023.106793. Available: https://doi.org/10.1016/j.eneco.2023.106793

[46] H. Naseer, K. C. Desouza, S. B. Maynard, and A. Ahmad, "Enabling cybersecurity incident response agility through dynamic capabilities: the role of real-time analytics," *European Journal of Information Systems*, pp. 1–21, Sep. 2023, doi: 10.1080/0960085x.2023.2257168. Available: https://doi.org/10.1080/0960085x.2023.2257168

[47] R. Borum, J. Felker, S. Kern, K. Dennesen, and T. Feyes, "Strategic cyber intelligence," *Information & Computer Security*, Jul. 2015, doi: 10.1108/ics-09-2014-0064. Available: https://doi.org/10.1108/ics-09-2014-0064

[48] A.R. Sinha, K. Singla, and T. M. M. Victor, "Artificial intelligence and machine learning for cybersecurity applications and challenges," in *Advances in information security, privacy, and ethics book series*, 2023, pp. 109–146. doi: 10.4018/978-1-6684-9317-5.ch007. Available: https://doi.org/10.4018/978-1-6684-9317-5.ch007

[49] H. Alqahtani and G. Kumar, "Machine learning for enhancing transportation security: A comprehensive analysis of electric and flying vehicle systems," *Engineering Applications of Artificial Intelligence*, vol. 129, p. 107667, Mar. 2024, doi: 10.1016/j.engappai.2023.107667. Available: https://doi.org/10.1016/j.engappai.2023.107667

[50] A.Padhi, A. Ashwini, S. K. Saxena, and C. D. S. Katoch, "Transforming clinical virology with AI, machine learning and deep learning: a comprehensive review and outlook," *VirusDisease*, vol. 34, no. 3, pp. 345–355, Sep. 2023, doi: 10.1007/s13337-023-00841-y. Available: https://doi.org/10.1007/s13337-023-00841-y

[51] Bécue, I. Praça, and J. Gama, "Artificial intelligence, cyber-threats and Industry 4.0: challenges and opportunities," *Artificial Intelligence Review*, vol. 54, no. 5, pp. 3849–3886, Feb. 2021, doi: 10.1007/s10462-020-09942-2. Available: https://doi.org/10.1007/s10462-020-09942-2

[52] Patcha and J. Park, "An overview of anomaly detection techniques: Existing solutions and latest technological trends," *Computer Networks*, vol. 51, no. 12, pp. 3448–3470, Aug. 2007, doi: 10.1016/j.comnet.2007.02.001. Available: https://doi.org/10.1016/j.comnet.2007.02.001

[53] P. Sharma, K. K. Sarma, and N. E. Mastorakis, "Artificial intelligence aided Electronic Warfare systems- recent trends and evolving applications," *IEEE Access*, vol. 8, pp. 224761–

224780, Jan. 2020, doi: 10.1109/access.2020.3044453. Available: https://doi.org/10.1109/access.2020.3044453

[54] F. Medjek, D. Tandjaoui, N. Djedjig, and I. Romdhani, "Fault-tolerant AI-driven intrusion detection system for the internet of things," *International Journal of Critical Infrastructure Protection*, vol. 34, p. 100436, Sep. 2021, doi: 10.1016/j.ijcip.2021.100436. Available: https://doi.org/10.1016/j.ijcip.2021.100436

[55] S. Cui, "AI-driven methods for resiliency and security assessment: the case for autonomous driving system and HPC storage system," Apr. 14, 2021. Available: https://www.ideals.illinois.edu/items/118498

[56] A.K. Kassem, "Intelligent system using machine learning techniques for security assessment and cyber intrusion detection," Jul. 23, 2021. Available: https://theses.hal.science/tel-03522384/

[57] D. Dunsin, M. C. Ghanem, K. Ouazzane, and V. Vassilev, "A comprehensive analysis of the role of artificial intelligence and machine learning in modern digital forensics and incident response," *arXiv (Cornell University)*, Sep. 2023, doi: 10.48550/arxiv.2309.07064. Available: https://arxiv.org/abs/2309.07064

[58] T. Johnson, "The AI-cyber security nexus," in *Manchester University Press eBooks*, 2021. doi: 10.7765/9781526145062.00016. Available: https://doi.org/10.7765/9781526145062.00016

[59] Johnson, "The AI-cyber nexus: implications for military escalation, deterrence and strategic stability," *Journal of Cyber Policy*, vol. 4, no. 3, pp. 442–460, Sep. 2019, doi: 10.1080/23738871.2019.1701693. Available: https://doi.org/10.1080/23738871.2019.1701693

[60] A.Tsamados *et al.*, "The Ethics of Algorithms: Key problems and solutions," in *Philosophical studies series*, 2021, pp. 97–123. doi: 10.1007/978-3-030-81907-1_8. Available: https://doi.org/10.1007/978-3-030-81907-1_8

[61] N. Díaz-Rodríguez, J. Del Ser, M. Coeckelbergh, M. L. De Prado, E. Herrera-Viedma, and F. Herrera, "Connecting the dots in trustworthy Artificial Intelligence: From AI principles, ethics, and key requirements to responsible AI systems and regulation," *Information Fusion*, vol. 99, p. 101896, Nov. 2023, doi: 10.1016/j.inffus.2023.101896. Available: https://doi.org/10.1016/j.inffus.2023.101896

[62] L. Dhirani, N. Mukhtiar, B. S. Chowdhry, and T. Newe, "Ethical Dilemmas and Privacy Issues in Emerging Technologies: A review," *Sensors*, vol. 23, no. 3, p. 1151, Jan. 2023, doi: 10.3390/s23031151. Available: https://doi.org/10.3390/s23031151

[63] S. Al-Mansoori, "The role of artificial intelligence and machine learning in shaping the future of cybersecurity: trends, applications, and ethical considerations," Sep. 21, 2023. Available: https://norislab.com/index.php/ijsa/article/view/36

[64] A.Nassar, "Machine Learning and Big Data analytics for Cybersecurity Threat Detection: A Holistic review of techniques and case studies," Feb. 06, 2021. Available: https://journals.sagescience.org/index.php/jamm/article/view/97

[65] F. Bouchama, "Enhancing Cyber Threat Detection through Machine Learning-Based Behavioral Modeling of Network Traffic Patterns," Sep. 03, 2021. Available: https://research.tensorgate.org/index.php/IJBIBDA/article/view/76

[66] Z. T. Sworna, Z. Mousavi, and M. A. Babar, "NLP methods in host-based intrusion detection systems: A systematic review and future directions," *Journal of Network and Computer*

*Applications*, vol. 220, p. 103761, Nov. 2023, doi: 10.1016/j.jnca.2023.103761. Available: https://doi.org/10.1016/j.jnca.2023.103761

[67] Arazzi *et al.*, "NLP-Based Techniques for Cyber Threat Intelligence," *arXiv (Cornell University)*, Nov. 2023, doi: 10.48550/arxiv.2311.08807. Available: https://arxiv.org/abs/2311.08807

[68] P. Singh, "Artificial intelligence: the backbone of national security in 21st century," *Tuijin Jishu/Journal of Propulsion Technology*, vol. 44, no. 4, pp. 2022–2038, Oct. 2023, doi: 10.52783/tjjpt.v44.i4.1174. Available: https://doi.org/10.52783/tjjpt.v44.i4.1174

[69] "EBSCOhost | 173039437 | Enhancing Cyber Forensics with AI and Machine Learning: A Study on Automated Threat Analysis and Classification." Available: https://web.s.ebscohost.com/abstract?direct=true&profile=ehost&scope=site&authtype=crawler&jrnl=20419031&AN=173039437&h=5iQkNHBpCMiVG9ta1Jv69Sg8KstTubHRtdtyvWESEDnFFBD4EyXv3YFDeqzRde%2fGc2L35djsioth7byX02l%2brg%3d%3d&crl=c&resultNs=AdminWebAuth&resultLocal=ErrCrlNotAuth&crlhashurl=login.aspx%3fdirect%3dtrue%26profile%3dehost%26scope%3dsite%26authtype%3dcrawler%26jrnl%3d20419031%26AN%3d173039437

[70] S. Rangaraju, "SECURE BY INTELLIGENCE: ENHANCING PRODUCTS WITH AI-DRIVEN SECURITY MEASURES," *EPH - International Journal of Science and Engineering*, vol. 9, no. 3, pp. 36–41, Dec. 2023, doi: 10.53555/ephijse.v9i3.212. Available: https://doi.org/10.53555/ephijse.v9i3.212

[71] Kaloudi and J. Li, "The AI-Based cyber threat landscape," *ACM Computing Surveys*, vol. 53, no. 1, pp. 1–34, Feb. 2020, doi: 10.1145/3372823. Available: https://doi.org/10.1145/3372823

[72] V. Lai, C. Chen, Q. V. Liao, A. Smith-Renner, and C. Tan, "Towards a Science of Human-AI Decision Making: A Survey of Empirical Studies," *arXiv (Cornell University)*, Dec. 2021, doi: 10.48550/arxiv.2112.11471. Available: http://arxiv.org/abs/2112.11471

[73] T. Heilig and I. Scheer, *Decision intelligence: Transform Your Team and Organization with AI-Driven Decision-Making*. John Wiley & Sons, 2023.

[74] M. Yungaicela-Naula, J. A. P. Díaz, M. Zareei, and C. Vargas-Rosales, "Towards security automation in Software Defined Networks," *Computer Communications*, vol. 183, pp. 64–82, Feb. 2022, doi: 10.1016/j.comcom.2021.11.014. Available: https://doi.org/10.1016/j.comcom.2021.11.014

[75] HuYupeng *et al.*, "Artificial intelligence Security: Threats and countermeasures," *ACM Computing Surveys*, vol. 55, no. 1, pp. 1–36, Nov. 2021, doi: 10.1145/3487890. Available: https://doi.org/10.1145/3487890

[76] E. Lin-Greenberg, "Allies and Artificial Intelligence: Obstacles to Operations and Decision-Making (Spring 2020)," *Texas School Work*, Jan. 2020, doi: 10.26153/tsw/8866. Available: https://repositories.lib.utexas.edu/handle/2152/81858

[77] N. Rane, "Role and challenges of CHATGPT and similar generative artificial intelligence in human resource Management," *Social Science Research Network*, Jan. 2023, doi: 10.2139/ssrn.4603230. Available: https://doi.org/10.2139/ssrn.4603230

[78] V. V. Vegesna, "AI-Enabled Blockchain Solutions for Sustainable Development, Harnessing Technological Synergy towards a Greener Future," Dec. 06, 2023. Available: https://ijsdai.com/index.php/IJSDAI/article/view/23

[79]     Hillier, "Turning the Hunted into the Hunter via Threat Hunting: Life Cycle, Ecosystem, Challenges and the Great Promise of AI," *arXiv.org*, Apr. 23, 2022. Available: https://arxiv.org/abs/2204.11076

[80]     R. Puzis, "ATHAFI: Agile threat hunting and Forensic investigation," *arXiv.org*, Mar. 07, 2020. Available: https://arxiv.org/abs/2003.03663

[81]     Sharma and B. Dash, "Impact of Big Data Analytics and ChatGPT on Cybersecurity," *IEEE*, Mar. 2023, doi: 10.1109/i3cs58314.2023.10127411. Available: https://doi.org/10.1109/i3cs58314.2023.10127411

[82]     Muralidhara, "THE EVOLUTION OF CLOUD COMPUTING SECURITY: ADDRESSING EMERGING THREATS," Dec. 06, 2017. Available: http://ijcst.com.pk/index.php/IJCST/article/view/237

[83]     Y. Diogenes and E. Ozkaya, *Cybersecurity – attack and defense strategies: Counter modern threats and employ state-of-the-art tools and techniques to protect your organization against cybercriminals*. Packt Publishing Ltd, 2019.

[84]     A. Parisi, *Hands-On Artificial Intelligence for Cybersecurity: Implement smart AI systems for preventing cyber attacks and detecting threats and network anomalies*. Packt Publishing Ltd, 2019.

[85]     M. A. Lozano, I. P. Llopis, and M. Domingo, "Threat hunting architecture using a machine learning approach for critical infrastructures protection," *Big Data and Cognitive Computing*, vol. 7, no. 2, p. 65, Mar. 2023, doi: 10.3390/bdcc7020065. Available: https://doi.org/10.3390/bdcc7020065

[86]     M. A. Lozano, I. P. Llopis, and M. Domingo, "Threat hunting architecture using a machine learning approach for critical infrastructures protection," *Big Data and Cognitive Computing*, vol. 7, no. 2, p. 65, Mar. 2023, doi: 10.3390/bdcc7020065. Available: https://doi.org/10.3390/bdcc7020065

[87]     M. A. Lozano, I. P. Llopis, and M. Domingo, "Threat hunting architecture using a machine learning approach for critical infrastructures protection," *Big Data and Cognitive Computing*, vol. 7, no. 2, p. 65, Mar. 2023, doi: 10.3390/bdcc7020065. Available: https://doi.org/10.3390/bdcc7020065

[88]     "AIS Electronic Library (AISeL) - ICIS 2023 Proceedings: Unsupervised Threat Hunting using Continuous Bag of Terms and Time (CBoTT)." Available: https://aisel.aisnet.org/icis2023/cyber_security/cyber_security/6/

[89]     Á. C. Bienzobas, "Threat Trekker: An approach to cyber threat hunting," *arXiv.org*, Oct. 06, 2023. Available: https://arxiv.org/abs/2310.04197

[90]     Unger *et al.*, "Securing the future railway system: technology forecast, security measures, and research demands," *Vehicles*, vol. 5, no. 4, pp. 1254–1274, Sep. 2023, doi: 10.3390/vehicles5040069. Available: https://doi.org/10.3390/vehicles5040069

[91]     P. F. Möller, "Threats and threat intelligence," in *Advances in information security*, 2023, pp. 71–129. doi: 10.1007/978-3-031-26845-8_2. Available: https://doi.org/10.1007/978-3-031-26845-8_2

[92]     Kaur, D. Gabrijelčič, and T. Klobučar, "Artificial intelligence for cybersecurity: Literature review and future research directions," *Information Fusion*, vol. 97, p. 101804, Sep. 2023, doi: 10.1016/j.inffus.2023.101804. Available: https://doi.org/10.1016/j.inffus.2023.101804

[93] "Threat intelligence and sharing for collaborative defense," *International Journal of Mechanical Engineering*, vol. 8, Jan. 2023, doi: 10.56452/7-5-200. Available: https://doi.org/10.56452/7-5-200

[94] A.Dehghantanha, M. Conti, and T. Dargahi, *Cyber Threat intelligence*. Springer, 2018.

[95] Kalla, "Advantages, Disadvantages and Risks Associated with ChatGPT and AI on Cybersecurity," Oct. 31, 2023. Available: https://papers.ssrn.com/sol3/papers.cfm?abstract_id=4619204

[96] Dehghantanha, M. Conti, and T. Dargahi, *Cyber Threat intelligence*. Springer, 2018.

[97] M. Repetto, "Adaptive monitoring, detection, and response for agile digital service chains," *Computers & Security*, vol. 132, p. 103343, Sep. 2023, doi: 10.1016/j.cose.2023.103343. Available: https://doi.org/10.1016/j.cose.2023.103343

[98] M. Sarfraz, *Cybersecurity Threats with New Perspectives*. BoD – Books on Demand, 2021.

[99] H. HaddadPajouh, A. Mohtadi, A. Dehghantanaha, H. Karimipour, X. Lin, and K. R. Choo, "A multikernel and metaheuristic feature selection approach for IoT malware threat hunting in the edge layer," *IEEE Internet of Things Journal*, vol. 8, no. 6, pp. 4540–4547, Mar. 2021, doi: 10.1109/jiot.2020.3026660. Available: https://doi.org/10.1109/jiot.2020.3026660

[100] Saeed, S. A. Suayyid, M. S. Al-Ghamdi, H. A. Almuhaisen, and A. M. Almuhaideb, "A Systematic Literature Review on Cyber Threat Intelligence for Organizational Cybersecurity Resilience," *Sensors*, vol. 23, no. 16, p. 7273, Aug. 2023, doi: 10.3390/s23167273. Available: https://doi.org/10.3390/s23167273

[101] P. Duy, N. H. Quyen, N. H. Khoa, T.-D. Tran, and V.-H. Pham, "FedChain-Hunter: A reliable and privacy-preserving aggregation for federated threat hunting framework in SDN-based IIoT," *Internet of Things*, vol. 24, p. 100966, Dec. 2023, doi: 10.1016/j.iot.2023.100966. Available: https://doi.org/10.1016/j.iot.2023.100966

[102] Alsamiri and K. Alsubhi, "Federated Learning for Intrusion Detection Systems in Internet of Vehicles: A General Taxonomy, Applications, and Future Directions," *Future Internet*, vol. 15, no. 12, p. 403, Dec. 2023, doi: 10.3390/fi15120403. Available: https://doi.org/10.3390/fi15120403

[103] M. Imran, H. U. R. Siddiqui, A. Raza, M. A. Raza, F. Rustam, and I. Ashraf, "A performance overview of machine learning-based defense strategies for advanced persistent threats in industrial control systems," *Computers & Security*, vol. 134, p. 103445, Nov. 2023, doi: 10.1016/j.cose.2023.103445. Available: https://doi.org/10.1016/j.cose.2023.103445

[104] R. Mamadaliev, "Artificial intelligence in cybersecurity: enhancing threat detection and mitigation," Jun. 08, 2023. Available: https://archive.interconf.center/index.php/conference-proceeding/article/view/3812

[105] Johnson, C. B. Jones, A. Chavez, and S. Hossain-McKenzie, "SOAR4DER: Security orchestration, automation, and response for distributed energy resources," in *Power systems*, 2023, pp. 387–411. doi: 10.1007/978-3-031-20360-2_16. Available: https://doi.org/10.1007/978-3-031-20360-2_16

[106] J. Chukwu, "Leveraging the MITRE ATT&CK Framework to Enhance Organizations Cyberthreat Detection Procedures," 2023. doi: 10.22215/etd/2023-15855. Available: https://doi.org/10.22215/etd/2023-15855

[107] F. Ilca, O. P. Lucian, and T. Bălan, "Enhancing Cyber-Resilience for Small and Medium-Sized Organizations with Prescriptive Malware Analysis, Detection and Response," *Sensors*, vol.

23, no. 15, p. 6757, Jul. 2023, doi: 10.3390/s23156757. Available: https://doi.org/10.3390/s23156757

[108] L. Stephens, "Research and technology challenges for human data analysts in future safety management systems," *NASA Technical Reports Server (NTRS)*, May 31, 2023. Available: https://ntrs.nasa.gov/citations/20230005989

[109] Mungoli, "Scalable, Distributed AI Frameworks: Leveraging cloud computing for enhanced deep learning performance and efficiency," *arXiv (Cornell University)*, Apr. 2023, doi: 10.48550/arxiv.2304.13738. Available: http://arxiv.org/abs/2304.13738

[110] "Optica Publishing Group." Available: https://opg.optica.org/jocn/fulltext.cfm?uri=jocn-15-8-C155&id=532131

[111] Rane, "Role and challenges of ChaTGPT and similar generative artificial intelligence in finance and accounting," *Social Science Research Network*, Jan. 2023, doi: 10.2139/ssrn.4603206. Available: https://doi.org/10.2139/ssrn.4603206

[112] J. Billings *et al.*, "Post-incident psychosocial interventions after a traumatic incident in the workplace: a systematic review of current research evidence and clinical guidance," *European Journal of Psychotraumatology*, vol. 14, no. 2, Nov. 2023, doi: 10.1080/20008066.2023.2281751. Available: https://doi.org/10.1080/20008066.2023.2281751

[113] S. K. Hassan and A. Ibrahim, "The role of Artificial Intelligence in Cyber Security and Incident Response," *International Journal for Electronic Crime Investigation*, vol. 7, no. 2, Jul. 2023, doi: 10.54692/ijeci.2023.0702154. Available: https://doi.org/10.54692/ijeci.2023.0702154

[114] Mohamed, "Current trends in AI and ML for cybersecurity: A state-of-the-art survey," *Cogent Engineering*, vol. 10, no. 2, Oct. 2023, doi: 10.1080/23311916.2023.2272358. Available: https://doi.org/10.1080/23311916.2023.2272358

[115] M. M. Saeed, R. A. Saeed, M. Abdelhaq, R. Alsaqour, M. K. Hasan, and R. A. Mokhtar, "Anomaly detection in 6G networks using machine learning methods," *Electronics*, vol. 12, no. 15, p. 3300, Jul. 2023, doi: 10.3390/electronics12153300. Available: https://doi.org/10.3390/electronics12153300

[116] Ali, "HuntGPT: Integrating Machine Learning-Based Anomaly Detection and Explainable AI with Large Language Models (LLMs)," *arXiv.org*, Sep. 27, 2023. Available: https://arxiv.org/abs/2309.16021

[117] M. Imran, H. U. R. Siddiqui, A. Raza, M. A. Raza, F. Rustam, and I. Ashraf, "A performance overview of machine learning-based defense strategies for advanced persistent threats in industrial control systems," *Computers & Security*, vol. 134, p. 103445, Nov. 2023, doi: 10.1016/j.cose.2023.103445. Available: https://doi.org/10.1016/j.cose.2023.103445

[118] Gamage, R. Dinalankara, J. Samarabandu, and A. Subasinghe, "A comprehensive survey on the applications of machine learning techniques on maritime surveillance to detect abnormal maritime vessel behaviors," *WMU Journal of Maritime Affairs*, vol. 22, no. 4, pp. 447–477, Jun. 2023, doi: 10.1007/s13437-023-00312-7. Available: https://doi.org/10.1007/s13437-023-00312-7

[119] S. El-Gendy, N. Abdelbaki, and M. A. Azer, "Deep Learning for Enhanced Security in the Internet of Nano Things: A Study on Data Classification for Normal and Abnormal Behavior," *IEEE*, Sep. 2023, doi: 10.1109/miucc58832.2023.10278380. Available: https://doi.org/10.1109/miucc58832.2023.10278380

[120] S. Kaur, A. Bhatia, and N. Tomer, "Using deep learning and big data analytics for managing Cyber-Attacks," in *Advances in systems analysis, software engineering, and high*

performance computing book series*, 2022, pp. 146–178. doi: 10.4018/978-1-6684-5722-1.ch008. Available: https://doi.org/10.4018/978-1-6684-5722-1.ch008

[121] A.K. Tyagi, V. Hemamalini, and G. Soni, "Digital health communication with Artificial Intelligence-Based Cyber Security," in *Advances in medical technologies and clinical practice book series*, 2023, pp. 178–213. doi: 10.4018/978-1-6684-8938-3.ch009. Available: https://doi.org/10.4018/978-1-6684-8938-3.ch009

[122] Z. A. E. Houda, "Cyber Threat Actors Review," in *CRC Press eBooks*, 2023, pp. 84–101. doi: 10.1201/9781003404361-5. Available: https://doi.org/10.1201/9781003404361-5

[123] Koloveas, T. Chantzios, S. Alevizopoulou, S. Skiadopoulos, and C. Tryfonopoulos, "InTIME: a Machine Learning-Based framework for gathering and leveraging web data to Cyber-Threat intelligence," *Electronics*, vol. 10, no. 7, p. 818, Mar. 2021, doi: 10.3390/electronics10070818. Available: https://doi.org/10.3390/electronics10070818

[124] A.K. Kar, P. S. Varsha, and S. Rajan, "Unravelling the impact of Generative Artificial intelligence (GAI) in Industrial Applications: A review of scientific and Grey literature," *Global Journal of Flexible Systems Management*, vol. 24, no. 4, pp. 659–689, Sep. 2023, doi: 10.1007/s40171-023-00356-x. Available: https://doi.org/10.1007/s40171-023-00356-x

[125] M. A. Ferrag, M. Debbah, and M. Al-Hawawreh, "Generative AI for Cyber Threat-Hunting in 6G-enabled IoT networks," *arXiv (Cornell University)*, Mar. 2023, doi: 10.48550/arxiv.2303.11751. Available: http://arxiv.org/abs/2303.11751

[126] X. Shu *et al.*, "Threat Intelligence Computing," *ACM DL*, Oct. 2018, doi: 10.1145/3243734.3243829. Available: https://doi.org/10.1145/3243734.3243829

[127] Chen, Q. V. Liao, J. Vaughan, and G. Bansal, "Understanding the Role of Human Intuition on Reliance in Human-AI Decision-Making with Explanations," *Proceedings of the ACM on Human-computer Interaction*, vol. 7, no. CSCW2, pp. 1–32, Sep. 2023, doi: 10.1145/3610219. Available: https://doi.org/10.1145/3610219

[128] Y. Wang, "Artificial intelligence in educational leadership: a symbiotic role of human-artificial intelligence decision-making," *Journal of Educational Administration*, vol. 59, no. 3, pp. 256–270, Feb. 2021, doi: 10.1108/jea-10-2020-0216. Available: https://doi.org/10.1108/jea-10-2020-0216

[129] S. Qiu, Q. Liu, S. Zhou, and C. Wu, "Review of Artificial intelligence adversarial attack and defense technologies," *Applied Sciences*, vol. 9, no. 5, p. 909, Mar. 2019, doi: 10.3390/app9050909. Available: https://doi.org/10.3390/app9050909

[130] M. Brundage, "The Malicious Use of Artificial intelligence: Forecasting, Prevention, and Mitigation," *arXiv.org*, Feb. 20, 2018. Available: https://arxiv.org/abs/1802.07228

[131] Björk, "Social and economic impacts of Maritime Automated Surface Ships," 2021. Available: https://hdl.handle.net/20.500.12380/302302

[132] "Threat hunting and active cyber defense - ProQuest." Available: https://www.proquest.com/docview/1900172347?pq-origsite=gscholar&fromopenview=true&sourcetype=Dissertations%20&%20Theses

[133] M. R. Fatemi and A. A. Ghorbani, "Threat hunting in Windows using big security log data," in *Advances in information security, privacy, and ethics book series*, 2020, pp. 168–188. doi: 10.4018/978-1-5225-9742-1.ch007. Available: https://doi.org/10.4018/978-1-5225-9742-1.ch007

[134] M. R. Fatemi, "Threat-hunting in Windows environment using host-based log data," 2019. Available: https://unbscholar.lib.unb.ca/items/de0a9682-d75b-4d3f-9c12-9078554140bf

[135] M. A. Lozano, I. P. Llopis, and M. Domingo, "Threat hunting architecture using a machine learning approach for critical infrastructures protection," *Big Data and Cognitive Computing*, vol. 7, no. 2, p. 65, Mar. 2023, doi: 10.3390/bdcc7020065. Available: https://doi.org/10.3390/bdcc7020065

[136] A.H. Nursidiq and C. Lim, "Cyber Threat Hunting to Detect Unknown Threats in the Enterprise Network," *IEEE*, Aug. 2023, doi: 10.1109/icocics58778.2023.10277438. Available: https://doi.org/10.1109/icocics58778.2023.10277438

[137] A.Islam, M. A. Babar, R. Croft, and H. Janicke, "SmartValidator: A framework for automatic identification and classification of cyber threat data," *Journal of Network and Computer Applications*, vol. 202, p. 103370, Jun. 2022, doi: 10.1016/j.jnca.2022.103370. Available: https://doi.org/10.1016/j.jnca.2022.103370

[138] M. N. Al-Mhiqani *et al.*, "A new intelligent multilayer framework for insider threat detection," *Computers & Electrical Engineering*, vol. 97, p. 107597, Jan. 2022, doi: 10.1016/j.compeleceng.2021.107597. Available: https://doi.org/10.1016/j.compeleceng.2021.107597

[139] Europe PMC, "Europe PMC." Available: https://europepmc.org/article/NBK/nbk525302

[140] Johansen, *Digital forensics and incident response: Incident response techniques and procedures to respond to modern cyber threats*. Packt Publishing Ltd, 2020.

[141] Y. Diogenes and E. Ozkaya, *Cybersecurity – attack and defense strategies: Counter modern threats and employ state-of-the-art tools and techniques to protect your organization against cybercriminals*. Packt Publishing Ltd, 2019.

[142] Steingartner, D. Galinec, and A. Kozina, "Threat Defense: Cyber Deception approach and Education for Resilience in Hybrid Threats model," *Symmetry*, vol. 13, no. 4, p. 597, Apr. 2021, doi: 10.3390/sym13040597. Available: https://doi.org/10.3390/sym13040597

[143] Rabieinejad, A. Yazdinejad, R. M. Parizi, and A. Dehghantanha, "Generative adversarial networks for cyber threat hunting in Ethereum blockchain," *Distributed Ledger Technologies Research and Practice*, vol. 2, no. 2, pp. 1–19, Jun. 2023, doi: 10.1145/3584666. Available: https://doi.org/10.1145/3584666

[144] Gupta, S. Tanwar, S. Tyagi, and N. Kumar, "Machine Learning Models for Secure Data Analytics: A taxonomy and threat model," *Computer Communications*, vol. 153, pp. 406–440, Mar. 2020, doi: 10.1016/j.comcom.2020.02.008. Available: https://doi.org/10.1016/j.comcom.2020.02.008

[145] Yuan and X. Wu, "Deep learning for insider threat detection: Review, challenges and opportunities," *Computers & Security*, vol. 104, p. 102221, May 2021, doi: 10.1016/j.cose.2021.102221. Available: https://doi.org/10.1016/j.cose.2021.102221

[146] P. F. Nardulli, S. L. Althaus, and M. Hayes, "A progressive supervised-learning approach to generating rich civil strife data," *Sociological Methodology*, vol. 45, no. 1, pp. 148–183, May 2015, doi: 10.1177/0081175015581378. Available: https://doi.org/10.1177/0081175015581378

[147] M. A. Lozano, I. P. Llopis, and M. Domingo, "Threat hunting architecture using a machine learning approach for critical infrastructures protection," *Big Data and Cognitive Computing*, vol. 7, no. 2, p. 65, Mar. 2023, doi: 10.3390/bdcc7020065. Available: https://doi.org/10.3390/bdcc7020065

[148]     M. Shaukat, R. Amin, M. M. A. Muslam, A. H. Alshehri, and J. Xie, "A hybrid approach for alluring ads phishing attack detection using machine learning," *Sensors*, vol. 23, no. 19, p. 8070, Sep. 2023, doi: 10.3390/s23198070. Available: https://doi.org/10.3390/s23198070

[149]     L. Li, "Application of Machine learning and data mining in Medicine: Opportunities and considerations," in *Artificial intelligence*, 2023. doi: 10.5772/intechopen.113286. Available: https://doi.org/10.5772/intechopen.113286

[150]     Md. A. Talukder *et al.*, "A dependable hybrid machine learning model for network intrusion detection," *Journal of Information Security and Applications*, vol. 72, p. 103405, Feb. 2023, doi: 10.1016/j.jisa.2022.103405. Available: https://doi.org/10.1016/j.jisa.2022.103405

[151]     J. P. Bharadiya, "Machine learning in cybersecurity: Techniques and challenges," *European Journal of Technology*, vol. 7, no. 2, pp. 1–14, Jun. 2023, doi: 10.47672/ejt.1486. Available: https://doi.org/10.47672/ejt.1486

[152]     V. O. Kayhan, M. Agrawal, and S. Shivendu, "Cyber threat detection: Unsupervised hunting of anomalous commands (UHAC)," *Decision Support Systems*, vol. 168, p. 113928, May 2023, doi: 10.1016/j.dss.2023.113928. Available: https://doi.org/10.1016/j.dss.2023.113928

[153]     "International Journal of Computing and Digital Systems," *International Journal of Computing and Digital Systems*, Feb. 2019, doi: 10.12785/ijcds. Available: https://doi.org/10.12785/ijcds

[154]     Apruzzese *et al.*, "The role of machine learning in cybersecurity," *Digital Threats*, vol. 4, no. 1, pp. 1–38, Mar. 2023, doi: 10.1145/3545574. Available: https://doi.org/10.1145/3545574

[155]     M. Muneer, "Cyber Security event detection using machine learning technique," Jun. 30, 2023. Available: http://ijcis.com/index.php/IJCIS/article/view/65

[156]     R. Konatham, "A secure and efficient IIoT anomaly detection approach using a hybrid deep learning technique," 2023. Available: https://etd.ohiolink.edu/acprod/odb_etd/etd/r/1501/10?clear=10&p10_accession_num=wright1693265848788066

[157]     H. Xie, S. Ma, H. Wang, N. Li, J. Zhu, and J. Wang, "Unsupervised clustering for the anomaly diagnosis of plunger lift operations," *Geoenergy Science and Engineering*, vol. 231, p. 212305, Dec. 2023, doi: 10.1016/j.geoen.2023.212305. Available: https://doi.org/10.1016/j.geoen.2023.212305

[158]     M. Bahri, F. Salutari, A. Putina, and M. Sozio, "AutoML: state of the art with a focus on anomaly detection, challenges, and research directions," *International Journal of Data Science and Analytics*, vol. 14, no. 2, pp. 113–126, Feb. 2022, doi: 10.1007/s41060-022-00309-0. Available: https://doi.org/10.1007/s41060-022-00309-0

[159]     R. Bhatia, S. Benno, J. Esteban, T. V. Lakshman, and J. Grogan, "Unsupervised machine learning for network-centric anomaly detection in IoT," *ACM Library*, Dec. 2019, doi: 10.1145/3359992.3366641. Available: https://doi.org/10.1145/3359992.3366641

[160]     A. G. Ribeiro, L. M. Matos, G. Moreira, A. Pilastri, and P. Cortez, "Isolation forests and deep autoencoders for industrial screw tightening anomaly detection," *Computers*, vol. 11, no. 4, p. 54, Apr. 2022, doi: 10.3390/computers11040054. Available: https://doi.org/10.3390/computers11040054

[161]     M. Guerreiro *et al.*, "Anomaly Detection in Automotive Industry Using Clustering Methods—A Case Study," *Applied Sciences*, vol. 11, no. 21, p. 9868, Oct. 2021, doi: 10.3390/app11219868. Available: https://doi.org/10.3390/app11219868

[162] E. G. Lopes and M. De Sevilha Gosling, "Cluster Analysis in Practice: Dealing with Outliers in Managerial Research," *RAC: Revista De Administração Contemporânea*, vol. 25, no. 1, Jan. 2021, doi: 10.1590/1982-7849rac2021200081. Available: https://doi.org/10.1590/1982-7849rac2021200081

[163] J. K. Chow, Z. Su, J. Wu, P. S. Tan, X. Mao, and Y.-H. Wang, "Anomaly detection of defects on concrete structures with the convolutional autoencoder," *Advanced Engineering Informatics*, vol. 45, p. 101105, Aug. 2020, doi: 10.1016/j.aei.2020.101105. Available: https://doi.org/10.1016/j.aei.2020.101105

[164] "Principal component analysis for special types of data," in *Springer eBooks*, 2006, pp. 338–372. doi: 10.1007/0-387-22440-8_13. Available: https://doi.org/10.1007/0-387-22440-8_13

[165] J. Zhang and M. Zulkernine, "Anomaly Based Network Intrusion Detection with Unsupervised Outlier Detection," *2006 IEEE International Conference on Communications*, Jan. 2006, doi: 10.1109/icc.2006.255127. Available: https://doi.org/10.1109/icc.2006.255127

[166] M. S. Chahal, "Harnessing AI and machine learning for intrusion detection in cyber security," *International Journal of Science and Research*, vol. 12, no. 5, pp. 2639–2645, May 2023, doi: 10.21275/sr231003163943. Available: https://doi.org/10.21275/sr231003163943

[167] M. Singh, B. M. Mehtre, S. Sangeetha, and V. Govindaraju, "User Behaviour based Insider Threat Detection using a Hybrid Learning Approach," *Journal of Ambient Intelligence and Humanized Computing*, vol. 14, no. 4, pp. 4573–4593, Mar. 2023, doi: 10.1007/s12652-023-04581-1. Available: https://doi.org/10.1007/s12652-023-04581-1

[168] M. I. Mihăilescu, S. L. Nita, M. Rogobete, and V. Mărăscu, "Unveiling Threats: Leveraging User Behavior Analysis for Enhanced Cybersecurity," *IEEE*, Jun. 2023, doi: 10.1109/ecai58194.2023.10194039. Available: https://doi.org/10.1109/ecai58194.2023.10194039

[169] A.Suresh and A. C. Jose, "Detection of malicious activities by AI-Supported Anomaly-Based IDS," in *Chapman and Hall/CRC eBooks*, 2023, pp. 79–93. doi: 10.1201/9781003346340-4. Available: https://doi.org/10.1201/9781003346340-4

[170] Liang *et al.*, "Outlier-based Anomaly Detection in Firewall Logs," *IEEE*, Oct. 2023, doi: 10.1109/ccci58712.2023.10290797. Available: https://doi.org/10.1109/ccci58712.2023.10290797

[171] A.Li and M. Qiu, *Reinforcement Learning for Cyber-Physical Systems: with Cybersecurity Case Studies*. 2019. Available: https://www.amazon.com/Reinforcement-Learning-Cyber-Physical-Systems-Cybersecurity/dp/1138543535

[172] Huang, L. Huang, and Q. Zhu, "Reinforcement Learning for feedback-enabled cyber resilience," *Annual Reviews in Control*, vol. 53, pp. 273–295, Jan. 2022, doi: 10.1016/j.arcontrol.2022.01.001. Available: https://doi.org/10.1016/j.arcontrol.2022.01.001

[173] Nguyen and V. J. Reddi, "Deep reinforcement learning for cyber security," *IEEE Transactions on Neural Networks and Learning Systems*, vol. 34, no. 8, pp. 3779–3795, Aug. 2023, doi: 10.1109/tnnls.2021.3121870. Available: https://doi.org/10.1109/tnnls.2021.3121870

[174] T. Liliengren, "Threat hunting, definition and framework," *DIVA*, 2018. Available: https://www.diva-portal.org/smash/record.jsf?pid=diva2%3A1205812&dswid=-2084

[175] U. Awan, "Cyber Attack Modelling using Threat Intelligence. An investigation into the use of threat intelligence to model cyber-attacks based on elasticsearch and honeypot data analysis," 2019. Available: https://bradscholars.brad.ac.uk/handle/10454/18672

[176]     A.Fuchs, A. Passarella, and M. Conti, "Modeling, Replicating, and Predicting Human Behavior: A survey," *ACM Transactions on Autonomous and Adaptive Systems*, vol. 18, no. 2, pp. 1–47, May 2023, doi: 10.1145/3580492. Available: https://doi.org/10.1145/3580492

[177]     K. M. Powell, D. Machalek, and T. Quah, "Real-time optimization using reinforcement learning," *Computers & Chemical Engineering*, vol. 143, p. 107077, Dec. 2020, doi: 10.1016/j.compchemeng.2020.107077. Available: https://doi.org/10.1016/j.compchemeng.2020.107077

[178]     "An application of reinforcement learning techniques in traditional pathfinding - ProQuest." Available: https://www.proquest.com/docview/2681415183?pq-origsite=gscholar&fromopenview=true&sourcetype=Dissertations%20&%20Theses

[179]     M. Gautam, "Deep Reinforcement learning for resilient power and energy Systems: progress, prospects, and future avenues," *Electricity*, vol. 4, no. 4, pp. 336–380, Dec. 2023, doi: 10.3390/electricity4040020. Available: https://doi.org/10.3390/electricity4040020

[180]     N. Kaloudi and J. Li, "The AI-Based cyber threat landscape," *ACM Computing Surveys*, vol. 53, no. 1, pp. 1–34, Feb. 2020, doi: 10.1145/3372823. Available: https://doi.org/10.1145/3372823

[181]     Guembe, A. A. Azeta, S. Misra, V. C. Osamor, L. F. Sanz, and V. Pospelova, "The emerging threat of AI-driven cyber attacks: a review," *Applied Artificial Intelligence*, vol. 36, no. 1, Mar. 2022, doi: 10.1080/08839514.2022.2037254. Available: https://doi.org/10.1080/08839514.2022.2037254

[182]     A.Ly and Y.-D. Yao, "A review of deep learning in 5G research: channel coding, massive MIMO, multiple access, resource allocation, and network security," *IEEE Open Journal of the Communications Society*, vol. 2, pp. 396–408, Jan. 2021, doi: 10.1109/ojcoms.2021.3058353. Available: https://doi.org/10.1109/ojcoms.2021.3058353

[183]     M. C. Ghanem and T. Chen, "Reinforcement learning for efficient network penetration testing," *Information*, vol. 11, no. 1, p. 6, Dec. 2019, doi: 10.3390/info11010006. Available: https://doi.org/10.3390/info11010006

[184]     Y. Bai, H. Zhao, X. Zhang, Z. Chang, R. Jäntti, and K. Yang, "Towards Autonomous Multi-UAV Wireless Network: A Survey of Reinforcement Learning-Based Approaches," *IEEE Communications Surveys and Tutorials*, vol. 25, no. 4, pp. 3038–3067, Jan. 2023, doi: 10.1109/comst.2023.3323344. Available: https://doi.org/10.1109/comst.2023.3323344

[185]     Y. Tang *et al.*, "Intelligent Attack Defense Method Based on Confrontational Training Deep Reinforcement Learning Agents," *SSRN*, Jan. 2023, doi: 10.2139/ssrn.4553041. Available: https://doi.org/10.2139/ssrn.4553041

[186]     S. Samtani, M. Abate, V. Benjamin, and W. Liu, "Cybersecurity as an industry: A Cyber Threat Intelligence perspective," in *Springer eBooks*, 2020, pp. 135–154. doi: 10.1007/978-3-319-78440-3_8. Available: https://doi.org/10.1007/978-3-319-78440-3_8

[187]     L. Ignaczak, G. Goldschmidt, C. A. Da Costa, and R. Da Rosa Righi, "Text mining in cybersecurity," *ACM Computing Surveys*, vol. 54, no. 7, pp. 1–36, Jul. 2021, doi: 10.1145/3462477. Available: https://doi.org/10.1145/3462477

[188]     A.Goyal, V. Gupta, and M. Kumar, "Recent Named Entity Recognition and Classification techniques: A systematic review," *Computer Science Review*, vol. 29, pp. 21–43, Aug. 2018, doi: 10.1016/j.cosrev.2018.06.001. Available: https://doi.org/10.1016/j.cosrev.2018.06.001

[189]    K. Pakhale, "Comprehensive overview of named Entity Recognition: Models, Domain-Specific applications and challenges," *arXiv.org*, Sep. 25, 2023. Available: https://arxiv.org/abs/2309.14084

[190]    O. Ukwen and M. Karabatak, "Review of NLP-based Systems in Digital Forensics and Cybersecurity," *IEEE*, Jun. 2021, doi: 10.1109/isdfs52919.2021.9486354. Available: https://doi.org/10.1109/isdfs52919.2021.9486354

[191]    T. Wu *et al.*, "Analysis of trending topics and text-based channels of information delivery in cybersecurity," *ACM Transactions on Internet Technology*, vol. 22, no. 2, pp. 1–27, Oct. 2021, doi: 10.1145/3483332. Available: https://doi.org/10.1145/3483332

[192]    Jo, Y. J. Lee, and S. Shin, "Vulcan: Automatic extraction and analysis of cyber threat intelligence from unstructured text," *Computers & Security*, vol. 120, p. 102763, Sep. 2022, doi: 10.1016/j.cose.2022.102763. Available: https://doi.org/10.1016/j.cose.2022.102763

[193]    Zhao, Q. Yan, J. Li, M. Shao, Z. He, and B. Li, "TIMiner: Automatically extracting and analyzing categorized cyber threat intelligence from social data," *Computers & Security*, vol. 95, p. 101867, Aug. 2020, doi: 10.1016/j.cose.2020.101867. Available: https://doi.org/10.1016/j.cose.2020.101867

[194]    Zhu and T. Dumitraş, "ChainSmith: Automatically Learning the Semantics of Malicious Campaigns by Mining Threat Intelligence Reports," *IEEE*, Apr. 2018, doi: 10.1109/eurosp.2018.00039. Available: https://doi.org/10.1109/eurosp.2018.00039

[195]    H. Sarker, H. Furhad, and R. Nowrozy, "AI-Driven Cybersecurity: An Overview, security intelligence modeling and research directions," *SN Computer Science*, vol. 2, no. 3, Mar. 2021, doi: 10.1007/s42979-021-00557-0. Available: https://doi.org/10.1007/s42979-021-00557-0

[196]    M. A. Wani, "AI and NLP-Empowered Framework for Strengthening Social Cyber Security," in *Advances in data mining and database management book series*, 2023, pp. 32–45. doi: 10.4018/978-1-6684-7216-3.ch002. Available: https://doi.org/10.4018/978-1-6684-7216-3.ch002

[197]    Jain, "Artificial Intelligence in the Cyber Security Environment," *Wiley*, pp. 101–117, Aug. 2021, doi: 10.1002/9781119760429.ch6. Available: https://doi.org/10.1002/9781119760429.ch6

[198]    Z. Kastrati, F. Dalipi, A. S. Imran, K. P. Nuçi, and M. A. Wani, "Sentiment Analysis of Students' Feedback with NLP and Deep Learning: A Systematic Mapping Study," *Applied Sciences*, vol. 11, no. 9, p. 3986, Apr. 2021, doi: 10.3390/app11093986. Available: https://doi.org/10.3390/app11093986

[199]    S. Peng *et al.*, "A survey on deep learning for textual emotion analysis in social networks," *Digital Communications and Networks*, vol. 8, no. 5, pp. 745–762, Oct. 2022, doi: 10.1016/j.dcan.2021.10.003. Available: https://doi.org/10.1016/j.dcan.2021.10.003

[200]    P. V. V. S. Srinivas, K. Gayathri, K. V. Bhavitha, Jahnavi, and K. Sarath, "BLIP-NLP Model for Sentiment Analysis," *IEEE*, Jul. 2023, doi: 10.1109/icecaa58104.2023.10212253. Available: https://doi.org/10.1109/icecaa58104.2023.10212253

[201]    A. Alishboyevich, "A methodological approach to understanding emotional states using textual data," Sep. 22, 2023. Available: https://universalpublishings.com/index.php/jusr/article/view/1988

[202]    L. Tan, C. P. Lee, and K. M. Lim, "A survey of sentiment analysis: Approaches, datasets, and future research," *Applied Sciences*, vol. 13, no. 7, p. 4550, Apr. 2023, doi: 10.3390/app13074550. Available: https://doi.org/10.3390/app13074550

[203] A.Anwar, I. U. Rehman, M. M. Nasralla, S. B. A. Khattak, and N. Khilji, "Emotions Matter: A Systematic Review and Meta-Analysis of the Detection and Classification of Students' Emotions in STEM during Online Learning," *Education Sciences*, vol. 13, no. 9, p. 914, Sep. 2023, doi: 10.3390/educsci13090914. Available: https://doi.org/10.3390/educsci13090914

[204] Kheiri, "SentimentGPT: Exploiting GPT for Advanced Sentiment Analysis and its Departure from Current Machine Learning," *arXiv.org*, Jul. 16, 2023. Available: https://arxiv.org/abs/2307.10234

[205] Punetha and G. Jain, "Optimizing Sentiment Analysis: A Cognitive Approach with Negation Handling via Mathematical Modelling," *Cognitive Computation*, Nov. 2023, doi: 10.1007/s12559-023-10227-3. Available: https://doi.org/10.1007/s12559-023-10227-3

[206] Chaudhary and M. Alam, *AI-Based Data Analytics: Applications for Business Management*. CRC Press, 2023.

[207] K. Kar, S. N. Tripathi, N. Malik, S. Gupta, and U. Sivarajah, "How does misinformation and capricious opinions impact the supply chain - A study on the impacts during the pandemic," *Annals of Operations Research*, vol. 327, no. 2, pp. 713–734, Nov. 2022, doi: 10.1007/s10479-022-04997-6. Available: https://doi.org/10.1007/s10479-022-04997-6

[208] S. Sarraf, A. K. Kushwaha, A. K. Kar, Y. K. Dwivedi, and M. Giannakis, "How did online misinformation impact stockouts in the e-commerce supply chain during COVID-19 – A mixed methods study," *International Journal of Production Economics*, vol. 267, p. 109064, Jan. 2024, doi: 10.1016/j.ijpe.2023.109064. Available: https://doi.org/10.1016/j.ijpe.2023.109064

[209] A.Baz, R. Ahmed, S. R. Khan, and S. Kumar, "Security Risk Assessment Framework for the Healthcare Industry 5.0," *Sustainability*, vol. 15, no. 23, p. 16519, Dec. 2023, doi: 10.3390/su152316519. Available: https://doi.org/10.3390/su152316519

[210] Sarkar and S. K. Shukla, "Behavioral analysis of cybercrime: Paving the way for effective policing strategies," *Journal of Economic Criminology*, vol. 2, p. 100034, Dec. 2023, doi: 10.1016/j.jeconc.2023.100034. Available: https://doi.org/10.1016/j.jeconc.2023.100034

[211] Khang, *AI-aided IoT technologies and applications for smart business and production*. 2024.

[212] Rane, "Integrating Building Information Modelling (BIM) and Artificial intelligence (AI) for smart construction schedule, cost, quality, and safety management: Challenges and opportunities," *Social Science Research Network*, Jan. 2023, doi: 10.2139/ssrn.4616055. Available: https://doi.org/10.2139/ssrn.4616055

[213] S. Parvez, Md. A. Uddin, Md. M. Islam, P. Bharman, and Md. A. Talukder, "Tomato leaf disease detection using convolutional neural network," *Research Square (Research Square)*, Nov. 2023, doi: 10.21203/rs.3.rs-3505828/v1. Available: https://doi.org/10.21203/rs.3.rs-3505828/v1

[214] Xiao, M. Zeng, J. Chen, M. Maimaiti, and Q. Liu, "Recognition of Wheat Leaf Diseases Using Lightweight Convolutional Neural Networks against Complex Backgrounds," *Life*, vol. 13, no. 11, p. 2125, Oct. 2023, doi: 10.3390/life13112125. Available: https://doi.org/10.3390/life13112125

[215] Li, X. Li, M. Chen, and X. A. Sun, "Deep Learning and Image Recognition," *IEEE*, Jul. 2023, doi: 10.1109/iceict57916.2023.10245041. Available: https://doi.org/10.1109/iceict57916.2023.10245041

[216] Y. Wang *et al.*, "Application of optimized convolutional neural networks for early aided diagnosis of essential tremor: Automatic handwriting recognition and feature analysis," *Medical*

*Engineering & Physics*, vol. 113, p. 103962, Mar. 2023, doi: 10.1016/j.medengphy.2023.103962. Available: https://doi.org/10.1016/j.medengphy.2023.103962

[217] Pomazan, "Development of an application for recognizing emotions using convolutional neural networks," 2023. Available: https://openarchive.nure.ua/items/4ed74819-57f4-48a7-942e-15d1bce8363c

[218] Özdemir, "Avg-topk: A new pooling method for convolutional neural networks," *Expert Systems With Applications*, vol. 223, p. 119892, Aug. 2023, doi: 10.1016/j.eswa.2023.119892. Available: https://doi.org/10.1016/j.eswa.2023.119892

[219] S. KM, V. Sowmya, S. K. P, and O. K. Sikha, "AI based rice leaf disease identification enhanced by Dynamic Mode Decomposition," *Engineering Applications of Artificial Intelligence*, vol. 120, p. 105836, Apr. 2023, doi: 10.1016/j.engappai.2023.105836. Available: https://doi.org/10.1016/j.engappai.2023.105836

[220] T. Xu, Y. Chen, Y. Wang, D. Zhang, and M. Zhao, "EMI threat assessment of UAV data link based on Multi-Task CNN," *Electronics*, vol. 12, no. 7, p. 1631, Mar. 2023, doi: 10.3390/electronics12071631. Available: https://doi.org/10.3390/electronics12071631

[221] A. Alabsi, M. Anbar, and A. Saleh, "CNN-CNN: Dual Convolutional Neural Network approach for feature selection and attack detection on Internet of things networks," *Sensors*, vol. 23, no. 14, p. 6507, Jul. 2023, doi: 10.3390/s23146507. Available: https://doi.org/10.3390/s23146507

[222] Moreno, A. Gomez, S. Altares-López, Á. Ribeiro, and D. Andújar, "Analysis of Stable Diffusion-derived fake weeds performance for training Convolutional Neural Networks," *Computers and Electronics in Agriculture*, vol. 214, p. 108324, Nov. 2023, doi: 10.1016/j.compag.2023.108324. Available: https://doi.org/10.1016/j.compag.2023.108324

[223] U. K. Lilhore *et al.*, "Hybrid CNN-LSTM model with efficient hyperparameter tuning for prediction of Parkinson's disease," *Scientific Reports*, vol. 13, no. 1, Sep. 2023, doi: 10.1038/s41598-023-41314-y. Available: https://doi.org/10.1038/s41598-023-41314-y

[224] Faruqui *et al.*, "SafetyMed: A novel IOMT intrusion detection system using CNN-LSTM hybridization," *Electronics*, vol. 12, no. 17, p. 3541, Aug. 2023, doi: 10.3390/electronics12173541. Available: https://doi.org/10.3390/electronics12173541

[225] Shafay, A. Ahmed, T. Hassan, J. Dias, and N. Werghi, "Programmable broad learning system for baggage threat recognition," *Multimedia Tools and Applications*, Jul. 2023, doi: 10.1007/s11042-023-16057-7. Available: https://doi.org/10.1007/s11042-023-16057-7

[226] H. L. Rettore, P. Zißner, M. Alkhowaiter, C. C. Zou, and P. Sevenich, "Military data space: challenges, opportunities, and use cases," *IEEE Communications Magazine*, pp. 1–7, Jan. 2023, doi: 10.1109/mcom.001.2300396. Available: https://doi.org/10.1109/mcom.001.2300396

[227] Fan, Y. Jiang, and A. Mostafavi, "Social Sensing in Disaster City Digital Twin: Integrated Textual–Visual–Geo Framework for Situational Awareness during Built Environment Disruptions," *Journal of Management in Engineering*, vol. 36, no. 3, May 2020, doi: 10.1061/(asce)me.1943-5479.0000745. Available: https://doi.org/10.1061/(asce)me.1943-5479.0000745

[228] Li, "Cyber security meets artificial intelligence: a survey," *Frontiers of Informaion Technology & Electronic Engineering*, vol. 19, no. 12, pp. 1462–1474, Dec. 2018, doi: 10.1631/fitee.1800573. Available: https://doi.org/10.1631/fitee.1800573

[229]  H. Salehinejad, "Recent advances in recurrent neural networks," *arXiv.org*, Dec. 29, 2017. Available: https://arxiv.org/abs/1801.01078

[230]  L. C. Jain and L. R. Medsker, *Recurrent neural networks*. 1999. doi: 10.1201/9781420049176. Available: https://doi.org/10.1201/9781420049176

[231]  S. Qureshi *et al.*, "A hybrid DL-Based detection mechanism for cyber threats in secure networks," *IEEE Access*, vol. 9, pp. 73938–73947, Jan. 2021, doi: 10.1109/access.2021.3081069. Available: https://doi.org/10.1109/access.2021.3081069

[232]  S. S. Oleiwi, G. A. Omran, and H. R. Abdulshaheed, "Detecting anomalies in computer networks recurrent neural networks," *Xinan Jiaotong Daxue Xuebao*, vol. 54, no. 5, Jan. 2019, doi: 10.35741/issn.0258-2724.54.5.12. Available: https://doi.org/10.35741/issn.0258-2724.54.5.12

[233]  U. AlHaddad, A. Basuhail, M. Khemakhem, F. Eassa, and K. Jambi, "Ensemble model based on hybrid deep learning for intrusion detection in smart grid networks," *Sensors*, vol. 23, no. 17, p. 7464, Aug. 2023, doi: 10.3390/s23177464. Available: https://doi.org/10.3390/s23177464

[234]  H. Hopfe, K. M. Lee, and C. Yu, "Short-term forecasting airport passenger flow during periods of volatility: Comparative investigation of time series vs. neural network models," *Journal of Air Transport Management*, vol. 115, p. 102525, Mar. 2024, doi: 10.1016/j.jairtraman.2023.102525. Available: https://doi.org/10.1016/j.jairtraman.2023.102525

[235]  A.Y. a. B. Ahmad, "E-Commerce Trend Analysis and Management for Industry 5.0 using User Data Analysis," Sep. 06, 2023. Available: https://www.ijisae.org/index.php/IJISAE/article/view/3441

[236]  B. Weerakody, K. W. Wong, G. Wang, and W. P. Ela, "A review of irregular time series data handling with gated recurrent neural networks," *Neurocomputing*, vol. 441, pp. 161–178, Jun. 2021, doi: 10.1016/j.neucom.2021.02.046. Available: https://doi.org/10.1016/j.neucom.2021.02.046

[237]  Z. C. Lipton, "A Critical Review of Recurrent Neural Networks for sequence Learning," *arXiv.org*, May 29, 2015. Available: https://arxiv.org/abs/1506.00019

[238]  M. Schmidt, "Recurrent Neural Networks (RNNs): A gentle Introduction and Overview," *arXiv.org*, Nov. 23, 2019. Available: https://arxiv.org/abs/1912.05911

[239]  A.K. Tyagi and A. Abraham, *Recurrent neural networks: Concepts and Applications*. CRC Press, 2022.

[240]  H. Li, H. Jiao, and Z. Yang, "Ship trajectory prediction based on machine learning and deep learning: A systematic review and methods analysis," *Engineering Applications of Artificial Intelligence*, vol. 126, p. 107062, Nov. 2023, doi: 10.1016/j.engappai.2023.107062. Available: https://doi.org/10.1016/j.engappai.2023.107062

[241]  A.H. Ribeiro, K. Tiels, L. A. Aguirre, and T. B. Schön, "Beyond exploding and vanishing gradients: analysing RNN training using attractors and smoothness," *arXiv (Cornell University)*, Jun. 2019, Available: http://export.arxiv.org/pdf/1906.08482

[242]  Z. Shi, M. Xu, Q. Pan, B. Yan, and H. Zhang, "LSTM-based Flight Trajectory Prediction," *IEEE*, Jul. 2018, doi: 10.1109/ijcnn.2018.8489734. Available: https://doi.org/10.1109/ijcnn.2018.8489734

[243]  Khan, M. A. Rawajbeh, L. K. Ramasamy, and S. Lim, "Context-Aware and click Session-Based Graph pattern mining with recommendations for smart EMS through AI," *IEEE Access*, vol. 11, pp. 59854–59865, Jan. 2023, doi: 10.1109/access.2023.3285552. Available: https://doi.org/10.1109/access.2023.3285552

[244]   Kumar, S. K. Sharma, and V. Dutot, "Artificial intelligence (AI)-enabled CRM capability in healthcare: The impact on service innovation," *International Journal of Information Management*, vol. 69, p. 102598, Apr. 2023, doi: 10.1016/j.ijinfomgt.2022.102598. Available: https://doi.org/10.1016/j.ijinfomgt.2022.102598

[245]   Menghani, "Efficient Deep Learning: A survey on making deep learning models smaller, faster, and better," *ACM Computing Surveys*, vol. 55, no. 12, pp. 1–37, Mar. 2023, doi: 10.1145/3578938. Available: https://doi.org/10.1145/3578938

[246]   M. Aach, E. Inanc, R. Sarma, M. Riedel, and A. Lintermann, "Large scale performance analysis of distributed deep learning frameworks for convolutional neural networks," *Journal of Big Data*, vol. 10, no. 1, Jun. 2023, doi: 10.1186/s40537-023-00765-w. Available: https://doi.org/10.1186/s40537-023-00765-w

[247]   Bonde, "Edge, Fog, and Cloud against Disease: The Potential of High-Performance Cloud Computing for Pharma Drug Discovery," in *Methods in molecular biology*, 2023, pp. 181–202. doi: 10.1007/978-1-0716-3449-3_8. Available: https://doi.org/10.1007/978-1-0716-3449-3_8

[248]   Mittal, R. L. Wittman, J. Gibson, J. Huffman, and H. Miller, "Providing a User Extensible Service-Enabled Multi-Fidelity Hybrid Cloud-Deployable SoS Test and Evaluation (T&E) Infrastructure: Application of Modeling and Simulation (M&S) as a Service (MSaaS)," *Information*, vol. 14, no. 10, p. 528, Sep. 2023, doi: 10.3390/info14100528. Available: https://doi.org/10.3390/info14100528

[249]   Z. Zhu, K.-J. Lin, A. K. Jain, and J. Zhou, "Transfer Learning in Deep Reinforcement Learning: a survey," *IEEE Transactions on Pattern Analysis and Machine Intelligence*, vol. 45, no. 11, pp. 13344–13362, Nov. 2023, doi: 10.1109/tpami.2023.3292075. Available: https://doi.org/10.1109/tpami.2023.3292075

[250]   Z. Wang, X. Liu, J. Yu, H. Wu, and H. Lyu, "A general deep transfer learning framework for predicting the flow field of airfoils with small data," *Computers & Fluids*, vol. 251, p. 105738, Jan. 2023, doi: 10.1016/j.compfluid.2022.105738. Available: https://doi.org/10.1016/j.compfluid.2022.105738

[251]   Chen, W. Ge, X. Liang, X. Jin, and Z. Du, "Lifelong learning with deep conditional generative replay for dynamic and adaptive modeling towards net zero emissions target in building energy system," *Applied Energy*, vol. 353, p. 122189, Jan. 2024, doi: 10.1016/j.apenergy.2023.122189. Available: https://doi.org/10.1016/j.apenergy.2023.122189

[252]   Di Ingegneria Dell'Informazione - Dei, "Methodological advancements in continual learning and industry 4.0 applications," Apr. 02, 2023. Available: https://www.research.unipd.it/handle/11577/3478852

[253]   A. Aldhaheri, F. Alwahedi, M. A. Ferrag, and A. Battah, "Deep learning for cyber threat detection in IoT networks: A review," *Internet of Things and Cyber-Physical Systems*, vol. 4, pp. 110–128, Jan. 2024, doi: 10.1016/j.iotcps.2023.09.003. Available: https://doi.org/10.1016/j.iotcps.2023.09.003

[254]   Notovich, H. C. Ben-Gal, and I. Ben-Gal, "Explainable Artificial Intelligence (XAI): motivation, terminology, and taxonomy," in *Springer eBooks*, 2023, pp. 971–985. doi: 10.1007/978-3-031-24628-9_41. Available: https://doi.org/10.1007/978-3-031-24628-9_41

[255]   Sado, C. K. Loo, W. S. Liew, M. Kerzel, and S. Wermter, "Explainable Goal-driven Agents and Robots - A comprehensive review," *ACM Computing Surveys*, vol. 55, no. 10, pp. 1–41, Feb. 2023, doi: 10.1145/3564240. Available: https://doi.org/10.1145/3564240

[256] R. Tiwari, "Explainable AI (XAI) and its Applications in Building Trust and Understanding in AI Decision Making," *Indian Scientific Journal of Research in Engineering and Management*, vol. 07, no. 01, Jan. 2023, doi: 10.55041/ijsrem17592. Available: https://doi.org/10.55041/ijsrem17592

[257] "Evaluation of trust in autonomous systems: Human trust sensing and trustworthy autonomous driving - ProQuest." Available: https://www.proquest.com/docview/2824645257?pq-origsite=gscholar&fromopenview=true&sourcetype=Dissertations%20&%20Theses

[258] V. Chamola, V. Hassija, R. S. A, D. Ghosh, D. Dhingra, and B. Sikdar, "A review of Trustworthy and Explainable Artificial Intelligence (XAI)," *IEEE Access*, vol. 11, pp. 78994–79015, Jan. 2023, doi: 10.1109/access.2023.3294569. Available: https://doi.org/10.1109/access.2023.3294569

[259] U. K. U. Khan, M. Ouaissa, M. Ouaissa, Z. A. E. Houda, and M. F. Ijaz, *Cyber Security for Next-Generation Computing Technologies*. CRC Press, 2024.

[260] L. Sánchez, "Decision support elements and enabling techniques to achieve a cyber defence situational awareness capability," 2023. doi: 10.4995/thesis/10251/194242. Available: https://doi.org/10.4995/thesis/10251/194242

[261] A. Habbal, M. Ali, and M. A. Abuzaraida, "Artificial Intelligence Trust, Risk and Security Management (AI TRiSM): Frameworks, applications, challenges and future research directions," *Expert Systems With Applications*, vol. 240, p. 122442, Apr. 2024, doi: 10.1016/j.eswa.2023.122442. Available: https://doi.org/10.1016/j.eswa.2023.122442

[262] Naithani and S. Tiwari, "Deep learning for the intersection of ethics and privacy in healthcare," in *Advances in systems analysis, software engineering, and high performance computing book series*, 2023, pp. 154–191. doi: 10.4018/978-1-6684-8531-6.ch008. Available: https://doi.org/10.4018/978-1-6684-8531-6.ch008

[263] N. A. Wani, R. Kumar, and J. Bedi, "DeepXplainer: An Interpretable Deep Learning Based Approach for Lung Cancer Detection using Explainable Artificial Intelligence," *Computer Methods and Programs in Biomedicine*, vol. 243, p. 107879, Jan. 2024, doi: 10.1016/j.cmpb.2023.107879. Available: https://doi.org/10.1016/j.cmpb.2023.107879

[264] Hassija *et al.*, "Interpreting Black-Box Models: A review on Explainable Artificial intelligence," *Cognitive Computation*, Aug. 2023, doi: 10.1007/s12559-023-10179-8. Available: https://doi.org/10.1007/s12559-023-10179-8

[265] A.-A. Bouramdane, "Cyberattacks in Smart Grids: Challenges and solving the Multi-Criteria Decision-Making for cybersecurity options, including ones that incorporate artificial intelligence, using an analytical hierarchy process," *Journal of Cybersecurity and Privacy*, vol. 3, no. 4, pp. 662–705, Sep. 2023, doi: 10.3390/jcp3040031. Available: https://doi.org/10.3390/jcp3040031

[266] Saeed and C. W. Omlin, "Explainable AI (XAI): A systematic meta-survey of current challenges and future opportunities," *Knowledge Based Systems*, vol. 263, p. 110273, Mar. 2023, doi: 10.1016/j.knosys.2023.110273. Available: https://doi.org/10.1016/j.knosys.2023.110273

[267] Dixit, "Assessing Methods to Make AI Systems More Transparent through Explainable AI (XAI)," Oct. 10, 2023. Available: https://ijmirm.com/index.php/ijmirm/article/view/48

[268] A.C. Timmons *et al.*, "A call to action on assessing and mitigating bias in artificial intelligence applications for mental health," *Perspectives on Psychological Science*, vol. 18, no. 5,

pp. 1062–1096, Dec. 2022, doi: 10.1177/17456916221134490. Available: https://doi.org/10.1177/17456916221134490

[269] N. A. Wani, R. Kumar, and J. Bedi, "DeepXplainer: An Interpretable Deep Learning Based Approach for Lung Cancer Detection using Explainable Artificial Intelligence," *Computer Methods and Programs in Biomedicine*, vol. 243, p. 107879, Jan. 2024, doi: 10.1016/j.cmpb.2023.107879. Available: https://doi.org/10.1016/j.cmpb.2023.107879

[270] A.Aldoseri, K. Al-Khalifa, and A. M. S. Hamouda, "Re-Thinking Data Strategy and Integration for Artificial intelligence: Concepts, opportunities, and challenges," *Applied Sciences*, vol. 13, no. 12, p. 7082, Jun. 2023, doi: 10.3390/app13127082. Available: https://doi.org/10.3390/app13127082

[271] Yang *et al.*, "Survey on Explainable AI: From approaches, limitations and applications aspects," *Human-Centric Intelligent Systems*, vol. 3, no. 3, pp. 161–188, Aug. 2023, doi: 10.1007/s44230-023-00038-y. Available: https://doi.org/10.1007/s44230-023-00038-y

[272] A.Memarian and T. Doleck, "Fairness, Accountability, Transparency, and Ethics (FATE) in Artificial Intelligence (AI) and higher education: A systematic review," *Computers & Education: Artificial Intelligence*, vol. 5, p. 100152, Jan. 2023, doi: 10.1016/j.caeai.2023.100152. Available: https://doi.org/10.1016/j.caeai.2023.100152

[273] Basir, E. D. Puspitasari, C. C. Aristarini, P. D. Sulastri, and A. M. A. Ausat, "Ethical use of CHATGPT in the context of leadership and strategic decisions," *Jurnal Minfo Polgan*, vol. 12, no. 1, pp. 1239–1246, Jul. 2023, doi: 10.33395/jmp.v12i1.12693. Available: https://doi.org/10.33395/jmp.v12i1.12693

[274] S. E. Whang, Y. Roh, H. Song, and L. Jae-Gil, "Data collection and quality challenges in deep learning: a data-centric AI perspective," *The VLDB Journal*, vol. 32, no. 4, pp. 791–813, Jan. 2023, doi: 10.1007/s00778-022-00775-9. Available: https://doi.org/10.1007/s00778-022-00775-9

[275] A.Brauneck *et al.*, "Federated Machine Learning, Privacy-Enhancing Technologies, and Data Protection Laws in Medical Research: Scoping Review," *Journal of Medical Internet Research*, vol. 25, p. e41588, Mar. 2023, doi: 10.2196/41588. Available: https://doi.org/10.2196/41588

[276] Á. Fernández-Quilez, "Deep learning in radiology: ethics of data and on the value of algorithm transparency, interpretability and explainability," *AI And Ethics*, vol. 3, no. 1, pp. 257–265, Apr. 2022, doi: 10.1007/s43681-022-00161-9. Available: https://doi.org/10.1007/s43681-022-00161-9

[277] S. Grimmelikhuijsen, "Explaining why the computer says no: Algorithmic transparency affects the perceived trustworthiness of automated Decision-Making," *Public Administration Review*, vol. 83, no. 2, pp. 241–262, Jun. 2022, doi: 10.1111/puar.13483. Available: https://doi.org/10.1111/puar.13483

[278] Roganović and M. Radenković, "ETHICAL USE OF AI IN DENTISTRY," in *IntechOpen eBooks*, 2023. doi: 10.5772/intechopen.1001828. Available: https://doi.org/10.5772/intechopen.1001828

[279] Giovanola and S. Tiribelli, "Beyond bias and discrimination: redefining the AI ethics principle of fairness in healthcare machine-learning algorithms," *AI & SOCIETY*, vol. 38, no. 2, pp. 549–563, May 2022, doi: 10.1007/s00146-022-01455-6. Available: https://doi.org/10.1007/s00146-022-01455-6

[280]	G. Vicente and H. Matute, "Humans inherit artificial intelligence biases," *Scientific Reports*, vol. 13, no. 1, Oct. 2023, doi: 10.1038/s41598-023-42384-8. Available: https://doi.org/10.1038/s41598-023-42384-8

[281]	M. Farayola, I. Tal, R. Connolly, T. Saber, and M. Bendechache, "Ethics and Trustworthiness of AI for Predicting the risk of Recidivism: A Systematic literature review," *Information*, vol. 14, no. 8, p. 426, Jul. 2023, doi: 10.3390/info14080426. Available: https://doi.org/10.3390/info14080426

[282]	S. Keleş, "Navigating in the moral landscape: analysing bias and discrimination in AI through philosophical inquiry," *AI And Ethics*, Nov. 2023, doi: 10.1007/s43681-023-00377-3. Available: https://doi.org/10.1007/s43681-023-00377-3

[283]	Mpu, "Bridging the Knowledge Gap on Special Needs Learner Support: The Use of Artificial Intelligence (AI) to Combat Digital Divide Post-COVID-19 Pandemic and beyond – A Comprehensive Literature Review," in *IntechOpen eBooks*, 2023. doi: 10.5772/intechopen.113054. Available: https://doi.org/10.5772/intechopen.113054

[284]	T. P. Pagano *et al.*, "Bias and Unfairness in Machine Learning Models: A Systematic review on datasets, tools, fairness metrics, and identification and mitigation methods," *Big Data and Cognitive Computing*, vol. 7, no. 1, p. 15, Jan. 2023, doi: 10.3390/bdcc7010015. Available: https://doi.org/10.3390/bdcc7010015

[285]	Li, "Ethical Considerations in Artificial Intelligence: A Comprehensive Disccusion from the Perspective of Computer Vision," *SHS Web of Conferences*, vol. 179, p. 04024, Jan. 2023, doi: 10.1051/shsconf/202317904024. Available: https://doi.org/10.1051/shsconf/202317904024

[286]	"The Ethics of Ai and Ml: Balancing Innovation and Responsibility in Business Applications," *European Electronic Letter*, Jan. 2023, doi: 10.52783/eel.v13i5.888. Available: https://doi.org/10.52783/eel.v13i5.888

[287]	M. K. Kamila and S. S. Jasrotia, "Ethical issues in the development of artificial intelligence: recognizing the risks," *International Journal of Ethics and Systems*, Jul. 2023, doi: 10.1108/ijoes-05-2023-0107. Available: https://doi.org/10.1108/ijoes-05-2023-0107

[288]	U. A. Usmani, A. Happonen, and J. Watada, "Human-Centered Artificial Intelligence: Designing for User Empowerment and Ethical Considerations," *IEEE*, Jun. 2023, doi: 10.1109/hora58378.2023.10156761. Available: https://doi.org/10.1109/hora58378.2023.10156761

[289]	A.Lakhani, "AI Revolutionizing Cyber security unlocking the Future of Digital Protection," *OSF*, Jul. 2023, doi: 10.31219/osf.io/cvqx3. Available: https://doi.org/10.31219/osf.io/cvqx3

[290]	I. Piraianu *et al.*, "Enhancing the Evidence with Algorithms: How Artificial Intelligence Is Transforming Forensic Medicine," *Diagnostics*, vol. 13, no. 18, p. 2992, Sep. 2023, doi: 10.3390/diagnostics13182992. Available: https://doi.org/10.3390/diagnostics13182992

[291]	R. Kunduru, "ARTIFICIAL INTELLIGENCE ADVANTAGES IN CLOUD FINTECH APPLICATION SECURITY," *CENTRAL ASIAN JOURNAL OF MATHEMATICAL THEORY AND COMPUTER SCIENCES - Central Asian Studies*, Aug. 14, 2023. Available: https://cajmtcs.centralasianstudies.org/index.php/CAJMTCS/article/view/492

[292]	Şengönül, R. Samet, Q. A. Al-Haija, A. Alqahtani, B. Alturki, and A. A. Alsulami, "An Analysis of Artificial Intelligence Techniques in Surveillance Video Anomaly Detection: A

Comprehensive survey," *Applied Sciences*, vol. 13, no. 8, p. 4956, Apr. 2023, doi: 10.3390/app13084956. Available: https://doi.org/10.3390/app13084956

[293] H. Sarker, "Machine learning for intelligent data analysis and automation in cybersecurity: Current and future Prospects," *Annals of Data Science*, vol. 10, no. 6, pp. 1473–1498, Sep. 2022, doi: 10.1007/s40745-022-00444-2. Available: https://doi.org/10.1007/s40745-022-00444-2

[294] "Machine learning based user modeling for enterprise security and privacy risk mitigation. - ProQuest." Available: https://www.proquest.com/docview/2315246769?pq-origsite=gscholar&fromopenview=true&sourcetype=Dissertations%20&%20Theses

[295] Pirc, D. DeSanto, and I. Davison, *Threat Forecasting: Leveraging big data for predictive analysis*. 2016. Available: https://bnt.execvox.com/book/88833323?_locale=fr

[296] Korte, "Measuring the quality of Open Source Cyber Threat Intelligence Feeds," *Theseus*, 2021. Available: https://www.theseus.fi/handle/10024/500534

[297] Tounsi and H. M. Rais, "A survey on technical threat intelligence in the age of sophisticated cyber attacks," *Computers & Security*, vol. 72, pp. 212–233, Jan. 2018, doi: 10.1016/j.cose.2017.09.001. Available: https://doi.org/10.1016/j.cose.2017.09.001

[298] H. Sarker, "Data-Driven Intelligence can Revolutionize Today's Cybersecurity World: A Position Paper," *arXiv.org*, Aug. 09, 2023. Available: https://arxiv.org/abs/2308.05126

[299] Lucas and B. Moeller, *The Effective Incident Response Team*. Addison-Wesley Professional, 2004.

[300] Kremer, "IC-SECURE: intelligent system for assisting security experts in generating playbooks for automated incident response," *arXiv.org*, Nov. 07, 2023. Available: https://arxiv.org/abs/2311.03825

[301] Islam, M. A. Babar, and S. Nepal, "A Multi-Vocal review of security orchestration," *ACM Computing Surveys*, vol. 52, no. 2, pp. 1–45, Apr. 2019, doi: 10.1145/3305268. Available: https://doi.org/10.1145/3305268

[302] Repetto, "Chaining Digital Services: Challenges to Investigate Cyber-Attacks at Run-Time," *IEEE Communications Magazine*, pp. 1–7, Jan. 2023, doi: 10.1109/mcom.002.2200942. Available: https://doi.org/10.1109/mcom.002.2200942

[303] Hasan, S. Shetty, and S. Ullah, "Artificial Intelligence Empowered Cyber Threat Detection and Protection for Power Utilities," *IEEE*, Dec. 2019, doi: 10.1109/cic48465.2019.00049. Available: https://doi.org/10.1109/cic48465.2019.00049

[304] M. Qumer and S. Ikrama, "Poppy Gustafsson: redefining cybersecurity through AI," *The Case for Women*, pp. 1–38, May 2022, doi: 10.1108/cfw.2022.000001. Available: https://doi.org/10.1108/cfw.2022.000001

[305] Kirov, "Cyber Security Risks and Opportunities of Artificial Intelligence: A Qualitative Study : How AI would form the future of cyber security," *DIVA*, 2023. Available: https://www.diva-portal.org/smash/record.jsf?pid=diva2%3A1775451&dswid=6601

[306] Ahmed, N. Moustafa, A. Barkat, and P. Haskell-Dowland, *Next-Generation Enterprise Security and Governance*. 2022. doi: 10.1201/9781003121541. Available: https://doi.org/10.1201/9781003121541

[307] Rangaraju, "AI SENTRY: REINVENTING CYBERSECURITY THROUGH INTELLIGENT THREAT DETECTION," *EPH - International Journal of Science and Engineering*, vol. 9, no. 3, pp. 30–35, Dec. 2023, doi: 10.53555/ephijse.v9i3.211. Available: https://doi.org/10.53555/ephijse.v9i3.211

[308] R. Montasari, F. Carroll, S. Macdonald, H. Jahankhani, A. Hosseinian-Far, and A. Daneshkhah, "Application of artificial intelligence and machine learning in producing actionable cyber threat intelligence," in *Advanced sciences and technologies for security applications*, 2020, pp. 47–64. doi: 10.1007/978-3-030-60425-7_3. Available: https://doi.org/10.1007/978-3-030-60425-7_3

[309] Langer and R. N. Landers, "The future of artificial intelligence at work: A review on effects of decision automation and augmentation on workers targeted by algorithms and third-party observers," *Computers in Human Behavior*, vol. 123, p. 106878, Oct. 2021, doi: 10.1016/j.chb.2021.106878. Available: https://doi.org/10.1016/j.chb.2021.106878

[310] Leyer and S. Schneider, "Decision augmentation and automation with artificial intelligence: Threat or opportunity for managers?," *Business Horizons*, vol. 64, no. 5, pp. 711–724, Sep. 2021, doi: 10.1016/j.bushor.2021.02.026. Available: https://doi.org/10.1016/j.bushor.2021.02.026

[311] Skopik, *Collaborative Cyber Threat Intelligence: Detecting and Responding to Advanced Cyber Attacks on National Level*. 2017.

[312] Skopik, G. Settanni, and R. Fiedler, "A problem shared is a problem halved: A survey on the dimensions of collective cyber defense through security information sharing," *Computers & Security*, vol. 60, pp. 154–176, Jul. 2016, doi: 10.1016/j.cose.2016.04.003. Available: https://doi.org/10.1016/j.cose.2016.04.003

[313] Felzmann, E. Fosch-Villaronga, C. Lutz, and A. Tamò-Larrieux, "Towards transparency by design for artificial intelligence," *Science and Engineering Ethics*, vol. 26, no. 6, pp. 3333–3361, Nov. 2020, doi: 10.1007/s11948-020-00276-4. Available: https://doi.org/10.1007/s11948-020-00276-4

[314] C. Robinson, "Trust, transparency, and openness: How inclusion of cultural values shapes Nordic national public policy strategies for artificial intelligence (AI)," *Technology in Society*, vol. 63, p. 101421, Nov. 2020, doi: 10.1016/j.techsoc.2020.101421. Available: https://doi.org/10.1016/j.techsoc.2020.101421

[315] J. Al-Saraireh and A. Masarweh, "A novel approach for detecting advanced persistent threats," *Egyptian Informatics Journal*, vol. 23, no. 4, pp. 45–55, Dec. 2022, doi: 10.1016/j.eij.2022.06.005. Available: https://doi.org/10.1016/j.eij.2022.06.005

[316] A.Khraisat, I. Gondal, P. Vamplew, and J. Kamruzzaman, "Survey of intrusion detection systems: techniques, datasets and challenges," *Cybersecurity*, vol. 2, no. 1, Jul. 2019, doi: 10.1186/s42400-019-0038-7. Available: https://doi.org/10.1186/s42400-019-0038-7

[317] Jacobsson, M. Boldt, and B. Carlsson, "A risk analysis of a smart home automation system," *Future Generation Computer Systems*, vol. 56, pp. 719–733, Mar. 2016, doi: 10.1016/j.future.2015.09.003. Available: https://doi.org/10.1016/j.future.2015.09.003

[318] H. H. Ngu, V. Metsis, S. Coyne, B. Chung, R. Pai, and J. Chang, "Personalized Fall Detection System," *IEEE*, Mar. 2020, doi: 10.1109/percomworkshops48775.2020.9156172. Available: https://doi.org/10.1109/percomworkshops48775.2020.9156172

[319] Wu *et al.*, "Cars Talk to Phones: A DSRC Based Vehicle-Pedestrian Safety System," *IEEE*, Sep. 2014, doi: 10.1109/vtcfall.2014.6965898. Available: https://doi.org/10.1109/vtcfall.2014.6965898

[320] P. Одарченко, M. Явич, G. Iashvili, S. Fedushko, and Y. Syerov, "Assessment of security KPIs for 5G network slices for special groups of subscribers," *Big Data and Cognitive Computing*,

vol. 7, no. 4, p. 169, Oct. 2023, doi: 10.3390/bdcc7040169. Available: https://doi.org/10.3390/bdcc7040169

[321] S. A. Chamkar, Y. Maleh, and N. Gherabi, "SOC Analyst Performance Metrics: Towards an optimal performance model," *EDPACS: The EDP Audit, Control, and Security Newsletter Online*, vol. 68, no. 3, pp. 16–29, Sep. 2023, doi: 10.1080/07366981.2023.2259046. Available: https://doi.org/10.1080/07366981.2023.2259046

[322] A.Staves, T. Anderson, H. Balderstone, B. Green, A. Gouglidis, and D. Hutchison, "A Cyber Incident response and recovery framework to support operators of industrial control systems," *International Journal of Critical Infrastructure Protection*, vol. 37, p. 100505, Jul. 2022, doi: 10.1016/j.ijcip.2021.100505. Available: https://doi.org/10.1016/j.ijcip.2021.100505

[323] Kaloudi and J. Li, "The AI-Based cyber threat landscape," *ACM Computing Surveys*, vol. 53, no. 1, pp. 1–34, Feb. 2020, doi: 10.1145/3372823. Available: https://doi.org/10.1145/3372823

[324] Hopkin, *Fundamentals of Risk management: Understanding, evaluating and implementing effective risk management*. 2010. Available: https://cds.cern.ch/record/1500855

[325] Hatzivasilis, I. Papaefstathiou, and C. Manifavas, "Software Security, Privacy, and Dependability: Metrics and Measurement," *IEEE Software*, vol. 33, no. 4, pp. 46–54, Jul. 2016, doi: 10.1109/ms.2016.61. Available: https://doi.org/10.1109/ms.2016.61

[326] Iganibo, M. Albanese, M. Mosko, E. A. Bier, and A. E. Brito, "An attack volume metric," *Security and Privacy*, vol. 6, no. 4, Jan. 2023, doi: 10.1002/spy2.298. Available: https://doi.org/10.1002/spy2.298

[327] S. Schmitt, F. Kandah, and D. Brownell, "Intelligent Threat Hunting in Software-Defined Networking," *IEEE*, Jan. 2019, doi: 10.1109/icce.2019.8661952. Available: https://doi.org/10.1109/icce.2019.8661952

[328] M. Muir and N. Moray, "Trust in automation. Part II. Experimental studies of trust and human intervention in a process control simulation," *Ergonomics*, vol. 39, no. 3, pp. 429–460, Mar. 1996, doi: 10.1080/00140139608964474. Available: https://doi.org/10.1080/00140139608964474

[329] M. Gollwitzer and P. Sheeran, "Implementation Intentions and goal achievement: a meta-analysis of effects and processes," in *Advances in Experimental Social Psychology*, 2006, pp. 69–119. doi: 10.1016/s0065-2601(06)38002-1. Available: https://doi.org/10.1016/s0065-2601(06)38002-1

[330] O'Brien, "Deployment Corrections: An incident response framework for frontier AI models," *arXiv.org*, Sep. 30, 2023. Available: https://arxiv.org/abs/2310.00328

[331] S. Ossenbuhl, J. Steinberger, and H. Baier, "Towards Automated Incident Handling: How to Select an Appropriate Response against a Network-Based Attack?," *IEEE*, May 2015, doi: 10.1109/imf.2015.13. Available: https://doi.org/10.1109/imf.2015.13

[332] A. Francis and B. B. Bekera, "A metric and frameworks for resilience analysis of engineered and infrastructure systems," *Reliability Engineering & System Safety*, vol. 121, pp. 90–103, Jan. 2014, doi: 10.1016/j.ress.2013.07.004. Available: https://doi.org/10.1016/j.ress.2013.07.004

[333] De Grenade *et al.*, "The nexus: reconsidering environmental security and adaptive capacity," *Current Opinion in Environmental Sustainability*, vol. 21, pp. 15–21, Aug. 2016, doi: 10.1016/j.cosust.2016.10.009. Available: https://doi.org/10.1016/j.cosust.2016.10.009

[334] A.PAGPB, R. MH, R. K, R. D, A. Senarathne, and K. Yapa, "WebGuardian: Holistic Approach to address Dynamic Web Application Threat Landscape," *International Research*

*Journal of Innovations in Engineering and Technology*, vol. 07, no. 07, pp. 37–42, Jan. 2023, doi: 10.47001/irjiet/2023.709004. Available: https://doi.org/10.47001/irjiet/2023.709004

[335]　Muhammad, "Integrative cybersecurity: merging zero trust, layered defense, and global standards for a resilient digital future," Nov. 30, 2022. Available: http://ijcst.com.pk/index.php/IJCST/article/view/274

[336]　Zhang *et al.*, "ResTune: Resource Oriented Tuning Boosted by Meta-Learning for Cloud Databases," *ACM*, Jun. 2021, doi: 10.1145/3448016.3457291. Available: https://doi.org/10.1145/3448016.3457291

[337]　H. Rong, H. Zhang, S. Xiao, C. Li, and C. Hu, "Optimizing energy consumption for data centers," *Renewable & Sustainable Energy Reviews*, vol. 58, pp. 674–691, May 2016, doi: 10.1016/j.rser.2015.12.283. Available: https://doi.org/10.1016/j.rser.2015.12.283

[338]　A.Hameed *et al.*, "A survey and taxonomy on energy efficient resource allocation techniques for cloud computing systems," *Computing*, vol. 98, no. 7, pp. 751–774, Jun. 2014, doi: 10.1007/s00607-014-0407-8. Available: https://doi.org/10.1007/s00607-014-0407-8

[339]　R. Najjar, "Redefining Radiology: A review of Artificial intelligence integration in medical imaging," *Diagnostics*, vol. 13, no. 17, p. 2760, Aug. 2023, doi: 10.3390/diagnostics13172760. Available: https://doi.org/10.3390/diagnostics13172760

[340]　M. Aoun, "Understanding the impact of AI-Driven automation on the workflow of radiologists in emergency care settings," Jun. 05, 2019. Available: https://questsquare.org/index.php/JOUNALICET/article/view/11

[341]　Sahiner *et al.*, "Deep learning in medical imaging and radiation therapy," *Medical Physics*, vol. 46, no. 1, Nov. 2018, doi: 10.1002/mp.13264. Available: https://doi.org/10.1002/mp.13264

[342]　Ker, L. Wang, J. P. Rao, and C. C. T. Lim, "Deep learning applications in medical image analysis," *IEEE Access*, vol. 6, pp. 9375–9389, Jan. 2018, doi: 10.1109/access.2017.2788044. Available: https://doi.org/10.1109/access.2017.2788044

[343]　J. Manhas, R. Gupta, and P. P. Roy, "A Review on Automated Cancer Detection in Medical Images using Machine Learning and Deep Learning based Computational Techniques: Challenges and Opportunities," *Archives of Computational Methods in Engineering*, vol. 29, no. 5, pp. 2893–2933, Nov. 2021, doi: 10.1007/s11831-021-09676-6. Available: https://doi.org/10.1007/s11831-021-09676-6

[344]　Kumar, A. Koul, R. Singla, and M. F. Ijaz, "Artificial intelligence in disease diagnosis: a systematic literature review, synthesizing framework and future research agenda," *Journal of Ambient Intelligence and Humanized Computing*, vol. 14, no. 7, pp. 8459–8486, Jan. 2022, doi: 10.1007/s12652-021-03612-z. Available: https://doi.org/10.1007/s12652-021-03612-z

[345]　J. Topol, "High-performance medicine: the convergence of human and artificial intelligence," *Nature Medicine*, vol. 25, no. 1, pp. 44–56, Jan. 2019, doi: 10.1038/s41591-018-0300-7. Available: https://doi.org/10.1038/s41591-018-0300-7

[346]　I. Razzak, S. Naz, and A. Zaib, "Deep learning for medical Image Processing: Overview, challenges and the future," in *Lecture notes in computational vision and biomechanics*, 2017, pp. 323–350. doi: 10.1007/978-3-319-65981-7_12. Available: https://doi.org/10.1007/978-3-319-65981-7_12

[347]　Zhou, M. B. Gotway, and J. Liang, "Interpreting medical images," in *Springer eBooks*, 2022, pp. 343–371. doi: 10.1007/978-3-031-09108-7_12. Available: https://doi.org/10.1007/978-3-031-09108-7_12

[348]     S. Rahi *et al.*, "Fungal infections in hematopoietic stem-cell transplant patients: a review of epidemiology, diagnosis, and management," *Therapeutic Advances in Infectious Disease*, vol. 8, p. 204993612110390, Jan. 2021, doi: 10.1177/20499361211039050. Available: https://doi.org/10.1177/20499361211039050

[349]     G. Curigliano *et al.*, "Cardiotoxicity of anticancer treatments: Epidemiology, detection, and management," *CA: A Cancer Journal for Clinicians*, vol. 66, no. 4, pp. 309–325, Feb. 2016, doi: 10.3322/caac.21341. Available: https://doi.org/10.3322/caac.21341

[350]     Lee and Y. J. Shin, "Machine learning for enterprises: Applications, algorithm selection, and challenges," *Business Horizons*, vol. 63, no. 2, pp. 157–170, Mar. 2020, doi: 10.1016/j.bushor.2019.10.005. Available: https://doi.org/10.1016/j.bushor.2019.10.005

[351]     Fan, Z. Yan, and S. Wen, "Deep Learning and Artificial intelligence in Sustainability: A review of SDGs, renewable energy, and Environmental health," *Sustainability*, vol. 15, no. 18, p. 13493, Sep. 2023, doi: 10.3390/su151813493. Available: https://doi.org/10.3390/su151813493

[352]     E. Edge and P. Sampaio, "A survey of signature based methods for financial fraud detection," *Computers & Security*, vol. 28, no. 6, pp. 381–394, Sep. 2009, doi: 10.1016/j.cose.2009.02.001. Available: https://doi.org/10.1016/j.cose.2009.02.001

[353]     Baesens, V. Van Vlasselaer, and W. Verbeke, *Fraud analytics using descriptive, predictive, and social network techniques*. 2015. doi: 10.1002/9781119146841. Available: https://doi.org/10.1002/9781119146841

[354]     Ryman-Tubb, P. Krause, and W. Garn, "How Artificial Intelligence and machine learning research impacts payment card fraud detection: A survey and industry benchmark," *Engineering Applications of Artificial Intelligence*, vol. 76, pp. 130–157, Nov. 2018, doi: 10.1016/j.engappai.2018.07.008. Available: https://doi.org/10.1016/j.engappai.2018.07.008

[355]     Y. Bao, G. Hilary, and B. Ke, "Artificial Intelligence and Fraud Detection," in *Springer series in supply chain management*, 2022, pp. 223–247. doi: 10.1007/978-3-030-75729-8_8. Available: https://doi.org/10.1007/978-3-030-75729-8_8

[356]     W. Hilal, S. A. Gadsden, and J. Yawney, "Financial Fraud: A review of anomaly detection techniques and recent advances," *Expert Systems With Applications*, vol. 193, p. 116429, May 2022, doi: 10.1016/j.eswa.2021.116429. Available: https://doi.org/10.1016/j.eswa.2021.116429

[357]     A.Nesvijevskaia, S. Ouillade, P. Guilmin, and J.-D. Zucker, "The accuracy versus interpretability trade-off in fraud detection model," *Data & Policy*, vol. 3, Jan. 2021, doi: 10.1017/dap.2021.3. Available: https://doi.org/10.1017/dap.2021.3

[358]     West and M. Bhattacharya, "Intelligent financial fraud detection: A comprehensive review," *Computers & Security*, vol. 57, pp. 47–66, Mar. 2016, doi: 10.1016/j.cose.2015.09.005. Available: https://doi.org/10.1016/j.cose.2015.09.005

[359]     Bhutoria, "Personalized education and Artificial Intelligence in the United States, China, and India: A systematic review using a Human-In-The-Loop model," *Computers & Education: Artificial Intelligence*, vol. 3, p. 100068, Jan. 2022, doi: 10.1016/j.caeai.2022.100068. Available: https://doi.org/10.1016/j.caeai.2022.100068

[360]     G. S. Bain, "The future of management education," *Journal of the Operational Research Society*, vol. 43, no. 6, pp. 557–561, Jun. 1992, doi: 10.1057/jors.1992.81. Available: https://doi.org/10.1057/jors.1992.81

[361] Rane, S. Choudhary, and J. Rane, "Education 4.0 and 5.0: Integrating Artificial Intelligence (AI) for personalized and adaptive learning," *Social Science Research Network*, Jan. 2023, doi: 10.2139/ssrn.4638365. Available: https://doi.org/10.2139/ssrn.4638365

[362] M. Rizvi, "Investigating AI-Powered Tutoring Systems that Adapt to Individual Student Needs, Providing Personalized Guidance and Assessments," *The Eurasia Proceedings of Educational and Social Sciences*, vol. 31, pp. 67–73, Oct. 2023, doi: 10.55549/epess.1381518. Available: https://doi.org/10.55549/epess.1381518

[363] G. Srinivasa, M. Kurni, and K. Saritha, "Harnessing the power of AI to education," in *Springer texts in education*, 2022, pp. 311–342. doi: 10.1007/978-981-19-6734-4_13. Available: https://doi.org/10.1007/978-981-19-6734-4_13

[364] Hooda, C. Rana, O. Dahiya, A. Rizwan, and S. Hossain, "Artificial intelligence for assessment and feedback to enhance student success in higher education," *Mathematical Problems in Engineering*, vol. 2022, pp. 1–19, May 2022, doi: 10.1155/2022/5215722. Available: https://doi.org/10.1155/2022/5215722

[365] Ouyang, L. Zheng, and P. Jiao, "Artificial intelligence in online higher education: A systematic review of empirical research from 2011 to 2020," *Education and Information Technologies*, vol. 27, no. 6, pp. 7893–7925, Feb. 2022, doi: 10.1007/s10639-022-10925-9. Available: https://doi.org/10.1007/s10639-022-10925-9

[366] J. M. Vytasek, A. Patzak, and P. H. Winne, "Analytics for student engagement," in *Intelligent systems reference library*, 2019, pp. 23–48. doi: 10.1007/978-3-030-13743-4_3. Available: https://doi.org/10.1007/978-3-030-13743-4_3

[367] J. Boaler, *Experiencing school mathematics : traditional and reform approaches to teaching and their impact on student learning*. 2002. Available: http://ci.nii.ac.jp/ncid/BA64863485

[368] R. Chaturvedi and S. Verma, "Opportunities and Challenges of AI-Driven Customer Service," in *Springer eBooks*, 2023, pp. 33–71. doi: 10.1007/978-3-031-33898-4_3. Available: https://doi.org/10.1007/978-3-031-33898-4_3

[369] Khan and M. Iqbal, "AI-Powered Customer Service: Does it Optimize Customer Experience?," *IEEE*, Jun. 2020, doi: 10.1109/icrito48877.2020.9198004. Available: https://doi.org/10.1109/icrito48877.2020.9198004

[370] A.Aldoseri, K. Al-Khalifa, and A. M. S. Hamouda, "A Roadmap for Integrating Automation with Process Optimization for AI-powered Digital Transformation," *Pre Print*, Oct. 2023, doi: 10.20944/preprints202310.1055.v1. Available: https://doi.org/10.20944/preprints202310.1055.v1

[371] Rane, "Role and challenges of ChatGPT and similar generative artificial intelligence in business Management," *Social Science Research Network*, Jan. 2023, doi: 10.2139/ssrn.4603227. Available: https://doi.org/10.2139/ssrn.4603227

[372] Castillo, A. I. Canhoto, and E. Said, "The dark side of AI-powered service interactions: exploring the process of co-destruction from the customer perspective," *Service Industries Journal*, vol. 41, no. 13–14, pp. 900–925, Jun. 2020, doi: 10.1080/02642069.2020.1787993. Available: https://doi.org/10.1080/02642069.2020.1787993

[373] S. Fox, M. Barbuceanu, and R. Teigen, "Agent-Oriented Supply-Chain Management," in *Springer eBooks*, 2001, pp. 81–104. doi: 10.1007/978-1-4615-1599-9_5. Available: https://doi.org/10.1007/978-1-4615-1599-9_5

[374]    Ritzer and R. Leidner, "Fast food, fast talk: service work and the routinization of everyday life.," *Contemporary Sociology*, vol. 23, no. 3, p. 430, May 1994, doi: 10.2307/2075368. Available: https://doi.org/10.2307/2075368

[375]    S. Vijay, M. Sharma, and R. Khanna, "Revolutionizing network management with an AI-driven intrusion detection system," *Multidisciplinary Science Journal*, vol. 5, p. 2023ss0313, Aug. 2023, doi: 10.31893/multiscience.2023ss0313. Available: https://doi.org/10.31893/multiscience.2023ss0313

[376]    Chiscop, F. Soro, and P. Smith, "AI-based detection of DNS misuse for network security," *ACM Digital Library*, Dec. 2022, doi: 10.1145/3565009.3569523. Available: https://doi.org/10.1145/3565009.3569523

[377]    F. Aldauiji, O. Batarfi, and M. Bayousef, "Utilizing cyber threat hunting techniques to find ransomware attacks: A survey of the state of the art," *IEEE Access*, vol. 10, pp. 61695–61706, Jan. 2022, doi: 10.1109/access.2022.3181278. Available: https://doi.org/10.1109/access.2022.3181278

[378]    Darabian *et al.*, "A multiview learning method for malware threat hunting: windows, IoT and android as case studies," *World Wide Web*, vol. 23, no. 2, pp. 1241–1260, Jan. 2020, doi: 10.1007/s11280-019-00755-0. Available: https://doi.org/10.1007/s11280-019-00755-0

[379]    M. Imran, H. U. R. Siddiqui, A. Raza, M. A. Raza, F. Rustam, and I. Ashraf, "A performance overview of machine learning-based defense strategies for advanced persistent threats in industrial control systems," *Computers & Security*, vol. 134, p. 103445, Nov. 2023, doi: 10.1016/j.cose.2023.103445. Available: https://doi.org/10.1016/j.cose.2023.103445

[380]    Shon and J. Moon, "A hybrid machine learning approach to network anomaly detection," *Information Sciences*, vol. 177, no. 18, pp. 3799–3821, Sep. 2007, doi: 10.1016/j.ins.2007.03.025. Available: https://doi.org/10.1016/j.ins.2007.03.025

[381]    M. Usama *et al.*, "Unsupervised machine learning for networking: techniques, applications and research challenges," *IEEE Access*, vol. 7, pp. 65579–65615, Jan. 2019, doi: 10.1109/access.2019.2916648. Available: https://doi.org/10.1109/access.2019.2916648

[382]    A.B. Nassif, M. A. Talib, Q. Nasir, and F. Dakalbab, "Machine Learning for Anomaly Detection: A Systematic Review," *IEEE Access*, vol. 9, pp. 78658–78700, Jan. 2021, doi: 10.1109/access.2021.3083060. Available: https://doi.org/10.1109/access.2021.3083060

[383]    Howard, "Artificial intelligence: Implications for the future of work," *American Journal of Industrial Medicine*, vol. 62, no. 11, pp. 917–926, Aug. 2019, doi: 10.1002/ajim.23037. Available: https://doi.org/10.1002/ajim.23037

[384]    S. Nahavandi, "Trusted Autonomy between Humans and Robots: Toward Human-on-the-Loop in Robotics and Autonomous Systems," *IEEE Systems, Man, and Cybernetics Magazine*, vol. 3, no. 1, pp. 10–17, Jan. 2017, doi: 10.1109/msmc.2016.2623867. Available: https://doi.org/10.1109/msmc.2016.2623867

[385]    A.Yeboah-Ofori *et al.*, "Cyber Threat Predictive Analytics for improving cyber Supply chain security," *IEEE Access*, vol. 9, pp. 94318–94337, Jan. 2021, doi: 10.1109/access.2021.3087109. Available: https://doi.org/10.1109/access.2021.3087109

[386]    Pirc, D. DeSanto, and I. Davison, *Threat Forecasting: Leveraging big data for predictive analysis*. 2016. Available: https://bnt.execvox.com/book/88833323?_locale=fr

[387]    Liu, Y. Shi, R. Xiong, and P. Tang, "Quantifying the reliability of defects located by bridge inspectors through human observation behavioral analysis," *Developments in the Built*

*Environment*, vol. 14, p. 100167, Apr. 2023, doi: 10.1016/j.dibe.2023.100167. Available: https://doi.org/10.1016/j.dibe.2023.100167

[388]  B. Singh and R. Jindal, "Trust factor-based analysis of user behavior using sequential pattern mining for detecting intrusive transactions in databases," *The Journal of Supercomputing*, vol. 79, no. 10, pp. 11101–11133, Feb. 2023, doi: 10.1007/s11227-023-05090-w. Available: https://doi.org/10.1007/s11227-023-05090-w

[389]  O. Olaniyi, "Advancing Data-Driven Decision-Making in Smart Cities through Big Data Analytics: A Comprehensive Review of Existing Literature," Aug. 18, 2023. Available: https://papers.ssrn.com/sol3/papers.cfm?abstract_id=4546193

[390]  Dalaklis, N. Nikitakos, D. Papachristos, and A. Dalaklis, "Opportunities and challenges in relation to big data analytics for the shipping and port industries," in *Springer eBooks*, 2023, pp. 267–290. doi: 10.1007/978-3-031-25296-9_14. Available: https://doi.org/10.1007/978-3-031-25296-9_14

[391]  S. Pramanik and S. K. Bandyopadhyay, "Analysis of big data," in *IGI Global eBooks*, 2022, pp. 97–115. doi: 10.4018/978-1-7998-9220-5.ch006. Available: https://doi.org/10.4018/978-1-7998-9220-5.ch006

[392]  A. Naseer, H. Naseer, A. Ahmad, S. B. Maynard, and A. M. Siddiqui, "Moving towards agile cybersecurity incident response: a case study exploring the enabling role of big data analytics-embedded dynamic capabilities," *Computers & Security*, vol. 135, p. 103525, Dec. 2023, doi: 10.1016/j.cose.2023.103525. Available: https://doi.org/10.1016/j.cose.2023.103525

[393]  H. Mobarak *et al.*, "Scope of machine learning in materials research—A review," *Applied Surface Science Advances*, vol. 18, p. 100523, Dec. 2023, doi: 10.1016/j.apsadv.2023.100523. Available: https://doi.org/10.1016/j.apsadv.2023.100523

[394]  "OVERVIEW OF MACHINE LEARNING AND NEURAL NETWORKS," *International Research Journal of Modernization in Engineering Technology and Science*, Nov. 2023, doi: 10.56726/irjmets46037. Available: https://doi.org/10.56726/irjmets46037

[395]  P. Bharadiya, "The role of machine learning in transforming business intelligence," *International Journal of Computing and Artificial Intelligence*, vol. 4, no. 1, pp. 16–24, Jan. 2023, doi: 10.33545/27076571.2023.v4.i1a.60. Available: https://doi.org/10.33545/27076571.2023.v4.i1a.60

[396]  S. Shukla, "Synergizing machine learning and cybersecurity for robust digital protection.," *Research Square (Research Square)*, Nov. 2023, doi: 10.21203/rs.3.rs-3571854/v1. Available: https://doi.org/10.21203/rs.3.rs-3571854/v1

[397]  Wang, Q. Sun, and C. Zhou, "Insider threat detection based on deep clustering of Multi-Source behavioral events," *Applied Sciences*, vol. 13, no. 24, p. 13021, Dec. 2023, doi: 10.3390/app132413021. Available: https://doi.org/10.3390/app132413021

[398]  Cavallaro, V. Cutello, M. Pavone, and F. Zito, "Discovering anomalies in big data: a review focused on the application of metaheuristics and machine learning techniques," *Frontiers in Big Data*, vol. 6, Aug. 2023, doi: 10.3389/fdata.2023.1179625. Available: https://doi.org/10.3389/fdata.2023.1179625

[399]  S. Kurnaz, "Intrusion detection system in internet of things networks using machine learning techniques," 2023. Available: http://openaccess.altinbas.edu.tr/xmlui/handle/20.500.12939/4261

[400] H. Alqahtani and G. Kumar, "Machine learning for enhancing transportation security: A comprehensive analysis of electric and flying vehicle systems," *Engineering Applications of Artificial Intelligence*, vol. 129, p. 107667, Mar. 2024, doi: 10.1016/j.engappai.2023.107667. Available: https://doi.org/10.1016/j.engappai.2023.107667

[401] J. P. Bharadiya, "Leveraging machine learning for enhanced business intelligence," Jul. 06, 2023. Available: http://ijcst.com.pk/index.php/IJCST/article/view/234

[402] O. O. Adebiyi, "Exploring the impact of predictive analytics on accounting and auditing expertise: A regression analysis of LinkedIn survey data," *Social Science Research Network*, Jan. 2023, doi: 10.2139/ssrn.4626506. Available: https://doi.org/10.2139/ssrn.4626506

[403] O. A. Olaniyi, N. Shah, A. I. Abalaka, and F. G. Olaniyi, "Harnessing Predictive Analytics for Strategic Foresight: A comprehensive review of techniques and applications in transforming raw data to actionable insights," *Social Science Research Network*, Jan. 2023, doi: 10.2139/ssrn.4635189. Available: https://doi.org/10.2139/ssrn.4635189

[404] M. Al-Hawawreh, M. Alazab, M. A. Ferrag, and M. S. Hossain, "Securing the Industrial Internet of Things against ransomware attacks: A comprehensive analysis of the emerging threat landscape and detection mechanisms," *Journal of Network and Computer Applications*, p. 103809, Dec. 2023, doi: 10.1016/j.jnca.2023.103809. Available: https://doi.org/10.1016/j.jnca.2023.103809

[405] E. Badidi, "EDGE AI for Early Detection of Chronic Diseases and the Spread of Infectious Diseases: Opportunities, challenges, and future directions," *Future Internet*, vol. 15, no. 11, p. 370, Nov. 2023, doi: 10.3390/fi15110370. Available: https://doi.org/10.3390/fi15110370

[406] A. Khatoon, A. Ullah, and M. Yasir, "Machine Learning-Based Detection and Prevention Systems for IOE," in *Internet of things*, 2023, pp. 109–125. doi: 10.1007/978-3-031-45162-1_7. Available: https://doi.org/10.1007/978-3-031-45162-1_7

[407] Å. A. F. F. N. O. T. Datateknik, "Artificial Intelligence in Cybersecurity and Network security," *Doria*, 2021. Available: https://www.doria.fi/handle/10024/181168

[408] Y. Wu, Y. Qiao, Y. Ye, and B. Lee, "Towards Improved Trust in Threat Intelligence Sharing using Blockchain and Trusted Computing," *IEEE*, Oct. 2019, doi: 10.1109/iotsms48152.2019.8939192. Available: https://doi.org/10.1109/iotsms48152.2019.8939192

[409] F. Casino, E. Πολίτου, E. Alepis, and C. Patsakis, "Immutability and Decentralized Storage: An analysis of Emerging threats," *IEEE Access*, vol. 8, pp. 4737–4744, Jan. 2020, doi: 10.1109/access.2019.2962017. Available: https://doi.org/10.1109/access.2019.2962017

[410] D. M. Mena and B. Yang, "Decentralized actionable cyber threat intelligence for networks and the internet of things," *Iot*, vol. 2, no. 1, pp. 1–16, Dec. 2020, doi: 10.3390/iot2010001. Available: https://doi.org/10.3390/iot2010001

[411] B. Putz, M. Dietz, P. Empl, and G. Pernul, "EtherTwin: Blockchain-based secure Digital twin information Management," *Information Processing and Management*, vol. 58,

no. 1, p. 102425, Jan. 2021, doi: 10.1016/j.ipm.2020.102425. Available: https://doi.org/10.1016/j.ipm.2020.102425

[412] G. Rathee, A. Sharma, R. Kumar, and R. Iqbal, "A Secure Communicating Things Network Framework for Industrial IoT using Blockchain Technology," *Ad Hoc Networks*, vol. 94, p. 101933, Nov. 2019, doi: 10.1016/j.adhoc.2019.101933. Available: https://doi.org/10.1016/j.adhoc.2019.101933

[413] D. Berdik, S. Otoum, N. Schmidt, D. Porter, and Y. Jararweh, "A survey on Blockchain for information systems management and security," *Information Processing and Management*, vol. 58, no. 1, p. 102397, Jan. 2021, doi: 10.1016/j.ipm.2020.102397. Available: https://doi.org/10.1016/j.ipm.2020.102397

[414] A. K. Tyagi, T. T. George, and G. Soni, "Blockchain-Based cybersecurity in Internet of Medical Things (IOMT)-Based assistive systems," in *Advances in medical technologies and clinical practice book series*, 2023, pp. 22–53. doi: 10.4018/978-1-6684-8938-3.ch002. Available: https://doi.org/10.4018/978-1-6684-8938-3.ch002

[415] P. C. Schmid, A. Schaffhauser, and R. Kashef, "IOTBCHAIN: Adopting Blockchain Technology to Increase PLC Resilience in an IoT Environment," *Information*, vol. 14, no. 8, p. 437, Aug. 2023, doi: 10.3390/info14080437. Available: https://doi.org/10.3390/info14080437

[416] S. Dhar, A. Khare, A. D. Dwivedi, and R. Singh, "Securing IoT devices: A novel approach using blockchain and quantum cryptography," *Internet of Things*, vol. 25, p. 101019, Apr. 2024, doi: 10.1016/j.iot.2023.101019. Available: https://doi.org/10.1016/j.iot.2023.101019

[417] V. Parihar, A. Malik, B. Bhushan, P. Bhattacharya, and A. Shankar, "Innovative smart grid solutions for fostering data security and effective privacy preservation," in *Intelligent systems reference library*, 2023, pp. 351–380. doi: 10.1007/978-3-031-46092-0_19. Available: https://doi.org/10.1007/978-3-031-46092-0_19

[418] P. Shelke, N. Sable, S. Dedgaonkar, and R. Mirajkar, "Applications of Blockchain: A Healthcare Use Case," in *Springer link*, 2023, pp. 67–88. doi: 10.1007/978-3-031-45952-8_4. Available: https://doi.org/10.1007/978-3-031-45952-8_4

[419] G. N. Brijwani, P. E. Ajmire, and P. V. Thawani, "Future of quantum computing in cyber security," in *Advances in systems analysis, software engineering, and high performance computing book series*, 2023, pp. 267–298. doi: 10.4018/978-1-6684-6697-1.ch016. Available: https://doi.org/10.4018/978-1-6684-6697-1.ch016

[420] G. Kahanda, V. Patel, M. Parikh, M. Ippolito, M. Solanki, and S. M. M. Ahmed, "The future era of quantum computing," in *Advanced sciences and technologies for security applications*, 2023, pp. 469–484. doi: 10.1007/978-3-031-20160-8_27. Available: https://doi.org/10.1007/978-3-031-20160-8_27

[421] P. Radanliev, "Cyber-attacks on Public Key Cryptography," *Preprint*, Sep. 2023, doi: 10.20944/preprints202309.1769.v1. Available: https://doi.org/10.20944/preprints202309.1769.v1

[422] S. Khan, C. Jain, S. K. Rathi, P. K. Maravi, A. K. Jhapate, and D. Joshi, "Quantum Computing in Data Security," *Wiley*, pp. 369–393, Oct. 2023, doi: 10.1002/9781394167401.ch22. Available: https://doi.org/10.1002/9781394167401.ch22

[423] S. Bajrić, "Enabling secure and trustworthy quantum networks: current State-of-the-Art, key challenges, and potential solutions," *IEEE Access*, vol. 11, pp. 128801–128809, Jan. 2023, doi: 10.1109/access.2023.3333020. Available: https://doi.org/10.1109/access.2023.3333020

[424] R. U. Rasool, H. F. Ahmad, W. Rafique, A. Qayyum, J. Qadir, and Z. Anwar, "Quantum Computing for Healthcare: a review," *Future Internet*, vol. 15, no. 3, p. 94, Feb. 2023, doi: 10.3390/fi15030094. Available: https://doi.org/10.3390/fi15030094

[425] A. Z. Mexriddinovich, "SAFEGUARDING DIGITAL SECURITY: ADDRESSING QUANTUM COMPUTING THREATS," Oct. 06, 2023. Available: https://uzresearchers.com/index.php/RESMD/article/view/873

[426] M.-L. How and S.-M. Cheah, "Business Renaissance: Opportunities and challenges at the dawn of the Quantum Computing Era," *Businesses*, vol. 3, no. 4, pp. 585–605, Nov. 2023, doi: 10.3390/businesses3040036. Available: https://doi.org/10.3390/businesses3040036

[427] P. S. Aithal, "Advances and New Research Opportunities in Quantum Computing Technology by Integrating it with Other ICCT Underlying Technologies," *International Journal of Case Studies in Business, IT, and Education*, pp. 314–358, Sep. 2023, doi: 10.47992/ijcsbe.2581.6942.0304. Available: https://doi.org/10.47992/ijcsbe.2581.6942.0304

[428] A. Mishra, A. V. Jha, B. Appasani, A. K. Ray, D. K. Gupta, and A. N. Ghazali, "Emerging technologies and design aspects of next generation cyber physical system with a smart city application perspective," *International Journal of Systems Assurance Engineering and Management*, vol. 14, no. S3, pp. 699–721, Jan. 2022, doi: 10.1007/s13198-021-01523-y. Available: https://doi.org/10.1007/s13198-021-01523-y

[429] A. Biswas, K. K. Mondal, and D. G. Roy, "A study of smart evolution on AI-Based Cyber-Physical system using blockchain techniques," in *Springer eBooks*, 2023, pp. 327–346. doi: 10.1007/978-3-031-31952-5_14. Available: https://doi.org/10.1007/978-3-031-31952-5_14

[430] S. Nethani, L. Sivaranjani, M. Kumar, B. Lal, and M. Tiwari, "Recognition and Integration of AI with IoT for Innovative Decision Making Techniques in Cyber Physical Systems," *IEEE*, Aug. 2023, doi: 10.1109/icaiss58487.2023.10250493. Available: https://doi.org/10.1109/icaiss58487.2023.10250493

[431] N. Kaloudi and J. Li, "The AI-Based cyber threat landscape," *ACM Computing Surveys*, vol. 53, no. 1, pp. 1–34, Feb. 2020, doi: 10.1145/3372823. Available: https://doi.org/10.1145/3372823

[432] A. H. El-Kady, S. Z. Halim, M. M. El-Halwagi, and F. Khan, "Analysis of safety and security challenges and opportunities related to cyber-physical systems," *Process

Safety and Environmental Protection*, vol. 173, pp. 384–413, May 2023, doi: 10.1016/j.psep.2023.03.012. Available: https://doi.org/10.1016/j.psep.2023.03.012

[433] A. B. Arrieta *et al.*, "Explainable Artificial Intelligence (XAI): Concepts, taxonomies, opportunities and challenges toward responsible AI," *Information Fusion*, vol. 58, pp. 82–115, Jun. 2020, doi: 10.1016/j.inffus.2019.12.012. Available: https://doi.org/10.1016/j.inffus.2019.12.012

[434] Q. V. Liao, "Human-Centered Explainable AI (XAI): from algorithms to user experiences," *arXiv.org*, Oct. 20, 2021. Available: https://arxiv.org/abs/2110.10790

[435] D. L. Minh, H. X. Wang, Y. F. Li, and T. N. Nguyen, "Explainable artificial intelligence: a comprehensive review," *Artificial Intelligence Review*, vol. 55, no. 5, pp. 3503–3568, Nov. 2021, doi: 10.1007/s10462-021-10088-y. Available: https://doi.org/10.1007/s10462-021-10088-y

[436] M. U. Islam, Md. M. Mottalib, M. Hassan, Z. I. Alam, S. Zobaed, and Md. F. Rabby, "The Past, Present, and Prospective Future of XAI: A Comprehensive Review," in *Studies in computational intelligence*, 2022, pp. 1–29. doi: 10.1007/978-3-030-96630-0_1. Available: https://doi.org/10.1007/978-3-030-96630-0_1

[437] H. De Bruijn, M. Warnier, and M. Janssen, "The perils and pitfalls of explainable AI: Strategies for explaining algorithmic decision-making," *Government Information Quarterly*, vol. 39, no. 2, p. 101666, Apr. 2022, doi: 10.1016/j.giq.2021.101666. Available: https://doi.org/10.1016/j.giq.2021.101666

[438] H. Felzmann, E. Fosch-Villaronga, C. Lutz, and A. Tamò-Larrieux, "Transparency you can trust: Transparency requirements for artificial intelligence between legal norms and contextual concerns," *Big Data & Society*, vol. 6, no. 1, p. 205395171986054, Jan. 2019, doi: 10.1177/2053951719860542. Available: https://doi.org/10.1177/2053951719860542

[439] D. Kaur, S. Uslu, K. J. Rittichier, and A. Durresi, "Trustworthy Artificial Intelligence: a review," *ACM Computing Surveys*, vol. 55, no. 2, pp. 1–38, Jan. 2022, doi: 10.1145/3491209. Available: https://doi.org/10.1145/3491209

[440] M. N. O. Sadiku and S. M. Musa, *A primer on multiple intelligences*. 2021. doi: 10.1007/978-3-030-77584-1. Available: https://doi.org/10.1007/978-3-030-77584-1

[441] M. Schranz *et al.*, "Swarm Intelligence and cyber-physical systems: Concepts, challenges and future trends," *Swarm and Evolutionary Computation*, vol. 60, p. 100762, Feb. 2021, doi: 10.1016/j.swevo.2020.100762. Available: https://doi.org/10.1016/j.swevo.2020.100762

[442] O. Zedadra, A. Guerrieri, N. Jouandeau, G. Spezzano, H. Séridi, and G. Fortino, "Swarm intelligence-based algorithms within IoT-based systems: A review," *Journal of Parallel and Distributed Computing*, vol. 122, pp. 173–187, Dec. 2018, doi: 10.1016/j.jpdc.2018.08.007. Available: https://doi.org/10.1016/j.jpdc.2018.08.007

[443] M. Andronie, G. Lăzăroiu, M. Iatagan, C. Uță, R. Ștefănescu, and M. Cocoșatu, "Artificial Intelligence-Based Decision-Making algorithms, internet of things sensing

networks, and Deep Learning-Assisted smart process management in Cyber-Physical production systems," *Electronics*, vol. 10, no. 20, p. 2497, Oct. 2021, doi: 10.3390/electronics10202497. Available: https://doi.org/10.3390/electronics10202497

[444] K. Malhotra, "Application of artificial intelligence in IoT security for crop yield prediction," Oct. 28, 2022. Available: https://researchberg.com/index.php/rrst/article/view/150

[445] R. S. Judson *et al.*, "Aggregating data for computational toxicology applications: The U.S. Environmental Protection Agency (EPA) Aggregated Computational Toxicology Resource (ACTOR) system," *International Journal of Molecular Sciences*, vol. 13, no. 2, pp. 1805–1831, Feb. 2012, doi: 10.3390/ijms13021805. Available: https://doi.org/10.3390/ijms13021805

[446] D. A. Keim, F. Mansmann, J. Schneidewind, J. Thomas, and H. Ziegler, "Visual Analytics: scope and challenges," in *Lecture Notes in Computer Science*, 2008, pp. 76–90. doi: 10.1007/978-3-540-71080-6_6. Available: https://doi.org/10.1007/978-3-540-71080-6_6

[447] T. M. Braun, B. C. M. Fung, F. Iqbal, and B. Shah, "Security and privacy challenges in smart cities," *Sustainable Cities and Society*, vol. 39, pp. 499–507, May 2018, doi: 10.1016/j.scs.2018.02.039. Available: https://doi.org/10.1016/j.scs.2018.02.039

[448] T. Sobb, B. Turnbull, and N. Moustafa, "Supply Chain 4.0: A survey of cyber security challenges, solutions and future directions," *Electronics*, vol. 9, no. 11, p. 1864, Nov. 2020, doi: 10.3390/electronics9111864. Available: https://doi.org/10.3390/electronics9111864

[449] F. Skopik, G. Settanni, and R. Fiedler, "A problem shared is a problem halved: A survey on the dimensions of collective cyber defense through security information sharing," *Computers & Security*, vol. 60, pp. 154–176, Jul. 2016, doi: 10.1016/j.cose.2016.04.003. Available: https://doi.org/10.1016/j.cose.2016.04.003

[450] G. Meng, Y. Liu, J. Zhang, A. Pokluda, and R. Boutaba, "Collaborative Security," *ACM Computing Surveys*, vol. 48, no. 1, pp. 1–42, Jul. 2015, doi: 10.1145/2785733. Available: https://doi.org/10.1145/2785733

[451] M. B. S. Al-Shuhaib and H. O. Hashim, "Mastering DNA chromatogram analysis in Sanger sequencing for reliable clinical analysis," *Journal of Genetic Engineering and Biotechnology*, vol. 21, no. 1, Nov. 2023, doi: 10.1186/s43141-023-00587-6. Available: https://doi.org/10.1186/s43141-023-00587-6

[452] M. N. Zozus, M. G. Kahn, and N. G. Weiskopf, "Data quality in clinical research," in *Computers in health care*, 2023, pp. 169–198. doi: 10.1007/978-3-031-27173-1_10. Available: https://doi.org/10.1007/978-3-031-27173-1_10

[453] S. Thekdi and T. Aven, "A classification system for characterizing the integrity and quality of evidence in risk studies," *Risk Analysis*, Apr. 2023, doi: 10.1111/risa.14153. Available: https://doi.org/10.1111/risa.14153

[454] A. Choudhury and H. Shamszare, "Investigating the impact of user trust on the adoption and use of ChatGPT: Survey analysis," *Journal of Medical Internet Research*, vol. 25, p. e47184, Jun. 2023, doi: 10.2196/47184. Available: https://doi.org/10.2196/47184

[455] N. Rane, "ChatGPT and similar Generative Artificial Intelligence (AI) for building and construction industry: Contribution, Opportunities and Challenges of large language Models for Industry 4.0, Industry 5.0, and Society 5.0," *Social Science Research Network*, Jan. 2023, doi: 10.2139/ssrn.4603221. Available: https://doi.org/10.2139/ssrn.4603221

[456] A. A. Khan, A. A. Laghari, M. Rashid, H. Li, A. R. Javed, and T. R. Gadekallu, "Artificial intelligence and blockchain technology for secure smart grid and power distribution Automation: A State-of-the-Art Review," *Sustainable Energy Technologies and Assessments*, vol. 57, p. 103282, Jun. 2023, doi: 10.1016/j.seta.2023.103282. Available: https://doi.org/10.1016/j.seta.2023.103282

[457] A. Gautam, "The evaluating the impact of artificial intelligence on risk management and fraud detection in the banking sector," Nov. 11, 2023. Available: https://scicadence.com/index.php/AI-IoT-REVIEW/article/view/25

[458] A. K. Tyagi and S. Tiwari, "The future of artificial intelligence in blockchain applications," in *Advances in systems analysis, software engineering, and high performance computing book series*, 2023, pp. 346–373. doi: 10.4018/978-1-6684-8531-6.ch018. Available: https://doi.org/10.4018/978-1-6684-8531-6.ch018

[459] N. Kaloudi and J. Li, "The AI-Based cyber threat landscape," *ACM Computing Surveys*, vol. 53, no. 1, pp. 1–34, Feb. 2020, doi: 10.1145/3372823. Available: https://doi.org/10.1145/3372823

[460] J. R. Vacca, *Online terrorist propaganda, recruitment, and radicalization*. 2019. doi: 10.1201/9781315170251. Available: https://doi.org/10.1201/9781315170251

[461] K. A. O'Brien, "Managing national security and law enforcement intelligence in a globalised world," *Review of International Studies*, vol. 35, no. 4, pp. 903–915, Oct. 2009, doi: 10.1017/s0260210509990349. Available: https://doi.org/10.1017/s0260210509990349

[462] Y. Luo, "A general framework of digitization risks in international business," *Journal of International Business Studies*, vol. 53, no. 2, pp. 344–361, May 2021, doi: 10.1057/s41267-021-00448-9. Available: https://doi.org/10.1057/s41267-021-00448-9

[463] G. Towett, R. S. Snead, K. Grigoryan, and J. Marczika, "Geographical and practical challenges in the implementation of digital health passports for cross-border COVID-19 pandemic management: a narrative review and framework for solutions," *Globalization and Health*, vol. 19, no. 1, Dec. 2023, doi: 10.1186/s12992-023-00998-7. Available: https://doi.org/10.1186/s12992-023-00998-7

[464] A. E. Omolara, A. Alabdulatif, O. I. Abiodun, M. Alawida, W. H. Alshoura, and H. Arshad, "The internet of things security: A survey encompassing unexplored areas and

new insights," *Computers & Security*, vol. 112, p. 102494, Jan. 2022, doi: 10.1016/j.cose.2021.102494. Available: https://doi.org/10.1016/j.cose.2021.102494

[465] P. Malhotra, Y. Singh, P. Anand, D. K. Bangotra, P. Singh, and W. Hong, "Internet of Things: evolution, concerns and security challenges," *Sensors*, vol. 21, no. 5, p. 1809, Mar. 2021, doi: 10.3390/s21051809. Available: https://doi.org/10.3390/s21051809

[466] Y. T. Chua *et al.*, "Identifying Unintended Harms of Cybersecurity Countermeasures," *IEEE*, Nov. 2019, doi: 10.1109/ecrime47957.2019.9037589. Available: https://doi.org/10.1109/ecrime47957.2019.9037589

[467] H. Habibzadeh, B. Nussbaum, F. Anjomshoa, B. Kantarcı, and T. Soyata, "A survey on cybersecurity, data privacy, and policy issues in cyber-physical system deployments in smart cities," *Sustainable Cities and Society*, vol. 50, p. 101660, Oct. 2019, doi: 10.1016/j.scs.2019.101660. Available: https://doi.org/10.1016/j.scs.2019.101660

[468] M. Abdel-Basset, H. Hawash, and K. M. Sallam, "Federated Threat-Hunting Approach for Microservice-Based Industrial Cyber-Physical System," *IEEE Transactions on Industrial Informatics*, vol. 18, no. 3, pp. 1905–1917, Mar. 2022, doi: 10.1109/tii.2021.3091150. Available: https://doi.org/10.1109/tii.2021.3091150

[469] O. Ibitoye, "The threat of adversarial attacks on machine learning in network Security -- a survey," *arXiv.org*, Nov. 06, 2019. Available: https://arxiv.org/abs/1911.02621

[470] A. Kott, "Autonomous Intelligent Cyber-Defense Agent (AICA) Reference Architecture. release 2.0," *arXiv.org*, Mar. 28, 2018. Available: https://arxiv.org/abs/1803.10664

[471] L. Coppolino, S. D′Antonio, G. Mazzeo, and L. Romano, "Cloud security: Emerging threats and current solutions," *Computers & Electrical Engineering*, vol. 59, pp. 126–140, Apr. 2017, doi: 10.1016/j.compeleceng.2016.03.004. Available: https://doi.org/10.1016/j.compeleceng.2016.03.004

[472] S. N. Bhatt, P. K. Manadhata, and L. Zomlot, "The operational role of security information and event management systems," *IEEE Security & Privacy*, vol. 12, no. 5, pp. 35–41, Sep. 2014, doi: 10.1109/msp.2014.103. Available: https://doi.org/10.1109/msp.2014.103

[473] H. M. Farooq and N. M. Otaibi, "Optimal Machine Learning Algorithms for Cyber Threat Detection," *IEEE*, Mar. 2018, doi: 10.1109/uksim.2018.00018. Available: https://doi.org/10.1109/uksim.2018.00018

[474] A. Arfeen, S. B. Ahmed, M. A. Khan, and S. F. A. Jafri, "Endpoint Detection & Response: A Malware Identification Solution," *IEEE*, Nov. 2021, doi: 10.1109/iccws53234.2021.9703010. Available: https://doi.org/10.1109/iccws53234.2021.9703010

[475] A. Seetharaman, N. Patwa, V. Jadhav, A. Saravanan, and D. Sangeeth, "Impact of factors influencing cyber threats on autonomous vehicles," *Applied Artificial Intelligence*,

vol. 35, no. 2, pp. 105–132, Dec. 2020, doi: 10.1080/08839514.2020.1799149. Available: https://doi.org/10.1080/08839514.2020.1799149

[476] E. Yağdereli, C. Gemci, and A. Z. Aktaş, "A study on cyber-security of autonomous and unmanned vehicles," *The Journal of Defense Modeling and Simulation*, vol. 12, no. 4, pp. 369–381, Mar. 2015, doi: 10.1177/1548512915575803. Available: https://doi.org/10.1177/1548512915575803

[477] D. Mitchell *et al.*, "Symbiotic system of systems design for safe and resilient autonomous robotics in offshore wind farms," *IEEE Access*, vol. 9, pp. 141421–141452, Jan. 2021, doi: 10.1109/access.2021.3117727. Available: https://doi.org/10.1109/access.2021.3117727

[478] H. A. Abbass, "Social integration of artificial intelligence: functions, automation allocation logic and Human-Autonomy Trust," *Cognitive Computation*, vol. 11, no. 2, pp. 159–171, Jan. 2019, doi: 10.1007/s12559-018-9619-0. Available: https://doi.org/10.1007/s12559-018-9619-0

[479] 郑南宁 *et al.*, "Hybrid-augmented intelligence: collaboration and cognition," *Frontiers of Informaion Technology & Electronic Engineering*, vol. 18, no. 2, pp. 153–179, Feb. 2017, doi: 10.1631/fitee.1700053. Available: https://doi.org/10.1631/fitee.1700053

[480] A. M. Antoniadi *et al.*, "Current Challenges and Future Opportunities for XAI in Machine Learning-Based Clinical Decision Support Systems: A Systematic review," *Applied Sciences*, vol. 11, no. 11, p. 5088, May 2021, doi: 10.3390/app11115088. Available: https://doi.org/10.3390/app11115088

[481] Y. R. Shrestha, S. M. Ben-Menahem, and G. Von Krogh, "Organizational Decision-Making structures in the age of artificial Intelligence," *California Management Review*, vol. 61, no. 4, pp. 66–83, Jul. 2019, doi: 10.1177/0008125619862257. Available: https://doi.org/10.1177/0008125619862257

[482] R. Gupta, S. Tanwar, S. Tyagi, and N. Kumar, "Machine Learning Models for Secure Data Analytics: A taxonomy and threat model," *Computer Communications*, vol. 153, pp. 406–440, Mar. 2020, doi: 10.1016/j.comcom.2020.02.008. Available: https://doi.org/10.1016/j.comcom.2020.02.008

[483] D. D. Miller and E. W. Brown, "Artificial intelligence in medical practice: the question to the answer?," *The American Journal of Medicine*, vol. 131, no. 2, pp. 129–133, Feb. 2018, doi: 10.1016/j.amjmed.2017.10.035. Available: https://doi.org/10.1016/j.amjmed.2017.10.035

[484] M. Soori, B. Arezoo, and R. Dastres, "Artificial intelligence, machine learning and deep learning in advanced robotics, a review," *Cognitive Robotics*, vol. 3, pp. 54–70, Jan. 2023, doi: 10.1016/j.cogr.2023.04.001. Available: https://doi.org/10.1016/j.cogr.2023.04.001

[485] A. Ovalle, D. Liang, and A. E. Boyd, "Should they? Mobile Biometrics and Technopolicy Meet Queer Community Considerations," *ACM*, Oct. 2023, doi: 10.1145/3617694.3623255. Available: https://doi.org/10.1145/3617694.3623255

[486] D. Garcia, "Algorithms and Decision-Making in military artificial intelligence," *Global Society*, pp. 1–10, Oct. 2023, doi: 10.1080/13600826.2023.2273484. Available: https://doi.org/10.1080/13600826.2023.2273484

[487] M. A. Akbar, A. A. Khan, S. Mahmood, S. Rafi, and S. Demi, "Trustworthy artificial intelligence: A decision-making taxonomy of potential challenges," *Software: Practice and Experience*, May 2023, doi: 10.1002/spe.3216. Available: https://doi.org/10.1002/spe.3216

[488] M. K. Kamila and S. S. Jasrotia, "Ethical issues in the development of artificial intelligence: recognizing the risks," *International Journal of Ethics and Systems*, Jul. 2023, doi: 10.1108/ijoes-05-2023-0107. Available: https://doi.org/10.1108/ijoes-05-2023-0107

[489] S. R. Sindiramutty, N. Z. Jhanjhi, S. K. Ray, H. Jazri, N. A. Khan, and L. Gaur, "Metaverse," in *Advances in medical technologies and clinical practice book series*, 2023, pp. 93–158. doi: 10.4018/978-1-6684-9823-1.ch003. Available: https://doi.org/10.4018/978-1-6684-9823-1.ch003

[490] S. R. Sindiramutty *et al.*, "Metaverse," in *Advances in medical technologies and clinical practice book series*, 2023, pp. 24–92. doi: 10.4018/978-1-6684-9823-1.ch002. Available: https://doi.org/10.4018/978-1-6684-9823-1.ch002

[491] H. Azam *et al.*, "Wireless Technology Security and Privacy: A Comprehensive Study," *Preprint*, Nov. 2023, doi: 10.20944/preprints202311.0664.v1. Available: https://doi.org/10.20944/preprints202311.0664.v1

[492] H. Azam *et al.*, "Defending the digital Frontier: IDPS and the battle against Cyber threat," *International Journal of Emerging Multidisciplinaries Computer Science & Artificial Intelligence*, vol. 2, no. 1, Nov. 2023, doi: 10.54938/ijemdcsai.2023.02.1.253. Available: https://doi.org/10.54938/ijemdcsai.2023.02.1.253

[493] S. Krishnan, R. Thangaveloo, S.-E. B. A. Rahman, and S. R. Sindiramutty, "Smart Ambulance Traffic Control system," *Trends in Undergraduate Research*, vol. 4, no. 1, pp. c28-34, Jun. 2021, doi: 10.33736/tur.2831.2021. Available: https://doi.org/10.33736/tur.2831.2021

[494] H. Azam, M. I. Dulloo, M. H. Majeed, J. P. H. Wan, L. T. Xin, and S. R. Sindiramutty, "Cybercrime Unmasked: Investigating cases and digital evidence," *International Journal of Emerging Multidisciplinaries Computer Science & Artificial Intelligence*, vol. 2, no. 1, Nov. 2023, doi: 10.54938/ijemdcsai.2023.02.1.255. Available: https://doi.org/10.54938/ijemdcsai.2023.02.1.255

[495] F. F. Ananna, R. Nowreen, S. S. R. A. Jahwari, E. Costa, L. Angeline, and S. R. Sindiramutty, "Analysing Influential factors in student academic achievement: Prediction modelling and insight," *International Journal of Emerging Multidisciplinaries Computer Science & Artificial Intelligence*, vol. 2, no. 1, Nov. 2023, doi:

10.54938/ijemdcsai.2023.02.1.254. Available: https://doi.org/10.54938/ijemdcsai.2023.02.1.254

[496]    H. Azam *et al.*, "Innovations in Security: A study of cloud Computing and IoT," *International Journal of Emerging Multidisciplinaries Computer Science & Artificial Intelligence*, vol. 2, no. 1, Nov. 2023, doi: 10.54938/ijemdcsai.2023.02.1.252. Available: https://doi.org/10.54938/ijemdcsai.2023.02.1.252

[497]    V. E. Adeyemo, A. Abdullah, N. Z. Jhanjhi, M. Supramaniam, and A. O. Balogun, "Ensemble and Deep-Learning Methods for Two-Class and Multi-Attack Anomaly Intrusion Detection: An Empirical study," *International Journal of Advanced Computer Science and Applications*, vol. 10, no. 9, 2019. DOI: 10.14569/ijacsa.2019.0100969

[498]    S. Adhikari, T. K. Gangopadhayay, S. Pal, D. Akila, M. Humayun, M. Alfayad, and N. Z. Jhanjhi, "A Novel Machine Learning–Based Hand Gesture Recognition Using HCI on IoT Assisted Cloud Platform," *Computer Systems Science and Engineering*, vol. 46, no. 2, pp. 2123–2140, 2023. DOI: 10.32604/csse.2023.034431

[499]    D. K. Alferidah and N. Jhanjhi, "Cybersecurity Impact over Bigdata and IoT Growth," in *IEEE Explore 2020 International Conference on Computational Intelligence (ICCI)*, 2020. DOI: 10.1109/icci51257.2020.9247722

[500]    I. Ali, N. Z. Jhanjhi, and A. Laraib, "Cybersecurity and blockchain usage in contemporary business," in *Advances in Information Security, Privacy, and Ethics book series*, 2022, pp. 49–64. DOI: 10.4018/978-1-6684-5284-4.ch003

[501]    I. Ali, N. Z. Jhanjhi, F. Amsaad, and A. Razaque, "The role of Cutting-Edge Technologies in Industry 4.0," in *Chapman and Hall/CRC eBooks*, 2022, pp. 97–109. DOI: 10.1201/9781003203087-4

[502]    M. H. Alkinani, A. A. Almazroi, N. Z. Jhanjhi, and N. A. Khan, "5G and IoT Based Reporting and Accident Detection (RAD) System to Deliver First Aid Box Using Unmanned Aerial Vehicle," *Sensors*, vol. 21, no. 20, 2021. DOI: 10.3390/s21206905

[503]    T. Almeida and J. Hidalgo, "UCI Machine Learning Repository," *UC Irvine Machine Learning Repository*, 2012. [Online]. Available: https://archive.ics.uci.edu/dataset/228/sms+spam+collection

[504]    N. Almoysheer, M. Humayun, A. E. Ahmed, and N. Z. Jhanjhi, "Enhancing Cloud Data Security using Multilevel Encryption Techniques Enhancing Cloud Data Security using," *ResearchGate*, 2021. [Online]. Available: https://www.researchgate.net/publication/353237178_Enhancing_Cloud_Data_Security_using_Multilevel_Encryption_Techniques_Enhancing_Cloud_Data_Security_using_Multilevel_Encryption_Techniques

[505]    M. Almulhim, N. I. Islam, and N. Z. Jhanjhi, "A Lightweight and Secure Authentication Scheme for IoT Based E-Health Applications," *International Journal of Computer Science and Network Security*, vol. 19, no. 1, pp. 107–120, 2019. [Online]. Available: http://paper.ijcsns.org/07_book/201901/20190113.pdf

[506]    Z. A. Almusaylim and N. Z. Jhanjhi, "A review on smart home present state and challenges: linked to context-awareness internet of things (IoT)," *Wireless Networks*, vol. 25, no. 6, pp. 3193–3204, 2018. [Online]. Available: https://doi.org/10.1007/s11276-018-1712-5

[507] Z. A. Almusaylim, N. Z. Jhanjhi, and L. T. Jung, "Proposing A Data Privacy Aware Protocol for Roadside Accident Video Reporting Service Using 5G In Vehicular Cloud Networks Environment," *IEEE*, 2018. [Online]. Available: https://doi.org/10.1109/iccoins.2018.8510588

[508] R. Anandan, B. Deepak, G. Suseendran, and N. Z. Jhanjhi, "Internet of Things Platform for Smart Farming," *Wiley*, pp. 337–369, 2021. [Online]. Available: https://doi.org/10.1002/9781119752165.ch13

[509] C. Annadurai, I. Nelson, K. N. Devi, M. Ramachandran, N. Z. Jhanjhi, M. Masud, and A. M. Sheikh, "Biometric Authentication-Based Intrusion Detection using Artificial intelligence internet of things in smart city," *Energies*, vol. 15, no. 19, pp. 7430, 2022. [Online]. Available: https://doi.org/10.3390/en15197430

[510] S. N. Brohi, N. Z. Jhanjhi, N. N. Brohi, and M. N. Brohi, "Key Applications of State-of-the-Art Technologies to Mitigate and Eliminate COVID-19.pdf," *Authorea Preprints*, 2020. [Online]. Available: https://doi.org/10.36227/techrxiv.12115596.v1

[511] S. K. Chaurasiya, A. Biswas, A. Nayyar, N. Z. Jhanjhi, and R. Banerjee, "DEICA: A differential evolution-based improved clustering algorithm for IoT-based heterogeneous wireless sensor networks," *International Journal of Communication Systems*, vol. 36, no. 5, 2023. [Online]. Available: https://doi.org/10.1002/dac.5420

[512] I. A. Chesti, M. Humayun, N. U. Sama, and N. Z. Jhanjhi, "Evolution, Mitigation, and Prevention of Ransomware," in *2020 2nd International Conference on Computer and Information Sciences (ICCIS)*, 2020. [Online]. Available: https://doi.org/10.1109/iccis49240.2020.9257708

[513] V. E. Deyemo, A. Abdullah, N. Z. Jhanjhi, M. Supramaniam, and A. O. Balogun, "Ensemble and Deep-Learning Methods for Two-Class and Multi-Attack Anomaly Intrusion Detection: An Empirical study," *International Journal of Advanced Computer Science and Applications*, vol. 10, no. 9, 2019. [Online]. Available: https://doi.org/10.14569/ijacsa.2019.0100969

[514] C. Diwaker, P. Tomar, A. Solanki, A. Nayyar, N. Z. Jhanjhi, A. Abdullah, and M. Supramaniam, "A new model for predicting Component-Based software reliability using soft computing," *IEEE Access*, vol. 7, pp. 147191–147203, 2019. [Online]. Available: https://doi.org/10.1109/access.2019.2946862

[515] F.-T.-Zahra, N. Z. Jhanjhi, S. N. Brohi, N. A. Malik, and M. Humayun, "Proposing a Hybrid RPL Protocol for Rank and Wormhole Attack Mitigation using Machine Learning," in *2020 2nd International Conference on Computer and Information Sciences (ICCIS)*, 2020. [Online]. Available: https://doi.org/10.1109/iccis49240.2020.9257607

[516] L. Gaur and N. Z. Jhanjhi, "Digital Twins and Healthcare: Trends, Techniques, and Challenges: Trends, Techniques, and Challenges," *IGI Global*, 2022.

[517] L. Gaur, A. Afaq, A. Solanki, G. Singh, S. Sharma, N. Z. Jhanjhi, H. T. My, and D. Le, "Capitalizing on big data and revolutionary 5G technology: Extracting and visualizing ratings and reviews of global chain hotels," *Computers & Electrical Engineering*, vol. 95, pp. 107374, 2021. [Online]. Available: https://doi.org/10.1016/j.compeleceng.2021.107374

[518] L. Gaur, G. Singh, A. Solanki, N. Z. Jhanjhi, U. Bhatia, S. Sharma, S. Verma, Kavita, N. Petrović, F. I. Muhammad, and W. Kim, "Disposition of youth in predicting sustainable development goals using the Neuro-fuzzy and random forest algorithms," *Human-Centric Computing and Information Sciences*. [Online]. Available: https://repository.usp.ac.fj/12807/

[519] G. Ghosh, S. Verma, N. Z. Jhanjhi, and M. Talib, "Secure surveillance system using chaotic image encryption technique," *IOP Conference Series*, vol. 993, no. 1, pp. 012062, 2020. [Online]. Available: https://doi.org/10.1088/1757-899x/993/1/012062

[520] R. Gopi, V. Sathiyamoorthi, S. Selvakumar, M. Ramesh, P. Chatterjee, N. Z. Jhanjhi, and A. K. Luhach, "Enhanced method of ANN based model for detection of DDoS attacks on multimedia internet of things," *Multimedia Tools and Applications*, vol. 81, no. 19, pp. 26739–26757, 2021. [Online]. Available: https://doi.org/10.1007/s11042-021-10640-6

[521] W. Gouda, N. U. Sama, G. Al-Waakid, M. Humayun, and N. Z. Jhanjhi, "Detection of skin cancer based on skin lesion images using deep learning," *Healthcare*, vol. 10, no. 7, pp. 1183, 2022. [Online]. Available: https://doi.org/10.3390/healthcare10071183

[522] B. Hamid, N. Z. Jhanjhi, M. Humayun, A. Khan, and A. Alsayat, "Cyber security issues and challenges for smart cities: A survey," in *2019 13th International Conference on Mathematics, Actuarial Science, Computer Science and Statistics (MACS)*, 2019, pp. 1-7. [Online]. Available: IEEE.

[523] M. Humayun, M. Alsaqer, and N. Z. Jhanjhi, "Energy optimization for smart cities using IoT," *Applied Artificial Intelligence*, vol. 36, no. 1, 2022. [Online]. Available: https://doi.org/10.1080/08839514.2022.2037255

[524] M. Humayun, N. Z. Jhanjhi, A. Alsayat, and V. Ponnusamy, "Internet of things and ransomware: Evolution, mitigation and prevention," *Egyptian Informatics Journal*, vol. 22, no. 1, pp. 105–117, 2021. [Online]. Available: https://doi.org/10.1016/j.eij.2020.05.003

[525] M. Humayun, N. Z. Jhanjhi, B. Hamid, and G. Ahmed, "Emerging smart logistics and transportation using IoT and blockchain," *IEEE Internet of Things Magazine*, vol. 3, no. 2, pp. 58–62, 2020. [Online]. Available: https://doi.org/10.1109/iotm.0001.1900097

[526] M. Humayun, N. Jhanjhi, M. Talib, M. H. Shah, and G. Suseendran, "Cybersecurity for data science: issues, opportunities, and challenges," in *Lecture notes in networks and systems*, pp. 435–444, 2021. [Online]. Available: https://doi.org/10.1007/978-981-16-3153-5_46

[527] M. Humayun, M. Niazi, N. Z. Jhanjhi, M. Alshayeb, and S. Mahmood, "Cyber Security Threats and Vulnerabilities: A Systematic Mapping study," *Arabian Journal for Science and Engineering*, vol. 45, no. 4, pp. 3171–3189, 2020. [Online]. Available: https://doi.org/10.1007/s13369-019-04319-2

[528] . Hussain, S. J. Hussain, N. Z. Jhanjhi, and M. Humayun, "SYN Flood Attack Detection based on Bayes Estimator (SFADBE) For MANET," in *2019 International Conference on Computer and Information Sciences (ICCIS)*, 2019. [Online]. Available: https://doi.org/10.1109/iccisci.2019.8716416

[529] S. J. Hussain, U. Ahmed, H. Liaquat, S. Mir, N. Z. Jhanjhi, and M. Humayun, "IMIAD: Intelligent Malware Identification for Android Platform," in *2019 International

Conference on Computer and Information Sciences (ICCIS)*, 2019. [Online]. Available: https://doi.org/10.1109/iccisci.2019.8716471

[530]	N. Z. Jhanjhi, M. Humayun, and S. N. Almuayqil, "Cyber security and privacy issues in industrial internet of things," *Computer Systems Science and Engineering*, vol. 37, no. 3, pp. 361–380, 2021. [Online]. Available: https://doi.org/10.32604/csse.2021.015206

[531]	A. Khan, N. Z. Jhanjhi, and M. Humayun, "The Role of Cybersecurity in Smart Cities," in *Cyber Security Applications for Industry 4.0*, pp. 195-208, 2022.

[532]	A. Khan, N. Z. Jhanjhi, and M. Humayun, "The role of cybersecurity in smart cities," in *Chapman and Hall/CRC eBooks*, pp. 195–208, 2022. [Online]. Available: https://doi.org/10.1201/9781003203087-9

[533]	A. Khan, N. Z. Jhanjhi, and R. Sujatha, "Emerging Industry Revolution IR 4.0 Issues and Challenges," in *Cyber Security Applications for Industry 4.0*, pp. 151-169, 2022.

[534]	J. Khan, M. A. Khan, N. Jhanjhi, M. Humayun, and A. Alourani, "Smart-City-based data fusion algorithm for internet of things," *Computers, Materials & Continua*, vol. 73, no. 2, pp. 2407–2421, 2022. [Online]. Available: https://doi.org/10.32604/cmc.2022.026693

[535]	K. Khan, "A taxonomy for the use of quantum computing in drone video streaming technology," Zenodo (CERN European Organization for Nuclear Research), 2023. [Online]. Available: https://doi.org/10.5281/zenodo.8143307

[536]	M. A. Khan, I. Khattak, and P. Lorenz, "An efficient and secure Cross-Domain authenticated key agreement scheme for unmanned aerial vehicles," papers.ssrn.com, 2023. [Online]. Available: https://doi.org/10.2139/ssrn.4543069

[537]	N. A. Khan, S. N. Brohi, and N. Z. Jhanjhi, "UAV's Applications, Architecture, Security Issues and Attack Scenarios: A Survey," in *Lecture notes in networks and systems*, pp. 753–760, 2020. [Online]. Available: https://doi.org/10.1007/978-981-15-3284-9_81

[538]	N. A. Khan, N. Z. Jhanjhi, S. N. Brohi, A. A. Almazroi, and A. A. Almazroi, "A secure communication protocol for unmanned aerial vehicles," *Computers, Materials & Continua*, vol. 70, no. 1, pp. 601–618, 2022. [Online]. Available: https://doi.org/10.32604/cmc.2022.019419

[539]	N. A. Khan, N. Jhanjhi, S. N. Brohi, R. S. A. Usmani, and A. Nayyar, "Smart traffic monitoring system using Unmanned Aerial Vehicles (UAVs)," *Computer Communications*, vol. 157, pp. 434–443, 2020. [Online]. Available: https://doi.org/10.1016/j.comcom.2020.04.049

[540]	N. A. Khan, N. Jhanjhi, S. N. Brohi, and Z. A. Almusaylim, "Proposing an algorithm for UAVs interoperability: MAVLink to STANAG 4586 for securing communication," in *Lecture notes in networks and systems*, pp. 413–423, 2021. [Online]. Available: https://doi.org/10.1007/978-981-16-3153-5_44

[541]	N. A. Khan, N. Jhanjhi, S. N. Brohi, and A. Nayyar, "Emerging use of UAV's: secure communication protocol issues and challenges," in *Elsevier eBooks*, pp. 37–55, 2020. [Online]. Available: https://doi.org/10.1016/b978-0-12-819972-5.00003-3

[542]	P. W. Koh and P. Liang, "Understanding black-box predictions via influence functions," in *International Conference on Machine Learning*, pp. 1885–1894, 2017. [Online]. Available: http://proceedings.mlr.press/v70/koh17a/koh17a.pdf

[543]	S. Kok, A. Azween, and N. Z. Jhanjhi, "Evaluation metric for crypto-ransomware detection using machine learning," *Journal of Information Security and Applications*, vol. 55, p. 102646, 2020. [Online]. Available: https://doi.org/10.1016/j.jisa.2020.102646

[544]	M. S. Kumar, S. Vimal, N. Z. Jhanjhi, S. S. Dhanabalan, and H. Alhumyani, "Blockchain based peer to peer communication in autonomous drone operation," *Energy Reports*, vol. 7, pp. 7925–7939, 2021. [Online]. Available: https://doi.org/10.1016/j.egyr.2021.08.073

[545]	Kumar, T., Pandey, B., Mussavi, S. H., & Jhanjhi, N. Z. (2015). CTHS based energy efficient Thermal Aware Image ALU design on FPGA. *Wireless Personal Communications, 85*(3), 671–696. [DOI: 10.1007/s11277-015-2801-8](https://doi.org/10.1007/s11277-015-2801-8)

[546]	Lalouani, W. (2023). Sec-PUF: Securing UAV Swarms Communication with Lightweight Physical Unclonable Functions. *2023 19th International Conference on Wireless and Mobile Computing, Networking and Communications (WiMob)*. [DOI: 10.1109/wimob58348.2023.10187758](https://doi.org/10.1109/wimob58348.2023.10187758)

[547]	Lee, S., Abdullah, A., Jhanjhi, N. Z., & Kok, S. (2021). Classification of botnet attacks in IoT smart factory using honeypot combined with machine learning. *PeerJ, 7*, e350. [DOI: 10.7717/peerj-cs.350](https://doi.org/10.7717/peerj-cs.350)

[548]	Li, J., Goh, W. W., & Jhanjhi, N. Z. (2021). A design of iot-based medicine case for the multi-user medication management using drone in elderly centre. *ResearchGate*. [Link](https://www.researchgate.net/publication/355288512_A_design_of_iot-based_medicine_case_for_the_multi-user_medication_management_using_drone_in_elderly_centre)

[549]	Lim, M., Abdullah, A., & Jhanjhi, N. Z. (2021). Performance optimization of criminal network hidden link prediction model with deep reinforcement learning. *Journal of King Saud University - Computer and Information Sciences, 33*(10), 1202–1210. [DOI: 10.1016/j.jksuci.2019.07.010](https://doi.org/10.1016/j.jksuci.2019.07.010)

[550]	Lim, M., Abdullah, A., Jhanjhi, N. Z., & Khan, M. K. (2020). Situation-Aware Deep Reinforcement Learning link Prediction model for evolving criminal networks. *IEEE Access, 8*, 16550–16559. [DOI: 10.1109/access.2019.2961805](https://doi.org/10.1109/access.2019.2961805)

[551]	Lim, M., Abdullah, A., Jhanjhi, N. Z., & Supramaniam, M. (2019). Hidden link prediction in criminal networks using the deep Reinforcement learning technique. *Computers, 8*(1), 8. [DOI: 10.3390/computers8010008](https://doi.org/10.3390/computers8010008)

[552]	Muzafar, S., & Jhanjhi, N. Z. (2022). DDoS attacks on software defined Network: Challenges and issues. *2022 International Conference on Business Analytics for Technology and Security (ICBATS)*. [DOI: 10.1109/icbats54253.2022.9780662](https://doi.org/10.1109/icbats54253.2022.9780662)

[553]	Muzafar, S., Jhanjhi, N. Z., Khan, N. A., & Ashfaq, F. (2022). DDOS attack detection approaches in on software defined network. *2022 14th International Conference on Mathematics, Actuarial Science, Computer Science and Statistics (MACS)*. [DOI: 10.1109/macs56771.2022.10022653](https://doi.org/10.1109/macs56771.2022.10022653)
[554]	Muzammal, S. M., Murugesan, R. K., Jhanjhi, N. Z., & Jung, L. T. (2020). SMTrust: Proposing Trust-Based Secure Routing Protocol for RPL Attacks for IoT Applications. *2020 International Conference on Computational Intelligence (ICCI)*. [DOI: 10.1109/icci51257.2020.9247818](https://doi.org/10.1109/icci51257.2020.9247818}
[555]	Nanglia, S., Ahmad, M., Khan, F., & Jhanjhi, N. Z. (2022). An enhanced Predictive heterogeneous ensemble model for breast cancer prediction. *Biomedical Signal Processing and Control, 72*, 103279. [DOI: 10.1016/j.bspc.2021.103279](https://doi.org/10.1016/j.bspc.2021.103279)
[556]	Ponnusamy, V., Aun, Y., Jhanjhi, N. Z., Humayun, M., & Almufareh, M. F. (2022). IoT wireless intrusion detection and network Traffic Analysis. *Computer Systems Science and Engineering, 40*(3), 865–879. [DOI: 10.32604/csse.2022.018801](https://doi.org/10.32604/csse.2022.018801)
[557]	Ponnusamy, V., Humayun, M., Jhanjhi, N. Z., Aun, Y., & Almufareh, M. F. (2022). Intrusion detection systems in internet of things and mobile Ad-Hoc networks. *Computer Systems Science and Engineering, 40*(3), 1199–1215. [DOI: 10.32604/csse.2022.018518](https://doi.org/10.32604/csse.2022.018518)
[558]	Ponnusamy, V., Jhanjhi, N. Z., & Humayun, M. (2020). Fostering Public-Private partnership. In *IGI Global eBooks* (pp. 237–255). [DOI: 10.4018/978-1-7998-1851-9.ch012](https://doi.org/10.4018/978-1-7998-1851-9.ch012)
[559]	R. Sujatha, G. Prakash, Noor Zaman Jhanjhi (2022) *Cyber Security Applications for Industry 4.0*, Chapman and Hall/CRC Cyber-Physical Systems Series, CRC Press.
[560]	Saeed, S., Almuhaideb, A. M., Kumar, N., Jhanjhi, N., & Zikria, Y. B. (2022). Cybersecurity and Blockchain Usage in Contemporary Business. In *Handbook of Research on Cybersecurity Issues and Challenges for Business and FinTech Applications* (pp. 49–64). IGI Global. [Link](https://www.igi-global.com/chapter/cybersecurity-and-blockchain-usage-in-contemporary-business/314074)
[561]	Saleh, M., Jhanjhi, N. Z., Abdullah, A., & Fatima-Tuz-Zahra. (2020). Proposing a Privacy Protection Model in Case of Civilian Drone. *IEEE Explore International Conference on Advanced Communication Technology (ICACT)*. [DOI: 10.23919/icact48636.2020.9061508](https://doi.org/10.23919/icact48636.2020.9061508)
[562]	Sankar, S., Ramasubbareddy, S., Luhach, A. K., Deverajan, G. G., Alnumay, W. S., Jhanjhi, N. Z., Ghosh, U., & Sharma, P. K. (2020). Energy efficient optimal parent selection based routing protocol for Internet of Things using firefly optimization algorithm. *Transactions on Emerging Telecommunications Technologies, 32*(8). [DOI: 10.1002/ett.4171](https://doi.org/10.1002/ett.4171)
[563]	Seong, T. B., Ponnusamy, V., Jhanjhi, N. Z., Annur, R., & Talib, M. (2021). A comparative analysis on traditional wired datasets and the need for wireless datasets for IoT wireless intrusion detection. *Indonesian Journal of Electrical Engineering and

[564] SH, K., Abdullah, A., Jhanjhi, N., & Supramaniam, M. (2019). Ransomware, Threat and Detection Techniques: A Review. *Journal of Computer Science and Network Security, 19*(2), 136–146. [Link to PDF](http://paper.ijcsns.org/07_book/201902/20190217.pdf)

[565] M. Shafiq, H. Ashraf, A. Ullah, M. Masud, M. Azeem, N. Z. Jhanjhi, and M. Humayun, "Robust Cluster-Based routing Protocol for IoT-Assisted smart devices in WSN," *Computers, Materials & Continua*, vol. 67, no. 3, pp. 3505–3521, 2021. [Online]. Available: https://doi.org/10.32604/cmc.2021.015533

[566] I. A. Shah, N. Z. Jhanjhi, F. Amsaad, and A. Razaque, "The Role of Cutting-Edge Technologies in Industry 4.0," in *Cyber Security Applications for Industry 4.0*, Chapman and Hall/CRC, 2022, pp. 97–109.

[567] S. R. Sindiramutty, N. Z. Jhanjhi, S. K. Ray, H. Jazri, N. A. Khan, and L. Gaur, "Metaverse: Virtual Meditation," in *Metaverse Applications for Intelligent Healthcare*, IGI Global, 2024, pp. 93–158. [Online]. Available: https://doi.org/10.4018/978-1-6684-9823-1.ch003

[568] S. R. Sindiramutty et al., "Metaverse: Virtual Gyms and Sports," in *Metaverse Applications for Intelligent Healthcare*, IGI Global, 2024, pp. 24–92. [Online]. Available: https://doi.org/10.4018/978-1-6684-9823-1.ch003

[569] V. Singhal et al., "Artificial Intelligence Enabled Road Vehicle-Train Collision Risk Assessment Framework for Unmanned railway level crossings," *IEEE Access*, vol. 8, pp. 113790–113806, 2020. [Online]. Available: https://doi.org/10.1109/access.2020.3002416

[570] R. Sujatha, G. Prakash, and N. Z. Jhanjhi, Cyber Security Applications for Industry 4.0, CRC Press, 2022.

[571] I. Taj and N. Jhanjhi, "Towards Industrial Revolution 5.0 and Explainable Artificial Intelligence: Challenges and Opportunities," *International Journal of Computing and Digital Systems*, vol. 12, no. 1, pp. 295–320, 2022. [Online]. Available: https://doi.org/10.12785/ijcds/120128

[572] A. Ullah, M. Azeem, H. Ashraf, A. A. Alaboudi, M. Humayun, and N. Jhanjhi, "Secure Healthcare Data Aggregation and Transmission in IoT—A Survey," *IEEE Access*, vol. 9, pp. 16849–16865, 2021. [Online]. Available: https://doi.org/10.1109/access.2021.3052850

[573] S. Y. B., A. S., K. A. M., and Z. Jhanjhi, *Handbook of Research on Cybersecurity Issues and Challenges for Business and FinTech Applications*, IGI Global, 2022. [Online]. Available: https://doi.org/10.4018/978-1-6684-5284-4

[574] F. Zahra, N. Jhanjhi, N. A. Khan, S. N. Brohi, M. Masud, and S. Aljahdali, "Protocol-Specific and Sensor Network-Inherited Attack Detection in IoT Using Machine Learning," *Applied Sciences*, vol. 12, no. 22, p. 11598, 2022. [Online]. Available: https://doi.org/10.3390/app122211598

Computer Science, 22*(2), 1165. [DOI: 10.11591/ijeecs.v22.i2.pp1165-1176](https://doi.org/10.11591/ijeecs.v22.i2.pp1165-1176)

[575] "Cybersecurity Threats Fast-Forward 2030: Fasten your Security-Belt Before the Ride!," *ENISA*. Available: https://www.enisa.europa.eu/news/cybersecurity-threats-fast-forward-2030